\documentclass[12pt]{article}

\usepackage[x11names]{xcolor}
\usepackage{tikz}
\usepackage{cite}
\usepackage{comment}
\usepackage[colorlinks,allcolors=black]{hyperref}
\usetikzlibrary{decorations.pathreplacing,decorations.markings}
\usepgflibrary{arrows}
\usepgflibrary[arrows]
\usetikzlibrary{arrows}
\usetikzlibrary[arrows]
\usetikzlibrary{positioning}
\usepackage{graphicx}
\usepackage{amsmath}
\usepackage{amssymb}
\usepackage{subsupscripts}
\usepackage{braket}
\usepackage{hyperref}
\usepackage{amsthm}
\usepackage{xcolor}

\makeatletter

\@addtoreset{equation}{section}
\makeatother
\newcommand{\ba}{\begin{eqnarray}}
\newcommand{\ea}{\end{eqnarray}}

\newcommand{\bpsi}{\bar\psi}

\usetikzlibrary{decorations.pathreplacing,decorations.markings}
\tikzset{
  on each segment/.style={
    decorate,
    decoration={
      show path construction,
      moveto code={},
      lineto code={
        \path [#1]
        (\tikzinputsegmentfirst) -- (\tikzinputsegmentlast);
      },
      curveto code={
        \path [#1] (\tikzinputsegmentfirst)
        .. controls
        (\tikzinputsegmentsupporta) and (\tikzinputsegmentsupportb)
        ..
        (\tikzinputsegmentlast);
      },
      closepath code={
        \path [#1]
        (\tikzinputsegmentfirst) -- (\tikzinputsegmentlast);
      },
    },
  },
  mid arrow/.style={postaction={decorate,decoration={
        markings,
        mark=at position .5 with {\arrow[#1]{stealth}}
      }}},
}

\DeclareMathOperator*{\tox}{\longrightarrow}
\DeclareMathOperator*{\Sym}{\text{Sym}}

\DeclareMathOperator*{\End}{\text{End}}

\newcommand{\tr}{{\rm tr~}}

\newcommand{\dket}[1]{\ket{#1}\!\rangle}
\newcommand{\dbra}[1]{\langle\!\bra{#1}}
\newcommand{\superp}[2]{\genfrac{}{}{0pt}{}{#1}{#2}}

 \def\d{\delta}

 \def\p{\partial}
 
 \def\a{\alpha}
 \def\b{\beta}
 \def\g{\gamma}
 \def\d{\delta}
 \def\e{\varepsilon}
 
 \def\th{\theta}
 
 \def\k{\kappa}
 \def\l{\lambda}
 \def\m{\mu}

 \def\r{\rho}

 \def\t{\tau}
 \def\th{\theta}

 \def\D{\Delta}

 \def\L{\Lambda}

\def\CA{{\mathcal{A}}}
\def\CB{{\mathcal{B}}}

\def\CE{{\mathcal{E}}}
\def\CF{{\mathcal{F}}}
\def\CG{{\mathcal{G}}}

\def\CI{{\mathcal{I}}}

\def\CK{{\mathcal{K}}}
\def\CL{{\mathcal{L}}}

\def\CN{{\mathcal{N}}}

\def\CR{{\mathcal{R}}}
\def\CS{{\mathcal{S}}}

\def\CX{{\mathcal{X}}}
\def\CY{{\mathcal{Y}}}

\def\bsg{{\boldsymbol{g}}}

\def\bst{{\boldsymbol{t}}}

\def\bsw{{\boldsymbol{w}}}
\def\bsx{{\boldsymbol{x}}}
\def\bsy{{\boldsymbol{y}}}

\def\bbst{\bar{\boldsymbol{t}}}

\def\la{\left\langle}
\def\ra{\right\rangle}

\def\hf{\dfrac{1}{2}}
\def\hd{{\hat d}}
\def\bc{{\bar{c}}}

\def\implies{\quad\Rightarrow\quad}

\def\vphi{\varphi}

\def\CS{\mathcal{S}}

\def\qf{\mathfrak{q}}

\def\hq{\hat q}

\def\bQ{{\bar Q}}

\def\bd{{\bar d}}

\def\vac{\emptyset}
\def\res{\mathop{\text{Res}}}

\def\mZ{\mathbb{Z}}

\def\mC{\mathbb{C}}
\def\End{\text{End}}

\def\gl{\mathfrak{gl}}
\def\sl{\mathfrak{sl}}

\def\glinf{\widehat{\mathfrak{gl}}(\infty)}

\def\Abox{{\tikz[scale=0.007cm] \draw (0,0) rectangle (1,1);}}
\def\sAbox{{\tikz[scale=0.005cm] \draw (0,0) rectangle (1,1);}}

\def\bsxi{{\boldsymbol{\xi}}}
\def\bsg{\boldsymbol{\gamma}}

\def\Eq{\text{Eq}}
\def\tz{\tilde{z}}

\begin{document}
\begin{titlepage}

\begin{center}
{\Huge A $(q,t)$-deformation of the 2d Toda\\
\vspace{5mm}
integrable hierarchy}

\vskip 2cm
{\Large Jean-Emile Bourgine\footnote{jean-emile.bourgine@unimelb.edu.au}, Alexandr Garbali\footnote{alexandr.garbali@unimelb.edu.au}}\\

\vskip 1cm
{\it School of Mathematics and Statistics}\\
{\it University of Melbourne}\\
{\it Parkville, Victoria 3010, Australia}\\

\end{center}
\vfill
\begin{abstract}
A $(q,t)$-deformation of the 2d Toda integrable hierarchy is introduced by enhancing the underlying symmetry algebra $\glinf\simeq \text{q-W}_{1+\infty}$ to the quantum toroidal $\gl(1)$ algebra. The difference-differential equations of the hierarchy are obtained from the expansion of $(q,t)$-bilinear identities, and two equations refining the 2d Toda equation are found in this way. The derivation of the bilinear identities follows from the isomorphism between the Fock representation of level $(2,0)$ of the quantum toroidal $\gl(1)$ algebra and the tensor product of the q-deformed Virasoro algebra with a $u(1)$ Heisenberg algebra. It leads to identify the $(q,t)$-deformed Casimir with the screening charges of the deformed Virasoro algebra. Due to the non-trivial coproduct, equations of the hierarchy no longer involve a single tau-function, but instead relate a set of different tau functions. We then define the universal refined tau function using the $L$-matrix of the quantum toroidal $\gl(1)$ algebra and interpret it as the generating function of the deformed tau functions. The equations of the hierarchy, written in terms of the universal refined tau function, combine into two-term quadratic equations similar to the $RLL$ equations.

\end{abstract}
\vfill
\end{titlepage}

\setcounter{footnote}{0}

\newpage

\section{Introduction}
The 2d Toda hierarchy is an integrable hierarchy of difference-differential equations introduced in \cite{Ueno1984} as an extension of the KP hierarchy. The first equation of the hierarchy is the celebrated 2d Toda lattice equation \cite{Mikhailov1979}. A great deal of knowledge on this model has been accumulated over the last forty years \cite{Takasaki1984,Takebe1990}, and we refer to \cite{Takasaki2018} for a short review. Following the method developed by the Kyoto School of integrability \cite{Jimbo1983}, a large class of solutions can be constructed using the fermionic representation of the algebra $\glinf$. These solutions are indexed by the group elements $\widehat{GL(\infty)}$, and particular choices of elements produce notable solutions e.g. hypergeometric functions \cite{Orlov2001}, or generalizations of the well-known solitons and polynomial solutions of the KP hierarchy. The Toda hierarchy has been found to enter in a remarkably wide range of problems in mathematical physics, including 2d string theory \cite{Alexandrov2002}, Hurwitz numbers \cite{Okounkov2000}, ABJM matrix model \cite{Furukawa2019} and topological string theory \cite{Aganagic2003,Nakatsu2007}.

In this paper, we introduce a $(q,t)$-deformation of the 2d Toda integrable hierarchy. One of our motivations is the correspondence between integrable hierarchies and topological strings first noticed in \cite{Aganagic2003}. This correspondence is based on an equivalence between the algebras $\glinf$ and quantum $W_{1+\infty}$. The integrable properties of the hierarchy stem from the presence of these infinite dimensional algebras of symmetries. A refinement of topological strings theory has been introduced in \cite{Iqbal2007,Awata2008} motivated by the correspondence with non-perturbative partition functions of 5d $\CN=1$ gauge theories in the omega-deformed background $\mC_{\e_1}\times\mC_{\e_2}\times S^1$. Refined topological strings amplitudes depend on two parameters $(q_1,q_2)\equiv(e^{\e_1},e^{\e_2})$, they reduce to the original amplitudes in the self-dual limit $(q_1,q_2)\to (q^{-1},q)$ where $q=e^{g_\text{str}}$ is identified with the exponentiated strings coupling constant. Refined topological strings also possess an infinite dimensional symmetry algebra \cite{AFS,Mironov2016,Awata2016a,Bourgine2017b}, the quantum toroidal $\gl(1)$ algebra, which can be constructed as a deformation of the quantum $W_{1+\infty}$ to which it reduces in the self-dual limit \cite{Miki2007}. Following this chain of correspondences, one of the authors proposed in \cite{Bourgine2021b} to introduce a deformation of integrable hierarchies with the quantum toroidal $\gl(1)$ algebra as the underlying symmetry. This paper presents a concrete realization of this idea as a refinement of the Kyoto School formalism.

A second motivation for investigating this deformation of the Toda hierarchy is the well-known reduction of integrable hierarchies to Painlev\'e equations \cite{Tsuda2005,Tsuda2012}. It was observed in \cite{Gamayun2012} that solutions of the Painlev\'e VI equation can be constructed from Virasoro conformal blocks with central charge $c=1$, or 4d $\CN=2$ instanton partition functions in the self-dual omega-background $\e_1+\e_2=0$ following the celebrated AGT correspondence \cite{Alday2010}. Similar quantities built from the q-deformed Virasoro algebra with parameters $q=t$ were shown to produce solutions of the q-Painlev\'e VI equation in \cite{Jimbo2017}, these quantities being related to self-dual 5d $\CN=1$ instanton partition functions through the 5d AGT correspondence \cite{Awata2009}. This correspondence between q-Painlev\'e equations and conformal blocks / instantons partitions functions can be extended to generic parameters $(q,t)$ by introducing a deformation of the q-Painlev\'e VI equation called quantum q-Painlev\'e VI (or qq-Painlev\'e VI for short)\cite{Bershtein2018}. It is then natural to wonder if these deformed equations also follow from the periodic reduction of a $(q,t)$-deformed integrable hierarchy. While it is not clear yet if the formalism developed in this paper has any relation with qq-Painlev\'e equations, we observe that they are both built upon the same underlying symmetry algebra. In fact, the recent observation in \cite{Awata2022} of a connection between the qq-Painlev\'e VI equation and the non-stationary difference equation \cite{Shiraishi2019}, a generalization of the q-difference Toda equation, also points in this direction. The non-stationary Ruijsenaars functions solving this equation have deep connection to quantum toroidal symmetry \cite{Fukuda2020,Shakirov2021}, which is the key ingredient in our construction. 

Our third motivation is the connection with Yang--Baxter integrability. The $(q,t)$-deformation of the 2d Toda integrable hierarchy which we consider in this paper is intimately related to the integrable model which is defined on tensor products of Fock spaces of quantum toroidal $\gl(1)$ algebra in \cite{FJMM_BA,FJMM2}.


\paragraph{Deformed hierarchy}
Concretely, the deformation of the hierarchy is introduced using (a bosonization of) the free fermion formalism developed in \cite{Date1981,Date1981a,Jimbo1983,Kac1989} (see also the review \cite{Alexandrov2012} and the book \cite{Miwa2000}). Refined versions of the Toda tau functions indexed by elements of the quantum toroidal $\gl(1)$ algebra are introduced. A deformation of the bilinear identity is obtained from the vanishing commutation relation between the coproduct of the algebra elements (evaluated in the horizontal Fock representation), and the screening charges of the deformed Virasoro algebra. The vanishing of this commutator is a consequence of the isomorphism found in \cite{Feigin2009a} between the representation of levels $(2,0)$ of the quantum toroidal $\gl(1)$ algebra and the deformed Virasoro algebra tensored with a Heisenberg algebra (the so-called $u(1)$ factor). Equations of the hierarchy follow from the usual expansion of the bilinear integral equation in the difference of time parameters.

In practice, the refinement of the hierarchy leads to two main issues. The first one is the loss of the free fermion presentation, since the quantum toroidal $\gl(1)$ algebra has no fermionic representation. Instead, we need to use the horizontal Fock representation defined in \cite{FT-shuffle,SV-Hall} which acts on the free boson Fock space. Thus, all fermionic formulas have to be bosonized before introducing their $(q,t)$-deformation. For instance, the Casimir operator involved in the basic bilinear equation has the following fermionic expression
\begin{equation}
\Psi=\sum_{r\in\mZ+1/2}\bpsi_{-r}\otimes\psi_r,
\end{equation} 
it is replaced by one of the q-Virasoro screening charges constructed in \cite{Shiraishi1995}. The latter does indeed reduce in the self-dual limit to the bosonized formula for the Casimir $\Psi$. Fortunately, the derivation of the bilinear identity is based on the bosonic formalism and there is no major difficulty for the refinement.

The second issue is the deformation of the coproduct. Unlike the group elements of $\widehat{GL(\infty)}$ which are co-commutative, i.e. $\D(G)=G\otimes G$, the Drinfeld coproduct of a generic element $G$ of the toroidal algebra takes the form of a (possibly infinite) series
\begin{equation}\label{eq:cop_intro}
\D(G)=\sum_\a G_\a\otimes G'_\a.
\end{equation}
As a consequence, in general the bilinear integral equations (and so the difference-differential equations of the hierarchy) no longer involve a single tau function $\t(G)$, but instead they relate members of a set of refined tau functions indexed by elements $G\in\CX$ closed under coproduct $\D(\CX)\subset\CX\otimes\CX$.\footnote{Note that $\CX$ does not need to be an algebra, i.e. it does not have to be closed under the product.} In the simple example where $G$ is built from generators of the Cartan subalgebra, the coproduct is co-commutative (up to deformation by central elements), and $\CX$ contains only a single element. However, the resulting tau function is almost trivial. More interesting examples can be obtained as certain matrix elements of the R-matrix. In this case, the deformed hierarchy produces a closed set of equations for the refined tau functions which can be checked explicitly.

Our main results are the following bilinear integral equations obeyed by the refined tau functions $\t_u(\bst,\bbst|G)$ defined in \eqref{ref_canonical}
\begin{align}\label{Hir_I}
\begin{split}
&\sum_\a\oint_\infty \dfrac{dz}{2i\pi}z^{-n-1}e^{-\sum_{k>0}(\g^k t_k-t_k')\frac{1-t^k}{1-q^k}z^k}\t_u(\bst+[\g^{-1}z^{-1}],\bbst|G_\a)\t_{uq^{-n}t^{-1}}(\bst'-[\g^{-2}z^{-1}],\bbst'|G'_\a)\\
+&\sum_\a\oint_0\dfrac{dz}{2i\pi}z^{-n-1}e^{\sum_{k>0}(\g^kt_{-k}-t_{-k}')p^kz^{-k}}\t_{ut^{-1}}(\bst,\bbst-[\g z]_{q,t}|G_\a)\t_{uq^{-n}}(\bst',\bbst'+[z]_{q,t}|G'_\a)=0,\\
\end{split}
\end{align}
where $\bst$, $\bbst$ denotes the sets of time parameters, and $u$ deforms the discrete parameter of the Toda lattice, and $n\in\mZ$. The parameters $(q,t)$ of the deformed Virasoro algebra are identified with the quantum group parameters $(q_1,q_2)\equiv(t^{-1},q)$ and $\g=t^{1/2}q^{-1/2}$, and we have assumed that the coproduct $\D(G)$ takes the generic form \eqref{eq:cop_intro}. The shift of time parameters $\bst\pm [z]_{q,t}$ is defined in \eqref{def_shift_qt}, it reduces to the standard shift $t_k\pm z^k/k$ in the self-dual limit. A second equation with the role of $q$ and $t$ exchanged is also derived in \eqref{Ref_Hirota_2dToda} starting from a different screening charge.

The difference-differential equations of the refined hierarchy are obtained by expanding the integral equation \eqref{Hir_I} in the differences of time parameters $\g^{|k|}t_k-t'_k$. New equations which trivialize in the self-dual limit are found at low orders, they can be found in \eqref{eq_0}, \eqref{eq_1} and \eqref{eq_2}. More interestingly, two equations deforming the Toda lattice equation were found
\begin{align}
\begin{split}
&\sum_\a\left[\p_1\t_u(\bst,\bst'|G_\a)\p_{-1}\t_{uqt^{-1}}(\bsg\bst,\bsg\bst'|G'_\a)-qt^{-1}\t_u(\bst,\bst'|G_\a)\p_1\p_{-1}\t_{uqt^{-1}}(\bsg\bst,\bsg\bst'|G'_\a)\right]\\
=&-\sum_\a\t_{ut^{-1}}(\bst,\bst'|G_\a)\t_{uq}(\bsg\bst,\bsg\bst'|G'_\a),\\
&\sum_\a\left[\p_{-1}\t_u(\bst,\bst'|G_\a)\p_1\t_{uq^{-1}t}(\bsg\bst,\bsg\bst'|G'_\a)-qt^{-1}\t_u(\bst,\bst'|G_\a)\p_1\p_{-1}\t_{uq^{-1}t}(\bsg\bst,\bsg\bst'|G'_\a)\right]\\
=&-\sum_\a\t_{ut}(\bst,\bst'|G_\a)\t_{uq^{-1}}(\bsg\bst,\bsg\bst'|G'_\a),
\end{split}
\end{align}
with the time derivatives $\p_{\pm1}=\p/\p t_{\pm1}$, and $\bsg\bst$, $\bsg\bst'$ indicating that the times $t_k$ are rescaled by a factor $\g^{|k|}$. The fact that we find two inequivalent equations is related to the non-cocommutativity of the coproduct, i.e. in general the opposite coproduct $\D^{op}$ differs from $\D$. In the self-dual limit, both reduce to the co-commutative coproduct and the two previous equations reproduce the usual Toda equation \eqref{Toda_usual} for the tau function.

\paragraph{Refined tau functions and the universal R-matrix}
The presence of an additional quantum group structure is the main new feature of our $(q,t)$-deformed model. It leads to a new class of solutions defined in section \ref{sec:R} using the $L$-matrix of the quantum group (i.e. the universal $R$-matrix evaluated in specific representations). In the self-dual limit, the coproduct becomes co-commutative, the quantum group structure trivializes, and the model reduces to the usual Toda integrable hierarchy.

The bilinear equations \eqref{Hir_I} can be combined and written in a compact form in terms of the universal refined tau function $\t_u(\bst,\bbst)$ which can be viewed as the generator of $\t_u(\bst,\bbst|G)$. The universal refined tau function is a formal series in $\bst$ and $\bbst$ with coefficients in the quantum toroidal algebra $\gl(1)$. Using the definition of the universal $R$-matrix of the quantum toroidal algebra $\gl(1)$ we show that $\t_u(\bst,\bbst)$ can be defined via the $L$-matrix $\CL(u)$ \cite{FJMM_BA} associated to the horizontal Fock representation of $\gl(1)$ with the spectral parameter $u$. The bilinear equations in terms of $\CL$ take the form
\begin{align}\label{eq:X-tilde_intro}
(\gamma^{c n}\otimes \Phi^{+}_n)  \CL_{13}(q^{-n}t^{-1}u)\CL_{12}(u)&=\CL_{13}(q^{-n} u)\CL_{12}(t^{-1} u)(1\otimes \Phi^{+}_n),   \\
\label{eq:Y-tilde_intro}
(\gamma^{c m}\otimes \Phi^{-}_m)  \CL_{13}(t^m q u)\CL_{12}(u)&= \CL_{13}(t^m u)\CL_{12}(q u) (1\otimes \Phi_m^{-}),
\end{align}
where $n,m\in \mathbb Z$, $c$ a central element of the algebra and $\Phi^{\pm}_n$ are the modes of the screening operators without the zero modes (see \eqref{def_Spm}, \eqref{def_Psi_pm} and \eqref{eq:Phi}). The modes $\Phi^{\pm}_n$ act in the tensor product of two horizontal Fock representations of the quantum toroidal algebra $\gl(1)$. The $RLL$ equation of the algebra associated to this representation in our conventions reads
\begin{align}
    \label{eq:RLL_intro}
    \check {\rm R}_{23}(u_2/u_1)\CL_{13}(u_2) \CL_{12}(u_1)
    =
    \CL_{13}(u_1) \CL_{12}(u_2) \check {\rm R}_{23}(u_2/u_1),
\end{align}
where $\check {\rm R}(u)=P {\rm R}(u)$, $P$ being the permutation matrix and ${\rm R}(u)$ is the horizontal-horizontal R-matrix normalized such that its vacuum matrix element is equal to $1$. By looking at the special cases $n=0$, $m=0$ and $n=1$, $m=-1$ in \eqref{eq:X-tilde_intro} and \eqref{eq:Y-tilde_intro} one finds that these equations reproduce \eqref{eq:RLL_intro} at special values of $u_1$ and $u_2$ by identifying\footnote{For a similar connection between the R-matrix and screening charges in the affine Yangian case see \cite{Litv_Vilk}.}
\begin{align}
    \label{eq:Rch=Psi}
\check{\rm R}(t^{-1}) = \Phi^+_0,
    \qquad
\check {\rm R}(q) = \Phi^-_0,
\qquad
\underset{x=q t^{-1}}{\text{Res}} \check{\rm R}(x) = q t^{-1} (1-q t^{-1}) \Phi^+_1\Phi^-_{-1}.
\end{align}
By this consideration we establish a connection between refined tau functions and integrability objects of quantum toroidal $\gl(1)$.

\paragraph{Outline}
This paper is organized as follows. In Section \ref{sec:algebra}, we summarize the algebraic results needed for our analysis. Section \ref{sec:hierarchy} is the core of this paper, we define the refined tau functions, show that they obey two bilinear identities, and expand them to obtain the differential-difference equations of the hierarchy. In Sections \ref{sec:analysis} and \ref{sec:R}, we present the analysis of several classes of tau functions with Section \ref{sec:R} focusing on the universal refined tau function. We conclude with a brief discussion on the deformation of the known 2d Toda tau functions. In Appendix \ref{app:Fermions} we discuss the $(q,t)$-deformation of the fermionic formalism and define the associated Baker-Akhiezer wave functions. Appendix \ref{app:Bilin} provides the technical details of the derivation of the bilinear identities. Finally, a brief reminder on the intertwining operators of the quantum toroidal $\gl(1)$ algebra used to express the vertical-horizontal R-matrix is given in Appendix \ref{app:Intertwiners}.

\begin{figure}
\begin{center}
\begin{tabular}{|c|c|c|c|}
\hline
Fock space & Vacuum state & Heisenberg modes & Algebra (rep.)\\
\hline
$\CF_n$ & $\ket{n}=e^{nQ}\ket{\vac}$ & $J_k$ & qu. $W_{1+\infty}$ ($\rho_f$)\\
$\CF_u$ & $\ket{u}=u^{\hd}\ket{\vac}$ & $J_k$ & qu. tor. $\gl(1)$ ($\rho_u^{(1,0)}$)\\
$\CF_v$ & $\dket{\vac}$ & $-$ & qu. tor. $\gl(1)$ ($\rho_v^{(0,1)}$)\\
$\CF_{m,n}$ & $\ket{m,n}=e^{\g_{m,n}\hat q}\ket{\vac}$ & $\a_k$ & q-Virasoro \\
\hline
\end{tabular}
\end{center}
\caption{Summary of the different Fock spaces used in this paper.}
\label{table_Fock}
\end{figure}

\paragraph{Notations} For reference, a brief summary of the different Fock spaces used in this paper is provided in figure \ref{table_Fock}. We also recall here the dictionary between quantum group parameters and q-deformed Virasoro parameters which is used throughout the paper
\begin{equation}\label{id_param}
(q_1,q_2)=(t^{-1},q),\quad q_3=(q_1q_2)^{-1}=\g^2=p^{-1}.
\end{equation}

\section{Quantum toroidal $\gl(1)$ and $q$-deformed Virasoro algebras}\label{sec:algebra}
The main technical tool we use to introduce the refined tau functions is the deformation of the quantum $W_{1+\infty}$ symmetry algebra into the quantum toroidal $\gl(1)$ algebra, and a relation between the latter and the q-deformed Virasoro algebra. In this section, we review the definitions of these algebras, and the roles they play in the hierarchy.

\subsection{\texorpdfstring{$\glinf$}{gl(infty)} and \texorpdfstring{q-W$_{1+\infty}$}{q-W_{infty}} symmetries}
In this subsection, we recall the fermionic construction of 2d Toda tau functions, and highlight the role of the underlying symmetry algebras.

\paragraph{Free bosons} Let $J_k$ and $Q$ denote the modes of the free bosonic field $\phi(z)$ satisfying the Heisenberg algebra
\begin{equation}\label{exp_phi}
\phi(z)=Q+J_0\log z-\sum_{k\in\mZ^\times}\dfrac1kz^{-k}J_k,\quad [J_k,J_l]=k\d_{k+l,0},\quad [J_k,Q]=\d_{k,0}.
\end{equation}
Let $\ket{\vac}$ denote the (neutral) vacuum state annihilated by positive modes $J_{k\geq0}\ket{\vac}=0$, and $\ket{n}=e^{nQ}\ket{\vac}$ the charged vacuum state of charge $n\in\mZ$. By construction, $\ket{0}=\ket{\vac}$, $J_k\ket{n}=0$ for $k>0$ and $J_0\ket{n}=n\ket{n}$. Let $\CF_n$ and $\CF$ denote respectively the Fock space of charge $n$ and the total Fock space defined as
\begin{equation}
\CF_n=\mC[J_{-1},J_{-2},\cdots]\ket{n},\quad \CF=\bigoplus_{n\in\mZ} \CF_n.
\end{equation} 
The bosonic normal-ordering which consists in moving positive modes $J_{k\geq0}$ to the right will be denoted $:\cdots:$. The adjoint action is obtained as $J_{k}^\dagger=J_{-k}$, $Q^\dagger=-Q$, in particular the dual vacua $\bra{n}=\bra{\vac}e^{-Qn}$ are annihilated by negative modes, and $\bra{n}\!\!\!\ket{m}=\d_{n,m}$.

\paragraph{Free fermions} The Dirac fermionic fields $\bpsi(z)$, $\psi(z)$ are defined as 
\begin{equation}
\psi(z)=\sum_{r\in\mathbb{Z}+1/2}z^{-r-1/2}\psi_r,\quad \bpsi(z)=\sum_{r\in\mathbb{Z}+1/2}z^{-r-1/2}\bpsi_r,\quad \{\bpsi_r,\psi_s\}=\d_{r+s,0},
\end{equation} 
and $\{\psi_r,\psi_s\}=\{\bpsi_r,\bpsi_s\}=0$. Through the boson-fermion correspondence, these fields can be realized as vertex operators on the bosonic Fock space
\begin{align}\label{bosonization}
\begin{split}
&\bpsi(z):\CF_n\to\CF_{n+1},\quad \bar\psi(z)=:e^{\phi(z)}:,\\
&\psi(z):\CF_n\to\CF_{n-1},\quad \psi(z)=:e^{-\phi(z)}:.
\end{split}
\end{align}
Expanding their action on the vacuum state, we deduce that positive fermionic modes annihilate the vacuum $\ket{\vac}$, and more generally $\psi_r\ket{n}=0$ for $r>n$ and $\bpsi_r\ket{n}=0$ for $r>-n$. The normal-ordering which consists in moving the positive fermionic modes to the right will be denoted $\vdots\cdots\vdots$.

\paragraph{Tau functions} Let $\bst=(t_1,t_2,\cdots)$, $\bbst=(t_{-1},t_{-2},\cdots)$ denote two infinite sets of time parameters, and $G\in\End(\CF)$ an operator acting on the Fock space. The correlation function
\begin{equation}\label{tau_2dToda}
\t_n(\bst,\bbst|G)=\bra{n}e^{\sum_{k>0}t_kJ_k}Ge^{\sum_{k>0}t_{-k} J_{-k}}\ket{n}
\end{equation}
defines a tau function of the Toda hierarchy if the operator $G$ satisfies the basic bilinear condition
\begin{equation}\label{bbc}
[G\otimes G,\Psi]=0,\quad \Psi=\sum_{r\in\mZ+1/2}\bpsi_{-r}\otimes\psi_r.
\end{equation}
In particular, the tau functions obey a bilinear integral equation, called the \textit{bilinear identity}, from which all the difference-differential equations of the hierarchy can be deduced. We recall the derivation of the bilinear identity from the basic bilinear condition in Appendix \ref{app:Bilin}.

\paragraph{Fermionic representation of $\glinf$} The basic bilinear condition \eqref{bbc} admits for solutions operators of the form
\begin{equation}
G=\exp\left(\sum_{r,s\in\mZ+1/2}\a_{r,s}\vdots\bpsi_r\psi_s\vdots\right),\quad \a_{r,s}\in\mC.
\end{equation} 
They define a representation of the group $\widehat{GL}(\infty)$. The corresponding Lie algebra $\glinf$ is generated by the elements $E_{r,s}$ indexed by half-integer $r,s\in\mZ+\frac12$ with the commutation relations\footnote{As a general rule in this paper, indices $k,l,m,n$ are integer valued while indices $r,s,t,u$ are half-integer valued.}
\begin{equation}
\left[E_{r,s}, E_{t,u}\right]=\delta_{s+t}E_{r,u}-\delta_{r+u}E_{t,s} +c_1(\theta(r)-\theta(t))\delta_{s+t}\delta_{r+u},\quad r,s,t,u\in\mZ+\hf
\end{equation}
where we used the Heaviside function $\th(r)=1$ when $r>0$ and $\th(r)=0$ when $r<0$, and $c_1$ is a central element. This algebra is a central extension of the algebra $\gl(\infty)$ of matrices with infinite size. It has a fermionic representation of level $\rho_f(c_1)=1$ defined as $\rho_f(E_{r,s})=\vdots\bpsi_r\psi_s\vdots$ which can be bosonized to define a map $\rho_f:\glinf\to\End(\CF)$.

The algebra $\glinf$ is compatible with the co-commutative coproduct $\D(E_{r,s})=E_{r,s}\otimes1+1\otimes E_{r,s}$. This coproduct takes the form $\D(G)=G\otimes G$ for the group elements $G=e^{\sum_{r,s}\a_{r,s}E_{r,s}}$, and the basic bilinear relation \eqref{bbc} simply expresses the fact that $\Psi$ is a Casimir for the representation $\rho_f\otimes\rho_f$
\begin{equation}
[(\rho_f\otimes\rho_f)\D(E_{r,s}),\Psi]=0.
\end{equation} 
This simple remark is a starting point of the Kac-Wakimoto construction \cite{Kac1989}.

\paragraph{Quantum $W_{1+\infty}$} In order to introduce the $(q,t)$-deformation, we need to reformulate the $\glinf$ symmetry in the language of the quantum $W_{1+\infty}$ algebra \cite{Lebedev1992}. This algebra\footnote{We refer to \cite{SWM,Bourgine2021} for a brief summary of the main properties of this algebra. Note that we have flipped the parameter $q\to q^{-1}$ with respect to \cite{Bourgine2021}, in agreement with the new identification of the parameters $(q_1,q_2)=(q^{-1},q)$ in the self-dual limit.} is generated by elements $W_{m,n}$ indexed with two integers $m,n\in\mZ$, and two central elements $(c_1,c_2)$ with the following relations involving the quantum parameter $q\in\mC^\times$
\begin{equation}\label{def_qW}
[W_{m,n},W_{m',n'}]=(q^{-m'n}-q^{-mn'})\left(W_{m+m',n+n'}+c_1\dfrac{\d_{m+m',0}}{1-q^{-n-n'}}-c_2\dfrac{\d_{n+n',0}}{1-q^{-m-m'}}\right).
\end{equation}
At level $c_2=0$, this algebra is equivalent to the algebra $\glinf$ upon a simple linear transformation of the generators reminiscent of a Fourier transform
\begin{equation}\label{Ers}
W_{m,n}=\sum_{s\in\mZ+1/2}q^{(s+1/2)n}E_{m-s,s}.
\end{equation}
The algebraic relations \eqref{def_qW} are compatible with the co-commutative coproducts $\D(W_{m,n})=W_{m,n}\otimes1+1\otimes W_{m,n}$, and so is the linear relation \eqref{Ers} between the two algebras. Using this relation, the fermionic representation of $\glinf$ can be extended to the generators of quantum $W_{1+\infty}$
\begin{equation}\label{def_rho_f}
\rho_f(W_{m,n})=\sum_{s\in\mZ+1/2}q^{(s+1/2)n}\vdots\bpsi_{m-s}\psi_s\vdots=\oint{\dfrac{dz}{2i\pi}\vdots\bpsi(z)z^m\psi(q^{-n}z)\vdots}.
\end{equation} 
We observe that the dependence in the zero-mode $Q$ drops in the expression of $\rho_f(W_{m,n})$, and so $\rho_f(W_{m,n})\in\End(\CF_k)$. The Casimir $\Psi$ also characterizes the elements of the fermionic representation of quantum $W_{1+\infty}$ on $\End(\CF\otimes\CF)$ as
\begin{equation}\label{eq:WPsi}
[(\rho_f\otimes\rho_f)\D(W_{m,n}),\Psi]=0.
\end{equation} 
We would like to emphasize that the quantum $W_{1+\infty}$ algebra is not a symmetry of the Toda hierarchy per se, and the parameter $q$ does not enter in the equations of the hierarchy. Instead, it enters in some of the solutions of the hierarchy which are expressed using the generators $W_{m,n}$.

\subsection{Quantum toroidal $\gl(1)$ algebra}
The quantum toroidal $\gl(1)$ algebra $\CE$ depends on two quantum parameters $q_1,q_2\in\mC^\times$, and it is customary to introduce a third one $q_3$ such that $q_1q_2q_3=1$. We will also denote $\g=q_3^{1/2}$. In the Drinfeld presentation, the algebra is generated by the modes of the following currents
\begin{equation}\label{DIM_currents}
x^\pm(z)=\sum_{k\in\mathbb{Z}}z^{-k}x^\pm_k,\quad \psi^\pm(z)=\sum_{k\geq0}z^{\mp k}\psi_{\pm k}^\pm,
\end{equation}
together with the central elements $(c,\bc)$ such that $\psi_0^\pm=\g^{\mp\bc}$. These currents obey the algebraic relations
\begin{align}
\begin{split}\label{def_DIM}
&[\psi^\pm(z),\psi^\pm(w)]=0,\quad \psi^+(z)\psi^-(w)=\dfrac{g(\g^c z/w)}{g(\g^{-c}z/w)}\psi^-(w)\psi^+(z),\\
&\psi^+(z)x^\pm(w)=g(\g^{\pm c/2}z/w)^{\pm1}x^\pm(w)\psi^+(z),\quad \psi^-(z)x^\pm(w)=g(\g^{\mp c/2}z/w)^{\pm1}x^\pm(w)\psi^-(z),\\
&\prod_{\a=1,2,3}(z-q_\a^{\pm1}w)\ x^\pm(z)x^\pm(w)=\prod_{\a=1,2,3}(z-q_\a^{\mp1}w)\ x^\pm(w)x^\pm(z),\\
&[x^+(z),x^-(w)]=\k\left(\d(\g^{-c}z/w)\psi^+(\g^{c/2}w)-\d(\g^c z/w)\psi^-(\g^{-c/2}w)\right),\\
&\Sym_{z_1,z_2,z_3}\dfrac{z_2}{z_3}[x^\pm(z_1),[x^\pm(z_2),x^\pm(z_3)]]=0\quad \text{(Serre relations)},
\end{split}
\end{align}
where $\k$ is a $\mathbb{C}$-number and $g(z)$ a rational function, both depending on the parameters $q_1,q_2$
\begin{equation}\label{def_g}
\k=\dfrac{(1-q_1)(1-q_2)}{1-q_1q_2},\quad g(z)=\prod_{\a=1,2,3}\dfrac{1-q_\a z}{1-q_\a^{-1}z}.
\end{equation}
In \eqref{def_DIM}, $\d(z)=\sum_{k\in\mZ}z^k$ denotes the formal delta power series, we refer to the first chapter of \cite{Frenkel2004} for more details about its mathematical definition and main properties. In the $\psi\psi$ and $\psi x$ relations, the rational functions $g(\a z/w)$ are expanded in positive/negative powers of $z/w$, while the $xx$ relations lead to quadratic relations between the modes $x_k^\pm$. We should emphasize that the currents $\psi^\pm(z)$ bear no direct relation with the fermionic fields $\psi(z)$, $\bpsi(z)$.

The modes $\psi_{\pm k}^\pm$ together with the central elements generate the equivalent of a Cartan subalgebra. This subalgebra can also be presented using the modes $a_k$ obtained from a different expansion of the currents $\psi^\pm(z)$
\begin{equation}\label{eq:psi_currents}
\psi^+(z)=\psi_{0}^+ e^{\sum_{k>0}z^{-k}a_{k}},\quad \psi^-(z)=\psi_{0}^- e^{-\sum_{k>0}z^{k}a_{-k}}.
\end{equation}
We note that the modes $a_k$ and $x_k^\pm$ obey the following commutation relations
\begin{equation}\label{com_ak}
[a_k,a_l]=(\g^{ck}-\g^{-ck})c_k\d_{k+l,0},\quad [a_k,x_l^\pm]=\pm\g^{\mp |k|c/2}c_k x_{l+k}^\pm,\quad\text{with}\quad c_k=-\dfrac1{k}\prod_{\a=1,2,3}(1-q_\a^k).
\end{equation}

\paragraph{Gradings} The algebra $\CE$ can be supplemented by two grading operators $d$ and $\bd$ with the commutation relations
\begin{align}
\begin{split}\label{def_gradings}
&[d,x_k^\pm]=-kx_k^\pm,\quad [d,\psi_{\pm k}^\pm]=\mp k\psi_{\pm k}^\pm,\quad [d,c]=[d,\bc]=0,\\
&[\bd,x_k^\pm]=\pm x_k^\pm,\quad [\bd,\psi_{\pm k}^\pm]=0,\quad [\bd,c]=[\bd,\bc]=0.
\end{split}
\end{align}
This extension of $\CE$ will be denoted by $\CE'$. They can be used to define the following automorphisms indexed by $\a\in\mC^\times$
\begin{align}\label{def_autom}
\begin{split}
& \a^d x^\pm(z) \a^{-d}=x^\pm(\a z),\quad \a^d\psi^\pm(z)\a^{-d}=\psi^\pm(pz),\\
& \a^\bd x^\pm(z) \a^{-\bd}=\a^{\pm1}x^\pm(z),\quad \a^{\bd}\psi^\pm(z)\a^{-\bd}=\psi^\pm(z).
\end{split}
\end{align}

\paragraph{Coproduct} The algebra $\CE$ is a Hopf algebra equipped with the Drinfeld coproduct
\begin{align}
\begin{split}\label{coproduct}
&\D(x_k^+)=x^+_k\otimes 1+\sum_{l\geq0}\g^{-(c\otimes1)(k+l/2)}\ \psi^-_{-l}\otimes x^+_{k+l},\quad \D(x_k^-)=\sum_{l\geq0}\g^{-(1\otimes c)(k-l/2)}\ x^-_{k-l}\otimes \psi^+_l+1\otimes x_k^-,\\
&\D(a_k)=a_k\otimes \g^{-|k|c/2}+\g^{|k|c/2}\otimes a_k,\quad \D(c)=c\otimes 1+1\otimes c,\quad \D(\bc)=\bc\otimes1+1\otimes \bc.
\end{split}
\end{align}
It can be extended to include the grading elements that are co-commutative.

\paragraph{Self-dual limit} In the self-dual limit $(q_1,q_2)\to (q^{-1},q)$ (i.e. $q_3\to1$), the toroidal algebra $\CE$ reduces to the quantum $W_{1+\infty}$ algebra. The limit of the Drinfeld generators reads
\begin{align}\label{self-dual}
\begin{split}
&x_k^+\to (1-q^{-1})q^{-k/2}W_{k,1}+c_1\d_{k,0},\quad x_k^-\to (1-q)q^{k/2}W_{k,-1}+c_1\d_{k,0},\\
&\dfrac{ka_k}{(q_1^{k/2}-q_1^{-k/2})(q_3^{k/2}-q_3^{-k/2})}\to W_{k,0}-\dfrac{c_2}{1-q^{-k}},\quad (c,\bc)\to(c_1,-c_2),
\end{split}
\end{align}
and the algebraic relations \eqref{def_DIM} reduce to the commutation relations \eqref{def_qW}. Remarkably, in this limit the Drinfeld coproduct becomes co-commutative, in agreement with our previous observation. The quantum toroidal algebra is a deformation of the universal enveloping algebra of the quantum $W_{1+\infty}$ algebra, and group-like elements now correspond to quantum group elements of co-unit one.\footnote{For an element with coproduct $\D(G)=G\otimes G$, the antipode is $S(G)=G^{-1}$ and the co-unit $\e(G)=1$.}

The grading operators defined by the relation \eqref{def_gradings} reduce to the natural gradings of quantum $W_{1+\infty}$, i.e. with a slight abuse of notation,
\begin{equation}
[d,W_{m,n}]=-m W_{m,n},\quad [\bd,W_{m,n}]=n W_{m,n}.
\end{equation} 

\subsection{Horizontal Fock representation}\label{sec_Horiz}
The terminology \textit{Fock representations} refers to a set of representations $\rho$ of levels $(\rho(c),\rho(\bc))=(m,n)\in\mZ\times\mZ$ of the quantum toroidal algebra $\gl(1)$ acting on the bosonic Fock space $\CF$ or its tensor products. In this paper, we are mainly interested in the representations with $(m,n)=(1,n)$ (resp. $(m,n)=(0,1)$), which we call \textit{horizontal (resp. vertical) Fock representations} following \cite{AFS}. We denote these representations by $\rho^{(m,n)}$, but the symbol will often be omitted to lighten the expressions if no confusion ensues. In fact, our main construction is based on the horizontal representation $\rho^{(1,0)}$ which replaces the fermionic representation $\rho_f$ in the original construction. We will also consider in the next subsection the representation $\rho^{(2,0)}=(\rho^{(1,0)}\otimes\rho^{(1,0)})\D$ acting on the tensor product $\CF\otimes\CF$. The vertical Fock representation will only enter in the definition of a certain class of refined tau function in section \ref{sec_VH}, and so we decided to present it in Appendix \ref{app:Intertwiners} to lighten this section.

In the horizontal Fock representation $\rho^{(1,n)}$, the Cartan modes $a_k$ form a deformed Heisenberg algebra and they can be identified with the modes $J_k$ up to a factor depending on $q_1,q_2$
\begin{equation}\label{q-osc}
\rho^{(1,n)}(a_k)=-\dfrac{\g^{-k/2}}{k}(1-q_2^k)(1-q_3^k)J_k,\quad \rho^{(1,n)}(a_{-k})=-\dfrac{\g^{-k/2}}{k}(1-q_1^k)(1-q_3^k)J_{-k},\quad (k>0).
\end{equation}
On the other hand, the currents $x^\pm(z)$ are expressed as
\begin{equation}\label{eq:rho_x}
\rho^{(1,n)}(x^\pm(z))=z^{\mp n}\eta^\pm(z),
\end{equation} 
using the vertex operators 
\begin{align}
\begin{split}\label{def_eta}
&\eta^+(z)=\exp\left(\sum_{k>0}\dfrac{z^{k}}{k}(1-q_1^k)J_{-k}\right)\exp\left(-\sum_{k>0}\dfrac{z^{-k}}{k}(1-q_2^k)J_k\right),\\
&\eta^-(z)=\exp\left(-\sum_{k>0}\dfrac{z^{k}}{k}\g^{k}(1-q_1^k)J_{-k}\right)\exp\left(\sum_{k>0}\dfrac{z^{-k}}{k}\g^{k}(1-q_2^k)J_k\right).
\end{split}
\end{align}
These operators satisfy the normal ordering 
\begin{align}
\label{eq:eta_ord}
\eta^+(z) \eta^+(w)     
&=
\zeta \left( \frac zw \right)
:\eta^+(z) \eta^+(w):,
&\qquad
\eta^-(z)\eta^-(w)
&=
\zeta \left( \frac wz \right)
:\eta^-(z)\eta^-(w):,\\
\label{eq:eta_ord_mixes}
\eta^+(z) \eta^-(w)     
&=
\zeta \left( \frac{z}{\gamma w} \right)^{-1}
:\eta^+(z) \eta^-(w):,
&\qquad
\eta^-(z)\eta^+(w)
&=
\zeta \left( \frac{ w}{\gamma z} \right)^{-1}
:\eta^-(z)\eta^+(w):,
\end{align}
where
\begin{align}
\label{eq:zeta}
\zeta(x) := \frac {(x-1)(x-q_1q_2)}{(x-q_1)(x-q_2)}.
\end{align}
A complex parameter $u\in\mC^\times$ can be included as the weight of the representation as explained below. 

\paragraph{Gradings and zero-modes} The representation of the grading operators $(d,\bd)$ is a little subtle, so we first restrict ourselves to the level $n=0$. In the representation $\rho^{(1,0)}$, the grading operator $d$ is quadratic in the Heisenberg modes
\begin{equation}
\rho^{(1,0)}(d)=L_0\quad\text{with}\quad L_0=\sum_{k>0}J_{-k}J_k.
\end{equation}
Then, the automorphism defined in \eqref{def_autom} follows from the property
\begin{equation}\label{prop_L0}
p^{L_0} J_k p^{-L_0}=p^{-k}J_k,\quad p\in\mC^{\times}.
\end{equation} 

An easy way to realize the second grading operator $\bd$ in this representation is to define it as a derivative with respect to the weight $\log u$. However, it has been observed in \cite{Bourgine2021b} that the horizontal weight $u$ plays the role of charge $n$ in the definition \eqref{tau_2dToda} of the tau functions, i.e. we have to recover $u\to q^{n}$ in the self-dual limit. Thus, in this context it is natural to introduce the charged vacua $\ket{u}$. More precisely, we introduce two operators $\hat d$ and $\hat u$ satisfying\footnote{Equivalently, we have the quantum torus relation $p^{\hat d} \hat u=p \hat u p^{\hat d}$.}  
\begin{align*}
    \qquad[\hat d, \hat u]=\hat u,
\end{align*}
and modify the expressions \eqref{eq:rho_x} of the Drinfeld currents $x^\pm(z)$ by inserting the operator $\hat u$
\begin{equation}\label{eq:rho_u_x}
\rho^{(1,0)}(x^\pm(z))=\hat{u}^{\pm 1}\eta^\pm(z).
\end{equation}
Then we set $\rho^{(1,0)}(\bd)=\hat d$ which ensures the validity of the commutation relations \eqref{def_gradings}. The charged Fock spaces are defined on the vacua $\ket{u}$ in the usual way, $\CF_u=\mC[J_{-1},J_{-2},\cdots]\ket{u}$\footnote{In fact, since we are only considering here the representation of grading elements of the form $q^{m\bd/2}t^{n\bd/2}$ we can work with the total Fock space 
\begin{equation}
\CF=\bigoplus_{m,n\in\mZ}\CF_{uq^{n/2}t^{m/2}}.
\end{equation}} 
 with the actions
\begin{align*}
    \hat u\ket{u}=u\ket{u}, 
    \qquad
    \ket{u}=u^{-\hat d}\ket{\vac},
\end{align*}
and we identify $\ket{\vac}=\ket{1}$. The formulas \eqref{q-osc} and \eqref{eq:rho_u_x} then define the representations \begin{align}
    \label{eq:rho-u}
    \rho_u^{(1,0)}:\CE\to\End(\CF_u).
\end{align} 
This construction generalizes to horizontal representations of levels $(1,n)$, but the grading now contains a mixing term
\begin{equation}
\rho^{(1,n)}(d)=L_0-n\hat d,\quad \rho^{(1,n)}(\bd)=\hat d.
\end{equation}

\paragraph{Self-dual limit and fermionic representation} The horizontal representation $\rho^{(1,0)}$ of $\CE$ reduces in the self-dual limit to the bosonization of the fermionic representation of quantum $W_{1+\infty}$ defined previously. In particular, the charged Fock spaces $\CF_u$ reduce to the ones defined previously
\begin{equation}\label{sd_lim}
\hat u\to q^{J_0},\quad \hat d\to -\dfrac{Q}{\log q},\quad \ket{u}\to\ket{n},\quad \CF_u\to\CF_n.
\end{equation}
The Drinfeld currents $x^\pm(z)$ reduce to
\begin{equation}
x^+(z)\to :\bpsi(z)\psi(q^{-1}z):,\quad x^-(z)\to :\bpsi(q^{-1}z)\psi(z):,
\end{equation} 
in agreement with \eqref{self-dual} after taking into account the different normal-orderings.

\subsection{Free field representation of the deformed Virasoro algebra}\label{sec_qVir}
From our analysis of the role of the symmetry algebra in the construction of tau functions we deduced that the tensor product $G\otimes G$ in the basic bilinear condition \eqref{bbc} is replaced after $(q,t)$-deformation by the evaluation of an element $G\in\CE$ in the representation
\begin{equation}\label{def_rho2}
\rho_{u_1,u_2}^{(2,0)}:\CE\to \End(\CF_{u_1}\otimes\CF_{u_2}),\qquad \rho_{u_1,u_2}^{(2,0)}=(\rho_{u_1}^{(1,0)}\otimes\rho_{u_2}^{(1,0)})\D.
\end{equation}
It is helpful to keep in mind the representation $\rho^{(2,0)}$
\begin{align}
    \label{def_rho2_X}
\rho^{(2,0)}:\CE\to \End(\CF\otimes\CF),\qquad \rho^{(2,0)}=(\rho^{(1,0)}\otimes\rho^{(1,0)})\D.
\end{align}
In this representation instead of the scalars $u_1$ and $u_2$ we encounter invertible operators $\hat u^{(1)}=\hat u \otimes 1$ and $\hat u^{(2)}=1 \otimes  \hat u$.
In the view of \eqref{def_rho2}, the Casimir $\Psi$ in \eqref{eq:WPsi} should be replaced by operators commuting with $\rho_{u_1,u_2}^{(2,0)}(\CE)$ on $\End(\CF_{u_1}\otimes\CF_{u_2})$. In this subsection, we construct these operators for specific values of the ratio $u_1/u_2$ by exploiting the relation between the representation $\rho_{u_1,u_2}^{(2,0)}$ and the deformed Virasoro algebra \cite{Shiraishi1995} (see also \cite{FKSW}).

\paragraph{Deformed Virasoro algebra} The deformed Virasoro algebra (or q-Virasoro algebra) is an algebra depending on two complex parameters $(q,t)$, and generated by the modes $T_k$ of the current $T(z)=\sum_{k\in\mZ}z^{-k}T_k$ called the stress-energy tensor. This current satisfies the algebraic relation
\begin{equation}
f(w/z)T(z)T(w)-f(z/w)T(w)T(z)=-\dfrac{(1-q)(1-t^{-1})}{1-p}\left(\d(pw/z)-\d(p^{-1}w/z)\right),
\end{equation}
with the function
\begin{equation}
f(z)=\exp\left(\sum_{k>0}\dfrac1k\dfrac{(1-q^k)(1-t^{-k})}{1+p^k}z^k\right),
\end{equation} 
where we denoted $p=q/t$.

The free field representation defines an action of the stress-energy tensor on yet another set of Fock spaces. These Fock spaces $\CF_{m,n}$ are indexed by two integer charges $m,n\in\mZ$, they are built using the modes of a $(q,t)$-deformed Heisenberg algebra\footnote{To avoid the confusion with the Cartan modes of the toroidal algebra, we use here $\a_k$ instead of $a_k$.}
\begin{equation}
[\a_k,\a_l]=k\dfrac{1-q^{|k|}}{1-t^{|k|}}\d_{k+l,0},\quad [\a_0,\hq]=\b^{-1},
\end{equation}
where $\b=\log t/\log q$ (i.e. $t=q^\b$). From the neutral vacuum state $\ket{\vac}$ annihilated by $\a_k$ with $k\geq0$, we define the charged vacua $\ket{m,n}=e^{\g_{m,n}\hq}\ket{\vac}$ with $\g_{m,n}=(m+1)\b/2-(n+1)/2$, $m,n\in\mZ$, and the corresponding bosonic Fock spaces $\CF_{m,n}=\mC[\a_{-1},\a_{-2},\cdots]\ket{m,n}$. In this representation, the stress-energy tensor has the form 
\begin{equation}
T(z)=\L(z)+:\L(pz)^{-1}:,
\end{equation} 
with the vertex operator
\begin{equation}
\L(z)=p^{1/2}\exp\left(-\sum_{k>0}\dfrac1k\dfrac{1-t^{k}}{1+p^k}t^{-k}p^{-k/2}z^k\a_{-k}\right)\exp\left(-\sum_{k>0}\dfrac1k(1-t^k)p^{k/2}z^{-k}\a_k\right)q^{\b\a_0}.
\end{equation}

\paragraph{Screening currents} The representation of the q-Virasoro algebra on the Fock spaces $\CF_{m,n}$ can be characterized by the commutation of the operators with certain screening charges built from the screening currents
\begin{align}
\begin{split}
&S_+(z)=e^{\sum_{k>0}\frac{z^k}k\frac{1-t^k}{1-q^k}\a_{-k}}e^{-\sum_{k>0}\frac{z^{-k}}k\frac{1-t^k}{1-q^k}(1+p^k)\a_k}e^{\b\hq}z^{2\b\a_0},\\
&S_-(z)=e^{-\sum_{k>0}\frac{z^{k}}k\a_{-k}}e^{\sum_{k>0}\frac{z^{-k}}k(1+p^{-k})\a_k}e^{-\hq}z^{-2\a_0}.
\end{split}
\end{align}
Since $z^{2\b\a_0}\ket{m,n}=z^{(m+1)\b-(n+1)}\ket{m,n}$, the screening current $S_+(z)$ is expanded in integer powers of $z$ only when it acts on Fock spaces $\CF_{m,n}$ with $m=-1$. Similarly, since $z^{-2\a_0}\ket{m,n}=z^{-(m+1)+(n+1)/\b}\ket{m,n}$, the current $S_-(z)$ has an expansion in integer powers of $z$ only when it acts on Fock spaces $\CF_{m,n}$ with $n=-1$. When the currents have an expansion in integer powers, we can define the screening charges
\begin{equation}\label{def_Psi_pm}
\Psi_\pm=\oint_0\dfrac{dz}{2i\pi} S_\pm(z).
\end{equation}
Thus, our previous requirement forces us to restrict the action of these operators to the corresponding Fock spaces. It was shown in \cite{Shiraishi1995} that the screening charges commute with the deformed Virasoro algebra on these spaces, i.e. $[T(z),\Psi_+]=0$ (resp. $[T(z),\Psi_-]=0$) when acting on states belonging to the Fock space $\CF_{-1,n}$ (resp. $\CF_{m,-1}$). In the next paragraph, we will argue that these screening charges play the role of the Casimir $\Psi$ for the refined hierarchy.

Note that it is possible to consider Fock spaces with more general charges if we introduce the higher screening charges
\begin{equation}
\Psi_\pm^{(n)}=\oint\prod_{a=1}^n\dfrac{dz_a}{2i\pi}S_\pm(z_1)S_\pm(z_2)\cdots S_\pm(z_n).
\end{equation}
The derivation of the bilinear identity presented in the next section is expected to extend to these higher screening charges, and should produce bilinear integral equations involving several nested contours. However, we will not address this problem here and simply focus on the simpler case of the Fock spaces $\CF_{-1,n}$ and $\CF_{m,-1}$.

\paragraph{Relation with the quantum toroidal $\gl(1)$ algebra} A remarkable relation between the representations of levels $(2,n)$ of the quantum toroidal $\gl(1)$ algebra and the deformed Virasoro algebra has been observed in \cite{Feigin2009a}. With the application to integrable hierarchies in mind, we restrict ourselves to the representation $\rho_{u_1,u_2}^{(2,0)}$ defined in \eqref{def_rho2}. In this representation, the Cartan modes $a_k$ form a (twisted) Heisenberg algebra $\b_k=k\rho_{u_1,u_2}^{(2,0)}(a_k)$ that is sometimes called the \textit{$u(1)$-factor}
\begin{equation}
\b_k=-\g^{-k}(1-q_2^k)(1-q_3^k)(J_k^{(1)}+\g^kJ_k^{(2)}),\quad \b_{-k}=\g^{-k}(1-q_1^k)(1-q_3^k)(J_{-k}^{(1)}+\g^kJ_{-k}^{(2)}),
\end{equation}
with the shortcut notation $J_k^{(1)}=J_k\otimes 1$, $J_k^{(2)}=1\otimes J_k$. To build the isomorphism, the key observation is that this Heisenberg algebra can be factorized from the representation of the Drinfeld currents $x^\pm(z)$, and the remaining factor identified with the q-Virasoro stress-energy tensor $T(z)$
\begin{equation}\label{eq:x-T}
\rho^{(2,0)}(x^\pm(z))=V_b^\pm(z)T(z),\quad V_b^\pm(z)=(u^{(1)}u^{(2)})^{\pm1/2}:e^{\mp\sum_{k\neq0}\frac{\g^{\mp|k|}}{k(q_3^k-q_3^{-k})}z^{-k}\b_k}.
\end{equation}
This relation holds upon the identification of parameters \eqref{id_param} and the relations
\begin{align}
\label{dict_qVir_tor}
&\a_k=-\dfrac{1-q^{k}}{1-t^k}p^k\dfrac{\g^k J_k^{(1)}-J_k^{(2)}}{1+p^{k}},
&
&\a_{-k}=-(\g^k J_{-k}^{(1)}-J_{-k}^{(2)}),\\
\label{dict_qVir_tor_0}
&q^{1-\b+2\b\a_0}= \hat u^{(2)}/\hat u^{(1)},
&
&e^{\hat q} = q^{\hd^{(1)}-\hd^{(2)}}.
\end{align} 
where we suppressed the algebra homomorphism notation. The $u(1)$-factor decouples from the deformed Virasoro algebra since $[\a_k,\b_l]=0$, and we can write formally
\begin{equation}
\rho^{(2,0)}(\CE)=\left(\text{Heisenberg}\right)\times\left(\text{q-Virasoro}\right).
\end{equation}
As a consequence of this relation, the screening charges \eqref{def_Psi_pm} of the deformed Virasoro algebra define the Casimir of the representation $\rho_{u_1,u_2}^{(2,0)}$ for specific values of the ratio $u_2/u_1$ ($u_2/u_1=q^{m\b-n}$ in $\CF_{m,n}$ due to \eqref{dict_qVir_tor_0}). It means that the operators of $\rho_{u_1,u_2}^{(2,0)}(\CE)$ in $\End(\CF_{u_1}\otimes\CF_{u_2})$ are fully characterized by the fact that they commute with these screening charges (up to the adjustment of the zero modes as explained below). Using the relations \eqref{dict_qVir_tor} between the Heisenberg modes, we can rewrite the screening currents as
\begin{align}
\begin{split}\label{def_Spm}
&S_+(z)=\exp\left(-\sum_{k>0}\dfrac{z^k}k\dfrac{1-t^k}{1-q^k}(\g^k J_{-k}^{(1)}-J_{-k}^{(2)})\right)\exp\left(\sum_{k>0}\dfrac{z^{-k}}kp^k(\g^{k}J_{k}^{(1)}-J_{k}^{(2)})\right)e^{\b\hq}z^{2\b\a_0},\\
&S_-(z)=\exp\left(\sum_{k>0}\dfrac{z^k}k(\g^kJ_{-k}^{(1)}-J_{-k}^{(2)})\right)\exp\left(-\sum_{k>0}\dfrac{z^{-k}}k\dfrac{1-q^k}{1-t^k}(\g^kJ_{k}^{(1)}-J_{k}^{(2)})\right)e^{-\hq}z^{-2\a_0},
\end{split}
\end{align}
and for the zero modes the relations \eqref{dict_qVir_tor_0} are implied.
\paragraph{Zero modes}
From the correspondence between the representation $\rho_{u_1,u_2}^{(2,0)}$ and the free field representation of the deformed Virasoro algebra, we deduced the following identification of the vacuum states
\begin{equation}
\ket{u_1}\otimes\ket{u_2}=\ket{v_0}_\text{u(1)}\otimes\ket{m,n}_\text{qVir},\quad v_0=(u_1u_2)^{-1/2},\quad u_2=t^m q^{-n}u_1.
\end{equation}
The screening charges $\Psi_\pm$ are operators mapping Fock spaces of different charges
\begin{equation}\label{eq:Psi-F}
\Psi_+:\CF_{-1,n}\to\CF_{1,n},\quad \Psi_-:\CF_{m,-1}\to\CF_{m,1}.
\end{equation}
As a result, they intertwine representations of the toroidal algebra with different weights
\begin{equation}\label{com_Psi_rho}
\rho_{u_1',u_2'}^{(2,0)}(\CE)\Psi_\pm=\Psi_\pm\rho_{u_1,u_2}^{(2,0)}(\CE),
\end{equation}
%
with $u_2=t^{-1}q^{-n}u_1$ and $u'_2=tq^{-n}u'_1$ in the case of $\Psi_+$. Similarly, $u_2=qt^mu_1$ and $u'_2=q^{-1}t^mu'_1$ in the case of $\Psi_-$. On the other hand, the products of weights $u_1u_2$ and $u_1'u'_2$ are equal since $\Psi_\pm$ do not contain any dependence in the algebra associated to the $u(1)$-factor. These relations fix all the weights but one, and we take
\begin{align}
\begin{split}\label{weights}
&\Psi_+:\quad u_1=u,\quad u_2=t^{-1}q^{-n}u,\quad u'_1=t^{-1}u,\quad u'_2=q^{-n}u,\\
&\Psi_-:\quad u_1=u,\quad u_2=qt^mu,\quad u'_1=qu,\quad u'_2=t^mu.
\end{split}
\end{align}

The relation \eqref{com_Psi_rho} is the refined version of the basic bilinear condition \eqref{bbc}. In the self-dual limit, $S_+(z)$ (resp. $S_-(z)$) reduces to $\psi(z)\otimes\bpsi(z)$ (resp. $\bpsi(z)\otimes\psi(z)$).\footnote{In particular, for the zero modes $\hat q\to Q^{(2)}-Q^{(1)}$ and $2\a_0\to J_0^{(2)}-J_0^{(1)}$.} Thus $\Psi_-$ reduces to $\Psi$ and $\Psi_+$ to $\bar\Psi=\oint \psi(z)\otimes\bpsi(z)$. Since the coproduct $\D$ becomes co-commutative in this limit, the relation \eqref{com_Psi_rho} does indeed reduce to \eqref{bbc}. In fact, a $(q,t)$-analogue of the Dirac fermions can also be introduced using the expression of the screening currents, this idea is explored in the Appendix \ref{app:Fermions}.

\section{Refined Toda hierarchy}\label{sec:hierarchy}
In this section, we introduce a $(q,t)$-deformation of the Toda tau functions, and show that they obey a set of bilinear identities. Expanding these identities, we derive the first difference-differential equations of a refined Toda hierarchy.

\subsection{Refined tau functions}\label{sec_ref_tau}
Let $\bst=(t_1,t_2,\cdots)$ and $\bbst=(t_{-1},t_{-2},\cdots)$ denote two sets of time parameters, $u\in\mC^\times$ a complex parameter, and $G\in\CE$ an element of the quantum toroidal $\gl(1)$ algebra. We define the \textit{refined 2d Toda tau functions} $\t_u(\bst,\bbst|G)$ as the following correlation functions in the bosonic Fock space $\CF_u$ introduced in Section \ref{sec_Horiz}
\begin{equation}\label{ref_canonical}
\t_u(\bst,\bbst|G)=\bra{u}e^{\sum_{k>0}t_kJ_k}\rho_u^{(1,0)}(G)e^{\sum_{k>0}t_{-k}J_{-k}}\ket{u},\quad G\in\CE.
\end{equation}
These tau functions also depend on the quantum group parameters $(q_1,q_2)\equiv (t^{-1},q)$ but we will omit them in the notation since they are spectators in our discussion. In the self-dual limit $t\to q$, $\rho_u^{(1,0)}(G)$ reduces to an element of the quantum $W_{1+\infty}$ algebra in the (bosonized) fermionic representation, and the expression \eqref{ref_canonical} reproduces the usual definition \eqref{tau_2dToda} of the 2d Toda tau function $\tau_n(\bst,\bbst|G)$, with the complex parameter $u$ reducing to the integer parameter $n$ of the as $u\to q^n$.\footnote{In the self-dual limit, $q^n$ coincides with one of the two weights of the fermionic representation $\rho_f$ of $W_{1+\infty}$. To avoid introducing too many technicalities, we decided to omit these weights in the definition \ref{def_rho_f} of the representation, but they can be found in the section 2.1.2 of \cite{Bourgine2021b}.}

Toda tau functions have a natural expansion in Schur functions, and their $(q,t)$-deformations \eqref{ref_canonical} can be expanded in a similar way in Macdonald functions. Indeed, Macdonald functions are closely related to the horizontal Fock representation of the quantum toroidal $\gl(1)$ algebra \cite{Feigin2009a}. This is better understood by introducing the Macdonald states $\ket{P_\l,u}\in\CF_u$ obtained as the inverse image $\iota^{-1}(P_\l(\bsx))$ of Macdonald function under the isomorphism $\iota$ sending the Fock space $\CF_u$ to the ring of symmetric functions
\begin{equation}
J_{-k}\to p_k(\bsx),\quad  J_k\to k\dfrac{\p}{\p p_k},\quad \ket{u}\to 1,
\end{equation} 
where $p_k(\bsx)$ are the power sum symmetric functions $\sum_{\a=1}^\infty x_\a^k$. With this definition, the action of a certain commutative subalgebra $b_k\in\CE$ is diagonal on these states, with eigenvalues\footnote{This subalgebra can be defined as the image of the Cartan modes $a_k$ under Miki's automorphism, and in particular $b_{\pm1}=x_0^\pm$\cite{Miki2007,Bourgine2018a}.}
\begin{equation}
\rho_u^{(1,n)}(b_k)\ket{P_\l,u}=(-\g)^{-k}u^kc_k\left(\sum_{i,j\in\l}q_1^{(i-1)k}q_2^{(j-1)k}-\dfrac1{(1-q_1^k)(1-q_2^k)}\right)\ket{P_\l,u}.
\end{equation} 
The dual states $\bra{P_\l,u}$ can be introduced by identifying the inner product with Macdonald's scalar product
\begin{equation}
\bra{P_\l,u}\!\!\!\ket{P_\mu,u}=\d_{\l,\mu}\la P_\l,P_\l\ra_{q,t},\quad \la P_\l,P_\l\ra_{q,t}=\prod_{(i,j)\in\l}\dfrac{1-q^{\l_i-j+1}t^{\l'_j-i}}{1-q^{\l_i-j}t^{\l'_j-i+1}}.
\end{equation} 
where $\lambda'$ is the transposed partition. Inserting the Cauchy identity 
\begin{equation}
1=\sum_\l \dfrac{\ket{P_\l,u}\bra{P_\l,u}}{\la P_\l,P_\l\ra_{q,t}}
\end{equation} 
in the definition \eqref{ref_canonical} of the refined tau functions, and using the properties
\begin{equation}
P_\l(\bsx)=\bra{u}e^{\sum_{k>0}\frac{p_k(\bsx)}{k}J_k}\ket{P_\l,u}=\bra{P_\l,u}e^{\sum_{k>0}\frac{1-t^k}{1-q^k}\frac{p_k(\bsx)}{k}J_{-k}}\ket{u},
\end{equation} 
we end up with a double expansion over partitions
\begin{equation}
\t_u(\bst,\bbst|G)=\sum_{\l,\mu}\dfrac{P_\l(\bsx)P_\mu(\bsy)}{\la P_\l,P_\l\ra_{q,t}\la P_\mu,P_\mu\ra_{q,t}} \bra{P_\l,u}\rho_u^{(1,0)}(G)\ket{P_\mu,u}.
\end{equation}
where the time variables are related to the Miwa variables in the following way
\begin{equation}\label{Miwa}
t_k=\dfrac1kp_k(\bsx),\quad t_{-k}=\dfrac1k\dfrac{1-t^k}{1-q^k}p_k(\bsy).
\end{equation} 

\paragraph{Refined hypergeometric tau functions} When the operator $\rho^{(1,0)}(G)$ is diagonal on the Macdonald basis, the refined tau function can be expanded in a single sum over partitions\footnote{It was pointed to us by a referee that the tau function in \eqref{tau_hyper} can be interpreted as the expectation value of the operator $G$ with respect to the Macdonald measure \cite{For_Reins,Bor-Cor}.}
\begin{equation}\label{tau_hyper}
\t_u(\bst,\bbst|G)=\sum_{\l}\dfrac{P_\l(\bsx)P_\l(\bsy)}{\la P_\l,P_\l\ra_{q,t}} G_\l,\quad\text{with}\quad \rho^{(1,0)}(G)\ket{P_\l}=G_\l\ket{P_\l}.
\end{equation}
This is a natural $(q,t)$-deformation of a class of hypergeometric tau functions. In particular, this class contains the tau functions introduced by Okounkov in the context of Hurwitz numbers \cite{Okounkov2000}
\begin{equation}
\t_n(\bst,\bbst|\qf,\b)=\bra{n}e^{\sum_{k>0}t_kJ_k}\qf^{L_0}e^{\b W_0}e^{\sum_{k>0}t_{-k}J_{-k}}\ket{n},\quad W_0=\sum_{r\in\mZ+1/2}r^2\vdots\bpsi_{-r}\psi_r\vdots,
\end{equation} 
for which the $(q,t)$-deformation reads
\begin{equation}
\t_u(\bst,\bbst|\qf,\b)=\bra{u}e^{\sum_{k>0}t_kJ_k}\qf^{L_0}F^\b e^{\sum_{k>0}t_{-k}J_{-k}}\ket{u}=\sum_\l\dfrac{P_\l(\bsx)P_\l(\bsy)}{\la P_\l,P_\l\ra_{q,t}} \qf^{|\l|} \prod_{(i,j)\in\l}q_1^{(i-1)\b}q_2^{(j-1)\b}
\end{equation} 
where the operator $W_0$ has been replaced with the framing operator $\log F$ introduced in \cite{Bourgine2021b}\footnote{Strictly speaking, this operator $F$ does not belong to the representation of the algebra $\CE$, but it can be obtained in a certain limit of operators in $\CE$ from which we hope to be able to define its coproduct.}
\begin{equation}
F\ket{P_\l}=\prod_{(i,j)\in\l} q_1^{i-1}q_2^{j-1}\ket{P_\l}\tox_{s.d.} q^{W_0}\ket{s_\l}=\prod_{(i,j)\in\l}q^{j-i}\ket{s_\l}.
\end{equation}

This type of tau functions are closely related to the $(q,t)$-deformed matrix models studied in \cite{Zenkevich2015,Lodin2018}. For instance, choosing $G=e^{\a b_1}$ with $b_1=x_0^+=\oint (2i\pi z)^{-1}x^+(z)dz$, we have the associated eigenvalues 
\begin{equation}
G_\l=\exp\left(-\a u\g^{-1}(1-q_1)(1-q_3)\sum_{i=1}^\infty q_1^{i-1}q_2^{\l_i}\right).
\end{equation}
But, it is also possible to expand this $(q,t)$-tau function in the parameter $\a$ and exploit the normal-ordering properties of the vertex operator $\eta^+(z)$ to rewrite the expansion as a sum of integrals reminiscent of the $(q,t)$-deformed matrix model,
\begin{align}
\begin{split}
\t_u(\bst,\bbst|G)&=\sum_{n=0}^\infty \dfrac{\a^n u^n}{n!}\oint\dfrac{dz_1}{2i\pi z_1}\cdots \dfrac{dz_n}{2i\pi z_n}\bra{\vac}e^{\sum_{k>0}t_k J_k}\eta^+(z_1)\cdots\eta^+(z_n)e^{\sum_{k>0}t_{-k}J_{-k}}\ket{\vac}\\
&=\sum_{n=0}^\infty \dfrac{\a^n u^n}{n!}\oint\dfrac{dz_1}{2i\pi z_1}\cdots \dfrac{dz_n}{2i\pi z_n}\prod_{\superp{i,j=1}{i<j}}^n\zeta(z_i/z_j) \times e^{\sum_{i=1}^n\sum_{k>0}(1-q_1^k)t_k z_i^k}e^{-\sum_{i=1}^n\sum_{k>0}(1-q_2^k)t_{-k} z_i^{-k}},\\
&=\sum_{n=0}^\infty \dfrac{\a^n u^n}{n!}\oint\dfrac{dz_1}{2i\pi z_1}\cdots \dfrac{dz_n}{2i\pi z_n}\prod_{\superp{i,j=1}{i<j}}^n\zeta(z_i/z_j) \times\prod_{i=1}^n\prod_\a\dfrac{(1-t^{-1}z_ix_\a)(1-z_i^{-1}y_\a)}{(1-z_ix_\a)(1-tz_i^{-1}y_\a)},
\end{split}
\end{align}
where we introduced the Miwa variables \eqref{Miwa} to write the last line. Similar integral formulas can be derived for operators of the form $G=e^{\a b_k}$, albeit involving more integrations.

Furthermore, for a certain operator $G$, the tau functions \eqref{tau_hyper} reproduce the $\b$-Hurwitz tau functions studied in \cite{Bonzom2022} in the Jack limit $q=t^{\b}\to1$. Ultimately, we would like to derive the hierarchy of difference-differential equations satisfied by this type of tau functions. Unfortunately, because of the complexity of the coproduct for these operators, we will not be able to provide an explicit answer in this paper. We hope to come back to this particular problem in a future work.

\subsection{Refined bilinear identities}\label{sec_Hir_Toda}
Refined tau functions \eqref{ref_canonical} are related through a set of bilinear integral equations which stems from the commutation relation \eqref{com_Psi_rho} involving the screening charges $\Psi_\pm$ of the q-Virasoro algebra and the coproduct of the element $G$\footnote{Note that the summation over $\a$ may contain infinitely many terms.}
\begin{equation}\label{coproduct_G}
\D(G)=\sum_\a G_\a\otimes G'_\a.
\end{equation}
We call these relations \textit{refined bilinear identities}, they take the form
\begin{align}\label{Ref_Hirota_2dToda}
\begin{split}
&\sum_\a\oint_\infty \dfrac{dz}{2i\pi}z^{-n-1}e^{-\sum_{k>0}(\g^k t_k-t_k')\frac{1-t^k}{1-q^k}z^k}\t_u(\bst+[\g^{-1}z^{-1}],\bbst|G_\a)\t_{uq^{-n}t^{-1}}(\bst'-[\g^{-2}z^{-1}],\bbst'|G'_\a)\\
+&\sum_\a\oint_0\dfrac{dz}{2i\pi}z^{-n-1}e^{\sum_{k>0}(\g^kt_{-k}-t_{-k}')p^kz^{-k}}\t_{ut^{-1}}(\bst,\bbst-[\g z]_{q,t}|G_\a)\t_{uq^{-n}}(\bst',\bbst'+[z]_{q,t}|G'_\a)=0,\\
&\sum_\a\oint_\infty\dfrac{dz}{2i\pi}z^{-m-1}e^{\sum_{k>0}(\g^k t_k-t'_k)z^k}\t_{u}(\bst-[\g z^{-1}]_{t,q},\bbst|G_\a)\t_{uqt^m}(\bst'+[z^{-1}]_{t,q},\bbst'|G'_\a)\\
+&\sum_\a\oint_0\dfrac{dz}{2i\pi}z^{-m-1} e^{-\sum_{k>0}(\g^k t_{-k}-t'_{-k})\frac{1-q^k}{1-t^k}z^{-k}}\t_{uq}(\bst,\bbst+[\g z]|G_\a)\t_{ut^m}(\bst',\bbst'-[z]|G'_\a)=0,
\end{split}
\end{align}
where the first equation is obtained from the commutation relation involving $\Psi_+$ and the second one using $\Psi_-$. Each equation depends on a free integer parameter, $n\in\mZ$ in the first equation, and $m\in\mZ$ in the second one, which enters as a power of the integration variable, and in the shift of the parameter $u$. These equations involve 
the usual shift of time parameters $\bst\pm[z]=(t_1\pm z, t_2\pm z^{2}/2,\cdots)$, but also one of two $(q,t)$-deformations of this shift
\begin{equation}\label{def_shift_qt}
\bst\pm[z]_{q,t}=\left(t_1\pm\dfrac{1-t}{1-q}z,t_2\pm\dfrac{1}2\dfrac{1-t^2}{1-q^2}z^2,\cdots\right),\quad \bst\pm[z]_{t,q}=\left(t_1\pm \dfrac{1-q}{1-t}z,t_2\pm\dfrac{1}2\dfrac{1-q^2}{1-t^2}z^2,\cdots\right).
\end{equation}
The shifts of the negative times $\bbst$ have the same expression with $t_k$ replaced with $t_{-k}$. The equations \eqref{Ref_Hirota_2dToda} are one of the main results of this paper, their derivation is given in Appendix \ref{app:Bilin}.

In the self-dual limit, the coproduct \eqref{coproduct_G} becomes co-commutative, i.e. $\D(G)=G\otimes G$, and the summation over $\a$ disappears. The $(q,t)$-deformed shifts of the time parameters $\bst\pm[z]_{q,t}$ and $\bst\pm[z]_{t,q}$ both reduce to the usual one $\bst\pm[z]$. Thus, the two refined bilinear identities \eqref{Ref_Hirota_2dToda} reduce to the bilinear identities of the 2d Toda hierarchy\footnote{It is obtained directly for the second equation, and follows from the change of variable $z\to 1/z$ in the first equation.}
\begin{align}
\begin{split}\label{Hirota_2dToda}
&\oint_\infty\dfrac{dz}{2i\pi}z^{n-n'}e^{\sum_k(t_k-t'_k)z^k}\t_n(\bst-[z^{-1}],\bbst|G)\t_{n'}(\bst'+[z^{-1}],\bbst'|G)\\
+&\oint_0\dfrac{dz}{2i\pi}z^{n-n'}e^{-\sum_k(t_{-k}-t'_{-k})z^{-k}}\t_{n+1}(\bst,\bbst+[z]|G)\t_{n'-1}(\bst',\bbst'-[z]|G)=0.
\end{split}
\end{align}
The derivation of these identities from the basic bilinear condition \eqref{bbc} is also reproduced in Appendix \ref{app:Bilin} to allow for comparison. We note that this derivation only employs the bosonic expression of fermionic fields, which is why it extends so easily to the $(q,t)$-deformed framework.

\paragraph{Reduction to the mKP hierarchy} The reduction of the 2d Toda hierarchy to the mKP hierarchy is obtained by assuming that the operator $G$ obeys the property $J_kG=GJ_{-k}$ for $k>0$. In this case, the tau function only depends on a single set of time parameters, e.g.
\begin{equation}\label{tau_mKP}
\t_n(\bst|G)=\bra{n}e^{\sum_{k>0}t_kJ_k}G\ket{n}.
\end{equation}
The derivation of the bilinear identity is similar to the case of 2d Toda, but the treatment of the r.h.s. is simplified by the absence of the exponential of negative times.\footnote{Indeed, in this case we can use the property $\Psi(\ket{n}\otimes\ket{n'})=0$ for $n\geq n'$ which follows from the action of the fermionic modes on the vacua. This property is used to simplify the equation \eqref{equ_H_step1} into
\begin{align}
\begin{split}
&\oint_\infty\dfrac{dz}{2i\pi}\bra{n+1,n'-1}\left(e^{J_+(\bst)}\otimes e^{J_+(\bst')}\right)(\bpsi(z)\otimes\psi(z))(G\otimes G)\left(e^{J_-(\bbst)}\otimes e^{J_-(\bbst')}\right)\ket{n,n'}=0
\end{split}
\end{align}
for $n\geq n'$.}
In this case, the bilinear identity contains only a single integral
\begin{equation}\label{Hirota_mKP}
\oint_\infty dz z^{n-n'}e^{\sum_{k>0}z^k(t_k-t'_k)}\t_{n}(\bst-[z^{-1}]|G)\t_{n'}(\bst'+[z^{-1}]|G)=0,\quad n\geq n'.
\end{equation}

A similar reduction can be introduced in the refined setting, provided that all the operators $G_\a$, $G'_\a$ entering the coproduct of $G$ obey the property $J_k G_\a =\L_k G_\a J_{-k}$ for some fixed $\L_k\in\mC(q,t)$. If so, we can remove the exponential of negative times in the definition \eqref{ref_canonical} of the tau functions. Using the fact that $\Psi_+\ket{-1,n}=0$ for $n<0$ and $\Psi_-\ket{m,-1}=0$ for $m<0$, we find the following $(q,t)$-deformation of the KP bilinear identity
\begin{align}\label{Ref_Hirota_KP}
\begin{split}
&\sum_\a\oint_\infty \dfrac{dz}{2i\pi}z^{-n-1}e^{-\sum_{k>0}(\g^k t_k-t_k')\frac{1-t^k}{1-q^k}z^k}\t_u(\bst+[\g^{-1}z^{-1}]|G_\a)\t_{uq^{-n}t^{-1}}(\bst'-[\g^{-2}z^{-1}]|G'_\a)=0,\quad n<0.\\
&\sum_\a\oint_\infty\dfrac{dz}{2i\pi}z^{-m-1}e^{\sum_{k>0}(\g^k t_k-t'_k)z^k}\t_{u}(\bst-[\g z^{-1}]_{t,q}|G_\a)\t_{uqt^m}(\bst'+[z^{-1}]_{t,q}|G'_\a)=0,\quad m<0.
\end{split}
\end{align}

\subsection{Difference-differential equations of the refined hierarchy}\label{sec_equ_diff}
The difference-differential equations of the refined hierarchy are obtained from the bilinear identities \eqref{Hirota_2dToda} by setting
\begin{align}
    \label{eq:epsilon}
    \e_{k}=\g^{|k|}t_k-t'_k    
\end{align}
and expanding at small $\e_k$, with $k\in\mZ$. We focus here on the first bilinear identity, i.e. the one derived from $\Psi_+$, the treatment of the second identity being parallel with the role of $q$ and $t$ simply being flipped and the coproduct replaced by its opposite. The difference-differential equations obtained in this way will be labeled $\Eq_n^{(\mu,\bar\mu)}$ with two partitions $\mu$, $\bar\mu$ and the parameter $n\in\mZ$ coinciding with the integer parameter in the bilinear identity. The two partitions label the order $O(\e_\mu\e_{-\bar\mu})$ at which the equation appears, with $\e_\mu=\e_{\mu_1}\e_{\mu_2}\cdots\e_{\mu_\ell}$ and $\e_{-\bar\mu}=\e_{-\bar\mu_1}\e_{-\bar\mu_2}\cdots\e_{-\bar\mu_{\bar\ell}}$, where $\ell,\bar \ell$ denote the lengths of the corresponding partitions.

To shorten a bit our expressions, which tend to be rather lengthy, we introduce the notation $\p_k=\p/\p t_k$, and often drop the parentheses after derivation operators, taking the unusual convention $\p AB$ meaning $(\p A)B$. We also denote for short $\bsg\bst=(\g t_1,\g^2 t_2,\cdots)$, $\bsg\bbst=(\g t_{-1},\g^2 t_{-2},\cdots)$, and
\begin{equation}
\t_u(G_\a)=\t_u(\bst,\bbst|G_\a),\quad \t_u'(G_\a')=\t_u(\bsg\bst,\bsg\bbst|G'_\a).
\end{equation} 

The bilinear identities can be rewritten in terms of time differences $\e_k$ as
\begin{align}
\begin{split}
&\sum_\a\oint_\infty \dfrac{dz}{2i\pi}z^{-n-1}e^{-\sum_{k>0}r_k\e_kz^k}\t_u(\bst+[\g^{-1}z^{-1}],\bbst|G_\a)e^{-\sum_{k>0}\g^{-k}(\e_k\p_k+\e_{-k}\p_{-k})}\t_{uq^{-n}t^{-1}}(\bsg\bst-[\g^{-2}z^{-1}],\bsg\bbst|G'_\a)\\
+&\sum_\a\oint_0\dfrac{dz}{2i\pi}z^{-n-1}e^{\sum_{k>0}\e_{-k}p^kz^{-k}}\t_{ut^{-1}}(\bst,\bbst-[\g z]_{q,t}|G_\a)e^{-\sum_{k>0}\g^{-k}(\e_k\p_k+\e_{-k}\p_{-k})}\t_{uq^{-n}}(\bsg\bst,\bsg\bbst+[z]_{q,t}|G'_\a)=0,
\end{split}
\end{align}
with 
\begin{equation}\label{def_rk}
r_k=\dfrac{1-t^k}{1-q^k}.
\end{equation}
From the combinatorial relation\footnote{We denote $m_k(\l)=\sharp\{i\in\mZ^{>0}\diagup \l_i=k\}$ the multiplicity of columns with $k$ boxes in $\l$.}
\begin{equation}
e^{\sum_{k>0}z^k p_k}=\sum_{\l}\tz_{\l}^{-1}p_\l z^{|\l|},\quad p_\l=p_{\l_1}p_{\l_2}\cdots p_{\l_\ell},\quad \tz_\l=\prod_{k\geq 1}m_k(\l)!,
\end{equation} 
for any set of parameters $p_k$ indexed by positive integers $k$, we deduce the integral equation $\Eq_n^{(\mu,\bar\mu)}$ at order $O(\e_\mu,\e_{-\bar\mu})$
\begin{align}\label{exp_bilin}
\begin{split}
&\sum_\a\sum_{\superp{\l,\nu}{\l\cup\nu=\mu}}\frac{r_\l\g^{|\l|}}{\tz_\l \tz_\nu \tz_{\bar\mu}}\oint_\infty\dfrac{dz}{2i\pi}z^{-n-1+|\l|}\t_u(\bst+[\g^{-1}z^{-1}],\bbst|G_\a)\p_{\nu}\p_{-\bar\mu}\t_{uq^{-n}t^{-1}}(\bsg\bst-[\g^{-2}z^{-1}],\bsg\bbst|G'_\a)\\
+&\sum_\a\sum_{\superp{\bar\l,\bar\nu}{\bar\l\cup\bar\nu=\bar\mu}}\frac{(-1)^{|\bar\l|}\g^{-|\bar\l|}}{\tz_{\bar\l} \tz_{\bar\nu} \tz_\mu}\oint_0\dfrac{dz}{2i\pi}z^{-n-1-|\bar\l|}\t_{ut^{-1}}(\bst,\bbst-[\g z]_{q,t}|G_\a)\p_{\mu}\p_{-\bar\nu}\t_{uq^{-n}}(\bsg\bst,\bsg\bbst+[z]_{q,t}|G'_\a)=0,
\end{split}
\end{align}
In this expression, $\l\cup\nu$ denotes the partition combining the columns of both $\l$ and $\nu$, and so $\e_{\l}\e_{\nu}=\e_{\l\cup\nu}$. We also used the notation $r_\l=r_{\l_1}r_{\l_2}\cdots r_{\l_\ell}$, $\p_{\l}=\p_{\l_1}\p_{\l_2}\cdots \p_{\l_\ell}$, $\p_{-\l}=\p_{-\l_1}\p_{-\l_2}\cdots\p_{-\l_\ell}$. Note that the first line is vanishing if $n>|\mu|$, and the second one if $n<-|\bar\mu|$, and so the equation is always non-trivial. The difference-differential equations of the hierarchy are obtained by expanding the integrands in powers of $z^{\mp 1}$. In the remaining of this section, we examine in details the first few orders.


\paragraph{Order $O(1)$} At this order, the expansion of the bilinear identity for the Toda hierarchy brings only a trivial equation, but it is not the case for the refined hierarchy. Indeed, the integral equation at this order is
\begin{align}
\begin{split}
&\sum_\a\oint_\infty \dfrac{dz}{2i\pi}z^{-n-1}\t_u(\bst+[\g^{-1}z^{-1}],\bbst|G_\a)\t_{uq^{-n}t^{-1}}(\bsg\bst-[\g^{-2}z^{-1}],\bsg\bbst|G'_\a)\\
+&\sum_\a\oint_0\dfrac{dz}{2i\pi}z^{-n-1}\t_{ut^{-1}}(\bst,\bbst-[\g z]_{q,t}|G_\a)\t_{uq^{-n}}(\bsg\bst,\bsg\bbst+[z]_{q,t}|G'_\a)=0,\\
\end{split}
\end{align}
For definiteness, we restrict ourselves to the values $n=0$ and $n=\pm1$. Expanding the integrands, the contour integrals can be computed easily and we end up with the following three equations
\begin{align}\label{eq_0}
\begin{split}
\Eq_{-1}^{(\vac,\vac)}:\quad&\sum_\a\left[\p_{1}\t_u(G_\a)\t_{uqt^{-1}}'(G'_\a)-p\t_u(G_\a)\p_1\t_{uqt^{-1}}'(G'_\a)\right]=0,\\
\Eq_{0}^{(\vac,\vac)}:\quad&\sum_\a\left[\t_u(G_\a)\t_{ut^{-1}}'(G'_\a)-\t_{ut^{-1}}(G_\a)\t_{u}'(G'_\a)\right]=0,\\
\Eq_{1}^{(\vac,\vac)}:\quad&\sum_\a\left[\p_{-1}\t_{u}(G_\a)\t_{u q^{-1}t}'(G'_\a)-p\t_{u}(G_\a)\p_{-1}\t_{uq^{-1}t}'(G'_\a)\right]=0.
\end{split}
\end{align}
In the self-dual limit, the summation disappears, both $\t_u(G_\a)$ and $\t_u'(G_\a')$ tend to $\t_n(\bst,\bbst|G)$, and these three equations become trivial.

\paragraph{Order $O(\e_{1})$ and $O(\e_{-1})$} At this order, the expansion \eqref{exp_bilin} of the bilinear identity simplifies into respectively
\begin{align}
\begin{split}
&\sum_\a\oint_\infty \dfrac{dz}{2i\pi}z^{-n-1}\t_u(\bst+[\g^{-1}z^{-1}],\bbst|G_\a)\left(\g r_1z+\p_1\right)\t_{uq^{-n}t^{-1}}(\bsg\bst-[\g^{-2}z^{-1}],\bsg\bbst|G'_\a)\\
+&\sum_\a\oint_0\dfrac{dz}{2i\pi}z^{-n-1}\t_{ut^{-1}}(\bst,\bbst-[\g z]_{q,t}|G_\a)\p_1\t_{uq^{-n}}(\bsg\bst,\bsg\bbst+[z]_{q,t}|G'_\a)=0,\\
\end{split}
\end{align}
and
\begin{align}
\begin{split}
&\sum_\a\oint_\infty \dfrac{dz}{2i\pi}z^{-n-1}\t_u(\bst+[\g^{-1}z^{-1}],\bbst|G_\a)\p_{-1}\t_{uq^{-n}t^{-1}}(\bsg\bst-[\g^{-2}z^{-1}],\bsg\bbst|G'_\a)\\
+&\sum_\a\oint_0\dfrac{dz}{2i\pi}z^{-n-1}\t_{ut^{-1}}(\bst,\bbst-[\g z]_{q,t}|G_\a)\left(-\g^{-1}z^{-1}+\p_{-1}\right)\t_{uq^{-n}}(\bsg\bst,\bsg\bbst+[z]_{q,t}|G'_\a)=0,
\end{split}
\end{align}
Specializing to $n=0$, we can evaluate the integrals and write down the corresponding equations of the refined hierarchy
\begin{align}\label{eq_1}
\begin{split}
\Eq_{0}^{(\sAbox,\vac)}:\quad&\sum_\a\left[r_1\p_1\t_u(G_\a)\t_{ut^{-1}}'(G'_\a)+(1-pr_1)\t_u(G_\a)\p_1\t_{ut^{-1}}'(G'_\a)-\t_{ut^{-1}}(G_\a)\p_1\t_{u}'(G'_\a)\right]=0,\\
\Eq_{0}^{(\vac,\sAbox)}:\quad&\sum_\a\left[r_1\p_{-1}\t_{ut^{-1}}(G_\a)\t_{u}'(G'_\a)+(1-pr_1)\t_{ut^{-1}}(G_\a)\p_{-1}\t_{u}'(G'_\a)-\t_u(G_\a)\p_{-1}\t_{ut^{-1}}'(G'_\a)\right]=0.
\end{split}
\end{align}
In the self-dual limit, these two equations also trivialize. On the other hand, choosing $n=\pm1$, we find four equations. Two of them, namely
\begin{align}\label{eq_2}
\begin{split}
&\Eq_{-1}^{(\sAbox,\vac)}:\\
&\ \sum_\a\left[p\left(1-p\dfrac{r_1}2\right)\t_u(G_\a)\p_1^2\t_{up}'(G'_\a)-(1-pr_1)\p_1\t_u(G_\a)\p_1\t_{up}'(G'_\a)-\dfrac{r_1}2\p_1^2\t_u(G_\a)\t_{up}'(G'_\a)\right]\\
&=\hf r_1\g^{-1}\sum_\a\left[\p_2\t_u(G_\a)\t_{up}'(G'_\a)-\g^{-4}\t_u(G_\a)\p_2\t_{up}'(G'_\a)\right],\\
&\Eq_{1}^{(\vac,\sAbox)}:\\
&\ \sum_\a\left[p\left(1-p\dfrac{r_1}2\right)\t_u(G_\a)\p_{-1}^2\t_{up^{-1}}'(G'_\a)-(1-pr_1)\p_{-1}\t_u(G_\a)\p_{-1}\t_{up^{-1}}'(G'_\a)-\dfrac{r_1}{2}\p_{-1}^2\t_u(G_\a)\t_{up^{-1}}'(G'_\a)\right]\\
&=-\dfrac{r_2}{2r_1}\sum_\a\left[\p_{-2}\t_u(G_\a)\t_{up^{-1}}'(G'_\a)-\g^{-4}\t_u(G_\a)\p_{-2}\t_{up^{-1}}'(G'_\a)\right],
\end{split}
\end{align}
trivialize in the self-dual limit, but the other two
\begin{align}\label{Ref_Toda}
\begin{split}
\Eq_{-1}^{(\vac,\sAbox)}:\quad&\sum_\a\left[\p_1\t_u(G_\a)\p_{-1}\t_{up}'(G'_\a)-p\t_u(G_\a)\p_1\p_{-1}\t_{up}'(G'_\a)+\t_{ut^{-1}}(G_\a)\t_{uq}'(G'_\a)\right]=0,\\
\Eq_1^{(\sAbox,\vac)}:\quad &\sum_\a\left[\p_{-1}\t_u(G_\a)\p_1\t_{up^{-1}}'(G'_\a)-p\t_u(G_\a)\p_1\p_{-1}\t_{up^{-1}}'(G'_\a)+\t_{ut}(G_\a)\t_{uq^{-1}}'(G'_\a)\right]=0,
\end{split}
\end{align}
do not. Instead, they reduce to the 2d Toda equation for tau function $\t_n(\bst,\bbst|G)$. Indeed, noticing that in the limit $t\to q$, we have $p=\g=1$, and the sum over $\a$ reduces to a single term corresponding to $G_\a=G_\a'=G$, we end up with the equation
\begin{equation}
\p_1\t_u(G)\p_{-1}\t_{u}(G)-\t_u(G)\p_1\p_{-1}\t_{u}(G)+\t_{uq^{-1}}(G)\t_{uq}(G)=0.
\end{equation}
The sum of the first two terms is equal to $-\t_u(G)^2\p_1\p_{-1}\log(\t_{u}(G))$, and specializing to $u=q^n$, we find for $\t_{q^n}(G)=\t_n(\bst,\bbst|G)$\footnote{The unusual sign of the r.h.s. comes from the fact that we have flipped the sign of the negative time $t_{-k}\to-t_{-k}$ with respect to the more traditional conventions used for instance in \cite{Takasaki2018}.}
\begin{equation}\label{Toda_usual}
\p_1\p_{-1}\log\t_n(\bst,\bbst|G)=\dfrac{\t_{n-1}(\bst,\bbst|G)\t_{n+1}(\bst,\bbst|G)}{\t_n(\bst,\bbst|G)^2}.
\end{equation}
This equation implies the Toda equation
\begin{equation}\label{2d_Toda}
\p_1\p_{-1}\phi(n)=e^{\phi(n+1)-\phi(n)}-e^{\phi(n)-\phi(n-1)},\quad\text{for}\quad e^{\phi(n)}=\dfrac{\t_{n+1}(\bst,\bbst|G)}{\t_n(\bst,\bbst|G)}.
\end{equation}
Thus, the two equations \eqref{Ref_Toda} are a $(q,t)$-deformation of the Toda equation, and in consistence with our terminology we will call them \textit{refined 2d Toda equations}. The derivation of these equations is the main result of this section.

\section{Analysis of refined Toda $\t$-functions}\label{sec:analysis}
In this section, we specialize the operator $G$ to various elements of the algebra $\CE$ and analyze the corresponding refined tau functions. Our examples in this section are obtained with $G$ given by either an exponential of Cartan generators or by grading elements.

\subsection{Cartan generators}
In the simplest example, $G$ is defined as an exponential of the Cartan generators $a_k$ for which the coproduct \eqref{coproduct} is co-commutative (up to factors involving the central element $\g^c$). For definiteness, consider the case of a family of elements $G_\bsxi$ parameterized by a set of complex numbers $\xi_k\in\mC$ and defined as follows
\begin{equation}
G_\bsxi=e^{\sum_{k>0}\xi_k a_{-k}}\in\CE.
\end{equation}
The coproduct of these elements can be written in the compact form
\begin{equation}
\D(G_\bsxi)=G_{\bsg^{-c^{(2)}/2}\bsxi}\otimes G_{\bsg^{c^{(1)}/2}\bsxi},
\end{equation} 
where we introduced the notation $\bsg^{\a}\bsxi=\{\g^{\a k}\xi_k\}$ to absorb the central elements. For these simple elements, the correlation functions \eqref{ref_canonical} defining the refined tau functions are easy to compute, we find
\begin{equation}\label{def_tau_txi}
\t(\bst,\bbst|\bsxi)=e^{\sum_{k>0}kt_kt_{-k}}e^{-\sum_{k>0}(1-q_1^k)(1-q_3^k)\g^{-k/2}t_k\xi_k}.
\end{equation}
Note that in this case the refined tau functions are actually independent of the weight $u$, and we have simplified the previous notation $\t_u(\bst,\bbst|G_{\bsxi})$ into $\t(\bst,\bbst|\bsxi)$. 

\paragraph{Bilinear identities} Once again, we will focus on the first of the two bilinear identities \eqref{Ref_Hirota_2dToda}, and the same arguments apply to the second one as well. For this particular choice of elements $G_{\bsxi}$, the bilinear identity simplifies into
\begin{align}
\begin{split}
&\oint_\infty \dfrac{dz}{2i\pi}z^{-n-1}e^{-\sum_{k>0}(\g^k t_k-t_k')\frac{1-t^k}{1-q^k}z^k}\t(\bst+[\g^{-1}z^{-1}],\bbst|\bsg^{-1/2}\bsxi)\t(\bst'-[\g^{-2}z^{-1}],\bbst'|\bsg^{1/2}\bsxi)\\
+&\oint_0\dfrac{dz}{2i\pi}z^{-n-1}e^{\sum_{k>0}(\g^kt_{-k}-t_{-k}')p^kz^{-k}}\t(\bst,\bbst-[\g z]_{q,t}|\bsg^{-1/2}\bsxi)\t(\bst',\bbst'+[z]_{q,t}|\bsg^{-1/2}\bsxi)=0.
\end{split}
\end{align}
Using the expression of the refined tau function \eqref{def_tau_txi} derived previously, we can compute the products
\begin{align}
\begin{split}
&\t(\bst+[\g^{-1} z^{-1}],\bbst|\bsg^{-1/2}\bsxi)\t(\bst'-[\g^{-2}z^{-1}],\bbst'|\bsg^{1/2}\bsxi)=e^{\sum_{k>0}(\g^k t_{-k}-t'_{-k})p^kz^{-k}},\\
&\t(\bst,\bbst-[\g z]_{q,t}|\bsg^{-1/2}\bsxi)\t(\bst',\bbst'+[z]_{q,t}|\bsg^{1/2}\bsxi)=e^{-\sum_{k>0}(\g^k t_k-t_k')\frac{1-t^k}{1-q^k}z^k},
\end{split}
\end{align}
and we observe that they no longer depend on the parameters $\xi_k$. Inserting these expressions in the bilinear identity, we find that the two integrals have the same integrand, and the identity takes the form
\begin{align}
\begin{split}
&\left(\oint_\infty+\oint_0\right) \dfrac{dz}{2i\pi}z^{-n-1}e^{-\sum_{k>0}(\g^k t_k-t_k')\frac{1-t^k}{1-q^k}z^k}e^{\sum_{k>0}(\g^k t_{-k}-t'_{-k})p^kz^{-k}}=0.
\end{split}
\end{align}
This equation should be understood as the vanishing of every terms of the expansion in the time differences $\e_k=\g^{|k|}t_k-t'_{-k}$. It is not difficult to check that this is indeed the case.\footnote{For instance, consider the expansion in $\a,\b$ in the following expression
\begin{align}
\begin{split}
&\left(\oint_0+\oint_\infty \right)\dfrac{dz}{2i\pi} z^{-n-1} e^{\a z}e^{\b z^{-1}}=\sum_{k,l=0}^\infty \dfrac{\a^k\b^l}{k!l!}\left(\oint_\infty \dfrac{dz}{2i\pi} z^{k-l-n-1}+\oint_0 \dfrac{dz}{2i\pi} z^{k-l-n-1}\right)\\
&=\sum_{k,l=0}^\infty \dfrac{\a^k\b^l}{k!l!}\left(-\oint_0 \dfrac{dz}{2i\pi} z^{-k+l+n-1}+\oint_0 \dfrac{dz}{2i\pi} z^{k-l-n-1}\right)=\sum_{k,l=0}^\infty \dfrac{\a^k\b^l}{k!l!}\left(-\d_{n,k-l}+\d_{n,k-l}\right)=0.
\end{split}
\end{align}}
The simplicity of this problem comes from the fact that $\rho^{(2,0)}(G)$ is purely a $u(1)$ factor, i.e. it does not contain the modes $\a_k$ of the deformed Virasoro algebra, and so it trivially commutes with the screening charges $\Psi_\pm$. This is reflected in the simplicity of the refined tau functions \eqref{def_tau_txi} which are exponentials of an expression linear in each time $t_k$.

\paragraph{Difference-differential equations} It is also instructive to verify explicitly that the first equations of the refined hierarchy are satisfied by the tau functions computed in \eqref{def_tau_txi}. At first order, the equations $\Eq_{\pm1}^{(\vac,\vac)}$ read
\begin{equation}
\Eq_{\mp1}^{(\vac,\vac)}:\quad \p_{\pm1}\log\t(\bst,\bbst|\bsg^{-1/2}\bsxi)=p\p_{\pm1}\log\t(\bsg\bst,\bsg\bbst|\bsg^{1/2}\bsxi),
\end{equation}
while the equation $\Eq_{0}^{(\vac,\vac)}$ in \eqref{eq_0} is trivial since there is no dependence in $u$. These equations are satisfied as a consequence of the property
\begin{equation}\label{prop_tau_txi}
\p_k\log\t(\bst,\bbst|\bsxi)=\g^{2|k|}\p_k\log\t(\bsg\bst,\bsg\bbst|\bsg\bsxi)
\end{equation} 
obeyed by the tau functions.

At orders $O(\e_{\pm1})$, we observe that the equations $\Eq_{0}^{(\sAbox,\vac)}$ and $\Eq_{0}^{(\vac, \sAbox)}$ written in \eqref{eq_1} are automatically satisfied as a consequence of the zero order equations $\Eq_{\pm1}^{(\vac,\vac)}$. To write down the equations \eqref{eq_2} and \eqref{Ref_Toda} obtained for $n=\pm1$ in a concise form, we use the shortcut notations $\t=\t(\bst,\bbst|\bsg^{-1/2}\bsxi)$ and $\t'=\t(\bsg\bst,\bsg\bbst|\bsg^{1/2}\bsxi)$
\begin{align}
\begin{split}
\Eq_{-1}^{(\sAbox,\vac)}:\quad&p\left(1-p\dfrac{r_1}2\right)\t\p_1^2\t'-(1-pr_1)\p_1\t\p_1\t'-\dfrac{r_1}2\p_1^2\t\t'=\hf r_1\g^{-1}\left(\p_2\t\t'-\g^{-4}\t\p_2\t'\right),\\
\Eq_{1}^{(\vac,\sAbox)}:\quad&p\left(1-p\dfrac{r_1}2\right)\t\p_{-1}^2\t'-(1-pr_1)\p_{-1}\t\p_{-1}\t'-\dfrac{r_1}{2}\p_{-1}^2\t\t'=-\dfrac{r_2}{2r_1}\left(\p_{-2}\t\t'-\g^{-4}\t\p_{-2}\t'\right),\\
\Eq_{-1}^{(\vac,\sAbox)}:\quad&\p_1\t\p_{-1}\t'-p\t\p_1\p_{-1}\t'+\t\t'=0,\\
\Eq_{1}^{(\sAbox,\vac)}:\quad&\p_{-1}\t\p_1\t'-p\t\p_1\p_{-1}\t'+\t\t'=0.
\end{split}
\end{align}
The first two equations are trivial in the self-dual limit, while the last two correspond to the refined Toda equations \eqref{Ref_Toda}. We observe that the r.h.s. of the first two equations is vanishing as a consequence of the property \eqref{prop_tau_txi}. This property can also be used to simplify the last two equations, and we find\footnote{We also use
\begin{align}
\begin{split}
\dfrac{\p^2\t(\bst,\bbst|\bsg^{-1/2}\bsxi)}{\t(\bst,\bbst|\bsg^{-1/2}\bsxi)}=p^2\dfrac{\p^2\t(\bsg\bst,\bsg\bbst|\bsg^{1/2}\bsxi)}{\t(\bsg\bst,\bsg\bbst|\bsg^{1/2}\bsxi)}.
\end{split}
\end{align}}
\begin{align}
\begin{split}
\Eq_{-1}^{(\sAbox,\vac)},\Eq_{1}^{(\vac,\sAbox)}:\quad&(1-pr_1)\p_{\pm1}^2\log\t(\bst,\bbst|\bsg^{-1/2}\bsxi)=0,\\
\Eq_{1}^{(\sAbox,\vac)},\Eq_{-1}^{(\vac,\sAbox)}:\quad&p\p_1\p_{-1}\log\t(\bsg\bst,\bsg\bbst|\bsg^{1/2}\bsxi)=1.
\end{split}
\end{align}
These remaining equations can be checked directly using the expression of the tau function found in \eqref{def_tau_txi}.

\paragraph{Remark} A slightly less trivial example is obtained from the following choice of algebra elements
\begin{equation}
G_{\bsxi,w}=x^+(w)e^{\sum_{k>0}\xi_ka_{-k}}.
\end{equation}
We compute the coproduct of this elements using \eqref{coproduct}
\begin{equation}
\D(G_{\bsxi,w})=G_{\bsg^{-1/2}\bsxi,w}\otimes G_{\bsg^{1/2}\bsxi}+G_{\bsg^{-1/2}\bsxi-\g^{1/2}\bsw}\otimes G_{\bsg^{1/2}\bsxi,\g w},
\end{equation}
with the understanding $\bsg^{-1/2}\bsxi-\bsg^{1/2}\boldsymbol{w}=\{\g^{-k/2}\xi_k-\g^{k/2}w^{k}\}$, the bilinear identities are actually linear in this case, with coefficients involving the refined tau functions $\t(\bst,\bbst|\bsxi)$ studied previously.

\subsection{Grading operators}
The study of the case where $G$ is defined using the grading operators introduced in \eqref{def_gradings} is enlightening. In fact, this analysis will also be useful for our application to R-matrix tau functions discussed in the next subsection. Note, however, that the grading operators are not genuine elements of the quantum toroidal $\gl(1)$ algebra, and so we cannot apply our formalism directly. Instead, we need to check explicitly if their representation of level $(2,0)$ obey the refined basic bilinear condition \eqref{com_Psi_rho} with the screening charges $\Psi_\pm$. It turns out that the grading $\bd$ brings no difficulties since $\rho^{(2,0)}(\bd)=\hd^{(1)}+\hd^{(2)}$ is a purely $u(1)$ factor which commutes with $\hat q=(\hd^{(1)}-\hd^{(2)})\log q$, and so with $\Psi_\pm$ as well.\footnote{Alternatively, we can use the simpler argument saying that only the ratio of weights $u_1/u_2$ enters in the identification between the representation $\rho^{(2,0)}_{u_1,u_2}$ and the free field representation of q-Virasoro, and this ratio is invariant under $\rho^{(2,0)}_{u_1,u_2}(\bd)=u_1\p_{u_1}+u_2\p_{u_2}$.}

On the other hand, the case of the grading operator $d$ is more subtle, and more interesting. Indeed, we observe that $\rho^{(2,0)}_{u_1,u_2}(\a^d)=\a^{L_0}\otimes\a^{L_0}$ for some $\a\in\mC$ does not commute with $\Psi_\pm$ but instead it obeys the relations
\begin{align}
\begin{split}
&\rho_{u'_1,u'_2}^{(2,0)}(\a^d)S_+(z)=S_+(\a z)\a^{-2\b\a_0}\rho_{u_1,u_2}^{(2,0)}(\a^d)\implies \rho_{u'_1,u'_2}^{(2,0)}(\a^d)\Psi_+(z)=\a^{n}\Psi_+\rho_{u_1,u_2}^{(2,0)}(\a^d)\quad \text{on}\ \CF_{-1,n}\\
&\rho_{u'_1,u'_2}^{(2,0)}(\a^d)S_-(z)=S_-(\a z)\a^{2\a_0}\rho_{u_1,u_2}^{(2,0)}(\a^d)\implies \rho_{u'_1,u'_2}^{(2,0)}(\a^d)\Psi_-(z)=\a^{m}\Psi_-\rho_{u_1,u_2}^{(2,0)}(\a^d)\quad \text{on}\ \CF_{m,-1}
\end{split}
\end{align}
As a result, the bilinear identities \eqref{Hirota_2dToda} have to be modified for the choice $G=\a^d$ by the introduction of an extra power of $\a$ in front of the first integral
\begin{align}
\begin{split}
&\a^n\oint_\infty \dfrac{dz}{2i\pi}z^{-n-1}e^{-\sum_{k>0}(\g^k t_k-t_k')\frac{1-t^k}{1-q^k}z^k}\t_u(\bst+[\g^{-1}z^{-1}],\bbst|G)\t_{uq^{-n}t^{-1}}(\bst'-[\g^{-2}z^{-1}],\bbst'|G)\\
+&\oint_0\dfrac{dz}{2i\pi}z^{-n-1}e^{\sum_{k>0}(\g^kt_{-k}-t_{-k}')p^kz^{-k}}\t_{ut^{-1}}(\bst,\bbst-[\g z]_{q,t}|G)\t_{uq^{-n}}(\bst',\bbst'+[z]_{q,t}|G)=0,\\
&\a^m\oint_\infty\dfrac{dz}{2i\pi}z^{-m-1}e^{\sum_{k>0}(\g^k t_k-t'_k)z^k}\t_{u}(\bst-[\g z^{-1}]_{t,q},\bbst|G)\t_{uqt^m}(\bst'+[z^{-1}]_{t,q},\bbst'|G)\\
+&\oint_0\dfrac{dz}{2i\pi}z^{-m-1} e^{-\sum_{k>0}(\g^k t_{-k}-t'_{-k})\frac{1-q^k}{1-t^k}z^{-k}}\t_{uq}(\bst,\bbst+[\g z]|G)\t_{ut^m}(\bst',\bbst'-[z]|G)=0,
\end{split}
\end{align}
where we also used the simple coproduct of $G$ to remove the summations.

When $G=\a^d$, the refined tau function \eqref{ref_canonical} can be computed using the property \eqref{prop_L0} of $L_0$, it gives
\begin{equation}
\t(\bst,\bbst|G)=e^{\sum_{k>0}k\a^kt_kt_{-k}},
\end{equation}
which is again independent of the weight $u$. We can use this expression to check the bilinear identities. Focusing on the first one, we compute the factors
\begin{align}
\begin{split}
&\t(\bst+[\g^{-1} z^{-1}],\bbst|G)\t(\bst'-[\g^{-2}z^{-1}],\bbst'|G)=e^{\sum_{k>0}Q^k(\g^k t_{-k}-t'_{-k})p^kz^{-k}},\\
&\t(\bst,\bbst-[\g z]_{q,t}|G)\t(\bst',\bbst'+[z]_{q,t}|G)=e^{-\sum_{k>0}Q^k(\g^k t_k-t_k')\frac{1-t^k}{1-q^k}z^k},
\end{split}
\end{align}
which can then be inserted in the integrand of the bilinear identity
\begin{align}
\begin{split}
&\a^n\oint_\infty \dfrac{dz}{2i\pi}z^{-n-1}e^{-\sum_{k>0}(\g^k t_k-t_k')\frac{1-t^k}{1-q^k}z^k}e^{\sum_{k>0}\a^k(\g^k t_{-k}-t'_{-k})p^kz^{-k}}\\
+&\oint_0\dfrac{dz}{2i\pi}z^{-n-1}e^{-\sum_{k>0}\a^k(\g^k t_k-t_k')\frac{1-t^k}{1-q^k}z^k}e^{\sum_{k>0}(\g^kt_{-k}-t_{-k}')p^kz^{-k}}=0.
\end{split}
\end{align}
Upon the rescaling of the integration variable $z\to \a z$ in the first integral, we recover the situation of the previous subsection with the same integrand in both integrals. For the same reason, the expansion in $\e_k$ of these integrals is vanishing, in agreement with our general result.

\section{R-matrix and refined tau functions}\label{sec:R}
In this section we study the generating function $\hat\tau_u(\bst,\bbst)$ of all refined tau functions \eqref{ref_canonical} which is built using the universal R-matrix $\CR$ of the quantum toroidal $\gl(1)$ algebra. The bilinear identities for this generator take the form of a two-term quadratic relation involving non-commutative objects.

\paragraph{The R-matrix of the quantum toroidal $\gl(1)$ algebra}
The universal R-matrix takes values in a completion of $\CE'\otimes\CE'$, where $\CE'$ denotes the quantum toroidal $\gl(1)$ algebra $\CE$ complemented with the two grading operators $d$ and $\bd$. Consider a PBW basis $\mathcal{B}'$ of $\CE'$ with respect to a pairing $(a|b^*)=\d_{a,b}$ for $a,a^*,b,b^*\in \CE'$, where ${}^*$ denotes the dual element (see \cite{Ng_shuffle,Ng_R}). The universal R-matrix has the form
\begin{align}
    \label{eq:R_PBW}
    \mathcal{R} = \sum_{b\in \mathcal{B}'} b^*\otimes b.
\end{align}
The degree elements $d,\bar d$ have a non-zero pairing only with the central elements $c,\bar c$ respectively, thus we have
\begin{align}
    \label{eq:R_PBW_X}
    \mathcal{R} = 
    \g^{d\otimes c+c\otimes d} \g^{\bc\otimes\bd+\bd\otimes\bc}
    \sum_{b\in \mathcal{B}} b^*\otimes b.
\end{align}
where $\mathcal{B}$ is the set of PBW basis elements excluding the degree elements. 
Recall the coproduct $\D$, defined in \eqref{coproduct}, and  the opposite coproduct $\D^{op}$. The universal R-matrix $\CR$ of the quantum toroidal algebra satisfies\footnote{We use the standard notation from the literature on quantum integrable systems in which the subscripts $ij$ in $\CR_{ij}$ indicate that the $\CR$-matrix acts on the $i$-th and $j$-th component of the tensor product.}
\begin{align}
    \label{eq:RD}
    &\CR \D(G) = \D^{op}(G) \CR,
    \qquad \forall G\in \CE, \\
    \label{eq:RRD1}
    &(\D \otimes 1) \CR = \CR_{13} \CR_{23},
\\
    \label{eq:RRD2}
&(1 \otimes \D) \CR = \CR_{13} \CR_{12}.
\end{align}
As a consequence of these equations $\CR$ satisfies the Yang--Baxter equation
\begin{align}
    \label{eq:YB}
    \CR_{23}\CR_{13}\CR_{12}
    =
    \CR_{12} \CR_{13}\CR_{23}.
\end{align}
We can further isolate the Cartan part in the universal R-matrix and write it as
\begin{equation}\label{def_R}
\CR=\g^{d\otimes c+c\otimes d}
\g^{\bc\otimes\bd+\bd\otimes\bc}\CK\bar\CR,\quad \CK=e^{-\sum_{k>0}\frac{\g^{-k(c\otimes1-1\otimes c)/2}}{c_k}a_k\otimes a_{-k}},
\end{equation}
where $a_{\pm k}$ are the modes of the currents $\psi^{\pm}(z)$ \eqref{eq:psi_currents} and the coefficients $c_k\in\mC(q_1,q_2)$ are defined in \eqref{com_ak}. The factor $\bar\CR$ takes values in $\CE^-\otimes\CE^+$ where $\CE^\pm$ are the algebras generated multiplicatively by the modes $x_k^\pm$. When it comes to the horizontal representation $\rho^{(1,0)}_u$, the presence of the operators $\g^{\bc\otimes\bd+\bd\otimes\bc}$ in \eqref{def_R} forces us to consider Fock spaces $\CF_u$ in which the spectral parameter $u$ varies as discussed in Section \ref{sec_Horiz}. To avoid this issue, the $\CL$-operator is usually defined by removing this factor before evaluating the second tensor component in the horizontal representation 
\begin{align}\label{eq:L_mat}
\mathcal{L}(u):=\left(1\otimes\rho^{(1,0)}_u\right)\left(\g^{-\bc\otimes\bd-\bd\otimes\bc} \CR\right)=\gamma^{-\bar c\otimes \hat d}\left(1\otimes\rho^{(1,0)}_u\right) \CR.
\end{align}
This operator $\mathcal{L}(u)\in\CE'\otimes\End(\CF_u)$ was considered in \cite{FJMM_BA} where it played a key role in deriving the Bethe Ansatz of the associated integrable model defined on the physical space of horizontal representations. The transfer matrix of this integrable model is computed by taking a twisted trace of a product of the horizontal-horizontal R-matrices
\begin{align}
    \label{eq:R_hh}
    R(u_2/u_1):=\left(\rho_{u_1}^{(1,0)}\otimes \rho_{u_2}^{(1,0)}\right) \CR\in\End(\CF_{u_1}\otimes\CF_{u_2}).
\end{align}
The matrix $R(u_2/u_1)$ was studied in \cite{Feigin_Tsym,Fukuda2017,Garbali2021,G_Ng}, and it is known in the rational limit as the Maulik--Okounkov R-matrix (see \cite{MO2012,Proch,Smirn,Litv_Vilk}). Applying $1\otimes \left(\rho_{u_1}^{(1,0)}\otimes \rho_{u_2}^{(1,0)}\right)$ to the Yang--Baxter equation \eqref{eq:YB} produces the RLL equation
\begin{align}
    \label{eq:RLL}
    R_{23}(u_2/u_1)\CL_{13}(u_2) \CL_{12}(u_1)
    =
    \CL_{12}(u_1) \CL_{13}(u_2) R_{23}(u_2/u_1),
\end{align}
where we used the equations \eqref{eq:L_mat}, \eqref{eq:R_hh} and cancelled the common factor $\gamma^{\bar c (\hat d^{(1)}+\hat d^{(2)})}$ that commutes with $R(u_2/u_1)$.

\paragraph{Universal refined tau function} The universal refined tau function is defined as the following evaluation of the $\CL$-operator
\begin{equation}\label{def_hat_tau}
\hat\tau_u(\bst,\bbst)=
\left(1\otimes\bra{u}e^{\sum_{k>0} t_kJ_k}\right)
\CL(u)\left(1\otimes e^{\sum_{k>0}t_{-k}J_{-k}}\ket{u}\right).
\end{equation}
This tau function is represented graphically in figure \ref{fig_tau_R}. The tau function $\hat\tau_u(\bst,\bbst)$ encodes the monodromy matrix associated to the horizontal representation. A particular monodromy matrix element labelled by two partitions $\lambda,\mu$ can be accessed by taking the coefficient of $t_\lambda,t_{-\mu}$ in $\hat\tau_u(\bst,\bbst)$. Therefore this tau function is intimately related to the integrable model which was considered in \cite{FJMM_BA,FJMM2}.

In fact, this universal tau function can be seen as the generating function of all refined tau functions of the form \eqref{ref_canonical}. Indeed, by inserting the expansion \eqref{eq:R_PBW_X} of the universal R-matrix into \eqref{eq:L_mat}, the equation \eqref{def_hat_tau} can be rewritten as a sum of refined tau functions over the elements of the PBW basis (excluding grading elements)
\begin{align}
    \label{eq:tau_bb}
    \hat\tau_u(\bst,\bbst): = \g^{d}\sum_{b\in \mathcal{B}} b^*\t_u(\bsg^c \bst,\bbst|b)\in\CE[\bst,\bbst].
\end{align}
The rescaling of the time parameters $\bsg^{c}\bst=(\g^c t_1,  \g^{2c}t_2, \cdots)$ is a consequence of the presence of the grading, and it is not essential in our argumentation. Any element $G\in \CE$ can be expanded in the PBW basis as $G=\sum_{b\in \mathcal{B}} c_b b$ with some coefficients $c_b$. Then, the pairing of the algebra can be used to recover the corresponding tau function
\begin{equation}
\tau_u(\bst,\bbst|G)=\sum_{b\in\CB}c_b\tau_u(\bst,\bbst|b)=(G|\g^{-d}\hat\tau_u(\bsg^{-c}\bst,\bbst)).
\end{equation} 

Alternatively, the universal tau functions can be obtained as the limit $\k\to0$ of a transfer matrix dressed with exponential of the times $(\bst,\bbst)$
\begin{equation}
\hat\tau_u(\bst,\bbst)=\lim_{\k\to0}\hat\tau_u(\bst,\bbst|\k),\quad \hat\tau_u(\bst,\bbst|\k)=1\otimes\tr_{\CF_u}\left[\k^{L_0}e^{\sum_{k>0} t_kJ_k}
\CL(u)e^{\sum_{k>0}t_{-k}J_{-k}}\right].
\end{equation} 
These objects display some similarity with the master T-operators studied in \cite{Alexandrov2011a,Zabrodin2012,Alexandrov2013} which are tau functions of the KP hierarchy. However, the definitions are slightly different since the sum over representations defining the master T-operators is replaced here with a single horizontal Fock representation. As a result, contrary to master T-operators, our operators $\hat\tau_u(\bst,\bbst|\k)$ do not commute for different times. It would be interesting to study further these objects, but we will leave this question for a future work and focus instead on the universal tau functions $\hat\tau_u(\bst,\bbst)$.

\begin{figure}
\begin{center}
\begin{tikzpicture}[scale=.5]
\draw (-2,0) -- (2,0);
\draw (0,-2) -- (0,2);
\node[scale=1,right] at (2,0) {$e^{\sum_{k>0}t_{-k}J_{-k}}\ket{u}$};
\node[scale=1,left] at (-2,0) {$\bra{u}e^{\sum_{k>0}t_{k}J_{k}}$};
\end{tikzpicture}
\caption{Definition of the universal refined tau function from the $\CL$-operator. The $\CL$-operator is represented by the cross.}
\label{fig_tau_R}
\end{center}
\end{figure}
\begin{figure}
\begin{center}
\begin{tikzpicture}[scale=.5]
\draw (-2,0) -- (2,0);
\draw (-2,2) -- (2,2);
\draw (0,-2) -- (0,4);
\node[scale=1,right] at (2,0) {$e^{\sum_{k>0} t'_{-k}J_{-k}}\ket{u_2}$};
\node[scale=1,left] at (-2,0) {$\bra{u_2}e^{\sum_{k>0}t'_{k}J_{k}}$};
\node[scale=1,right] at (2,2) {$e^{\sum_{k>0}t_{-k}J_{-k}}\ket{u_1}$};
\node[scale=1,left] at (-2,2) {$\bra{u_1}e^{\sum_{k>0}t_{k}J_{k}}$};
\end{tikzpicture}
\caption{Element entering the basic bilinear condition.}
\label{fig_tau_R2}
\end{center}
\end{figure}

\paragraph{Bilinear identities} The universal refined tau function $\hat\tau_u(\bst,\bbst)$ also satisfies a set of bilinear identities of the form \eqref{Ref_Hirota_2dToda}, but the property \eqref{eq:RRD1} of the R-matrix allows us to replace the sum over elements $G_\a\otimes G_\a'$ coming from the coproduct by a product of universal tau functions
\begin{align}\label{Hirota_univ}
\begin{split}
&\g^{nc}\oint_\infty \dfrac{dz}{2i\pi}z^{-n-1}e^{-\sum_{k>0}(\g^k t_k-t_k')\frac{1-t^k}{1-q^k}z^k}\hat\t_{uq^{-n}t^{-1}}(\bst'-[\g^{-2}z^{-1}],\bbst')\hat\t_u(\bst+[\g^{-1}z^{-1}],\bbst)\\
+&\oint_0\dfrac{dz}{2i\pi}z^{-n-1}e^{\sum_{k>0}(\g^kt_{-k}-t_{-k}')p^kz^{-k}}\hat\t_{uq^{-n}}(\bst',\bbst'+[z]_{q,t})\hat\t_{ut^{-1}}(\bst,\bbst-[\g z]_{q,t})=0,\\
&\g^{mc}\oint_\infty\dfrac{dz}{2i\pi}z^{-m-1}e^{\sum_{k>0}(\g^k t_k-t'_k)z^k}\hat\t_{uqt^m}(\bst'+[z^{-1}]_{t,q},\bbst')\hat\t_{u}(\bst-[\g z^{-1}]_{t,q},\bbst)\\
+&\oint_0\dfrac{dz}{2i\pi}z^{-m-1} e^{-\sum_{k>0}(\g^k t_{-k}-t'_{-k})\frac{1-q^k}{1-t^k}z^{-k}}\hat\t_{ut^m}(\bst',\bbst'-[z])\hat\t_{uq}(\bst,\bbst+[\g z])=0.
\end{split}
\end{align}
The derivation of these identities follow the same lines as what was done in Appendix \ref{app:Bilin} to obtain the relations \eqref{Ref_Hirota_2dToda}. However, the presence of grading operators in the R-matrix renders the derivation a bit more complicated, and we will discuss it briefly below. The replacement of the coproduct with a product of universal tau-functions also occurs in the difference-differential equations, and for instance the $(q,t)$-deformed Toda equations \eqref{Ref_Toda} take the form
\begin{align}
\begin{split}
\Eq_{-1}^{(\vac,\sAbox)}:\quad&\p_{-1}\hat\t_{up}(\bsg\bst,\bsg\bbst)\p_1\hat\t_u(\bst,\bbst)-p\p_1\p_{-1}\hat\t_{up}(\bsg\bst,\bsg\bbst)\hat\t_u(\bst,\bbst)+\g^c\hat\t_{uq}(\bsg\bst,\bsg\bbst)\hat\t_{ut^{-1}}(\bst,\bbst)=0,\\
\Eq_1^{(\sAbox,\vac)}:\quad &\p_1\hat\t_{up^{-1}}(\bsg\bst,\bsg\bbst)\p_{-1}\hat\t_u(\bst,\bbst)-p\p_1\p_{-1}\hat\t_{up^{-1}}(\bsg\bst,\bsg\bbst)\hat\t_u(\bst,\bbst)+\g^{c}\hat\t_{uq^{-1}}(\bsg\bst,\bsg\bbst)\hat\t_{ut}(\bst,\bbst)=0.
\end{split}
\end{align}

While the expression for the universal R-matrix is known \cite{Ng_R}, it is not easy to use it to show directly that the universal refined tau function \eqref{def_hat_tau} satisfies the identities \eqref{Hirota_univ} (see Theorem 4.16 in \cite{Ng_R}). On the other hand, for certain choices of representations, more explicit expressions of the R-matrix have been obtained, and can be used to study these equations. In the next subsection, we will examine two important cases, namely the vertical-horizontal and horizontal-horizontal Fock R-matrices.

The starting point for the derivation of the bilinear identities \eqref{Hirota_univ} is again the $(q,t)$-deformed basic bilinear condition \eqref{com_Psi_rho}. This equation can be summed over the elements of the algebra to produce
\begin{align}\label{eq:com_Psi_bb}
 \sum_{b\in\mathcal{B}}b^* 
 \left( \Psi_{\pm} \rho_{u_1,u_2}^{(2,0)}(b)
 -
 \rho_{u_1',u_2'}^{(2,0)}(b) \Psi_{\pm}   \right)= 0.
\end{align}
We can rewrite this identity in terms of $\CR$ by recalling \eqref{eq:R_PBW_X}
\begin{align}\label{eq:com_R_Psi}
 (1\otimes \Psi_{\pm}) (1\otimes\rho_{u_1,u_2}^{(2,0)})
 \left(\g^{-d\otimes c-c\otimes d-\bc\otimes\bd-\bd\otimes\bc}\right)
 \CR  =(1\otimes\rho_{u_1',u_2'}^{(2,0)})
 \left(\g^{-d\otimes c-c\otimes d-\bc\otimes\bd-\bd\otimes\bc}\CR\right)
 (1\otimes \Psi_{\pm}).
\end{align}
After a careful treatment of the grading operators, we end up with\footnote{The evaluation of $\g^{d\otimes c+c\otimes d+\bc\otimes\bd+\bd\otimes\bc}$ under the representation $\rho^{(2,0)}_{u_1,u_2}$ brings the factors $\g^{\bar c(\hat d^{(1)}+\hat d^{(2)})}$ and $\g^{d\otimes2}$ which commute with $\Psi_\pm$, and $\g^{c\otimes\CL_0}$ which follows from
\begin{equation}
\rho^{(2,0)}_{u_1,u_2}(d)=\CL_0-\sum_{k>0}\dfrac{\b_{-k}\b_k}{kc_k(q_3^k-q_3^{-k})},\quad \CL_0=\sum_{k>0}\dfrac{1-t^k}{1-q^k}\a_{-k}\a_k,
\end{equation} 
the $u(1)$-part trivially commuting with $\Psi_\pm$. This extra factor $\g^{c\CL_0}$ obeys non-trivial commutation relations with the screening currents, namely $\g^{c\CL_0}S_+(z)\g^{-c\CL_0}=S_+(\g^c z)\g^{-2\b\a_0c}$ and $\g^{c\CL_0}S_-(z)\g^{-c\CL_0}=S_-(\g^c z)\g^{2\a_0c}$. After integrating out and evaluating over the Fock space $\CF_{-1,n}$ and $\CF_{m,-1}$, it brings the extra factors $\g^{nc}$ and $\g^{mc}$ in the following equations.}
\begin{align}\label{eq:com_R_Psi_p}
 (\g^{c n}\otimes \Psi_{+}) 
  (1\otimes\rho_{u_1,u_2}^{(2,0)})
 \CR  &=(1\otimes\rho_{u_1',u_2'}^{(2,0)})
 \CR (1\otimes \Psi_{+}), \\
 \label{eq:com_R_Psi_m}
 (\g^{c m}\otimes \Psi_{-}) 
  (1\otimes\rho_{u_1,u_2}^{(2,0)})
 \CR  &=(1\otimes\rho_{u_1',u_2'}^{(2,0)})
 \CR (1\otimes \Psi_{-}),
\end{align}
where $n$ and $m$ refer to the labels of the Fock spaces in \eqref{eq:Psi-F}. By recalling $\rho_{u_1,u_2}^{(2,0)}$ from \eqref{def_rho2}, the identity \eqref{eq:RRD2} for the universal R-matrix and the definition \eqref{eq:L_mat} of the operator $\CL(u)$ we can write
\begin{align}
(1\otimes\rho_{u_1,u_2}^{(2,0)})\CR=\gamma^{\bar c(\hat d^{(1)}+\hat d^{(2)})}\CL_{13}(u_2)\CL_{12}(u_1).
\end{align}
Inserting this on both sides of \eqref{eq:com_R_Psi_p} and \eqref{eq:com_R_Psi_m}, we find after cancelling the grading operators 
\begin{align}\label{eq:com_L_Psi_p}
(\gamma^{c n}\otimes \Psi_{+})  \CL_{13}(u_2)\CL_{12}(u_1)&=\CL_{13}(u_2')\CL_{12}(u_1') (1\otimes \Psi_{+}), \\
\label{eq:com_L_Psi_m}
(\gamma^{c m}\otimes \Psi_{-})  \CL_{13}(u_2)\CL_{12}(u_1)&=\CL_{13}(u_2')\CL_{12}(u_1') (1\otimes \Psi_{-}).
\end{align}
In order to deal with the zero modes of $\Psi_\pm$ it is useful to introduce two operators $\Phi_n^+$ and $\Phi_m^-$
\begin{align}
    \label{eq:Phi}
    \Phi^+_n := e^{-\beta \hat q} \Psi_+ = 
    t^{ \hat d^{(2)}-\hat d^{(1)}}\Psi_+ ,
    \qquad
    \Phi^-_m := e^{ \hat q} \Psi_- = q^{ \hat d^{(1)}-\hat d^{(2)}} \Psi_-.
\end{align}
These operators now have an index since they act in the Fock spaces labelled as follows
\begin{equation}\label{eq:Phi-F}
\Phi^+_n:\CF_{-1,n}\to\CF_{-1,n},\quad \Phi^-_m:\CF_{m,-1}\to\CF_{m,-1}.
\end{equation}
We replace $\Psi_\pm$ with these operators in \eqref{eq:com_L_Psi_p} and \eqref{eq:com_L_Psi_m} and use \eqref{weights}
\begin{align}\label{eq:X-tilde}
(\gamma^{c n}\otimes \Phi^{+}_n)  \CL_{13}(q^{-n}t^{-1}u)\CL_{12}(u)&=\CL_{13}(q^{-n} u)\CL_{12}(t^{-1} u)(1\otimes \Phi^{+}_n),\\
\label{eq:Y-tilde}
(\gamma^{c m}\otimes \Phi^{-}_m)  \CL_{13}(t^m q u)\CL_{12}(u)&= \CL_{13}(t^m u)\CL_{12}(q u) (1\otimes \Phi_m^{-}).
\end{align}
These two equations can be interpreted as RLL equations as mentioned in the introduction. The two equations \eqref{eq:X-tilde} and \eqref{eq:Y-tilde} give rise to two bilinear relations for the universal refined tau function \eqref{def_hat_tau}. We sandwich \eqref{eq:X-tilde} and \eqref{eq:Y-tilde} between the states (see figure \ref{fig_tau_R2}) \begin{align*}
    1\otimes \bra{u_1} e^{\sum_{k>0}t_{k}J_{k}} \otimes \bra{u_2} e^{\sum_{k>0}t'_{k}J_{k}}, 
    \qquad 
        1\otimes  e^{\sum_{k>0}t_{-k}J_{-k}}\ket{u_1}\otimes  e^{\sum_{k>0}t'_{-k}J_{-k}}\ket{u_2}.
\end{align*}
Finally, the action on these states with $\Phi^{\pm}$ can be calculated in the same way as what was done in Appendix \ref{app:Bilin}.

\paragraph{Remark} We can use the intertwining property $\CR\D=\D^{op}\CR$ of the R-matrix, to derive some interesting identities on the refined tau functions. For instance, using the two Cartan currents, we find
\begin{align}
\begin{split}
&\hat\tau_u(\bst,\bbst)\psi^+(z)=e^{\sum_{k>0}t_{-k}z^{-k}\g^{ck/2}(1-q_2^k)(1-q_3^k)}\psi^+(\g^{-1}z)\hat\tau_u(\bst+[\g^{2-c/2}z^{-1}]_2-[\g^{-c/2}z^{-1}]_2,\bbst),\\
&\hat\tau_u(\bst,\bbst+[\g^{c/2} z]_1-[\g^{2+c/2} z]_1)\psi^-(z)=e^{\sum_{k>0}t_{k}z^{k}\g^{-ck/2}(1-q_1^k)(1-q_3^k)}\psi^-(\g z)\hat\tau_u(\bst,\bbst),
\end{split}
\end{align}
with $\bst\pm[z]_\a=\bst\pm[z]\mp[q_\a z]=\{t_k\pm\frac1kz^k(1-q_\a^k)\}$, while the currents $x^\pm(z)$ produce the two relations
\begin{align}
\begin{split}
&\hat\tau_u(\bst,\bar\bst)x^+(z)-e^{\sum_{k>0}t_kz^k(1-q_1^k)(1-q_3^k)}x^+(\g z)\hat\tau_u(\bst,\bar\bst)\\
&=ue^{\sum_{k>0}t_kz^k(1-q_1^k)}\hat\tau_u(\bst-[z^{-1}]_2,\bar\bst)-ue^{-\sum_{k>0}t_{-k}z^{-k}\g^{-ck}(1-q_2^k)}\hat\tau_u(\bst,\bbst+[\g^{ck}z]_1,\bar\bst)\psi^-(\g^{c/2}z),\\
&x^-(z)\hat\tau_u(\bst,\bbst)-e^{-\sum_{k>0}t_{-k}z^{-k}\g^{-k}(1-q_2^k)(1-q_3^k)}\hat\tau_u(\bst,\bbst)x^-(\g z)\\
&=u^{-1}e^{\sum_{k>0}t_{-k}z^{-k}\g^k(1-q_2^k)}\hat\tau_u(\bst,\bbst-[\g z]_1)-u^{-1}e^{-\sum_{k>0}t_kz^k\g^{(1+c)k}(1-q_1^k)}\psi^+(\g^{c/2}z)\hat\tau_u(\bst+[\g^{1-c}z^{-1}]_2,\bbst).
\end{split}
\end{align}

\subsection{Vertical refined tau function}\label{sec_VH}
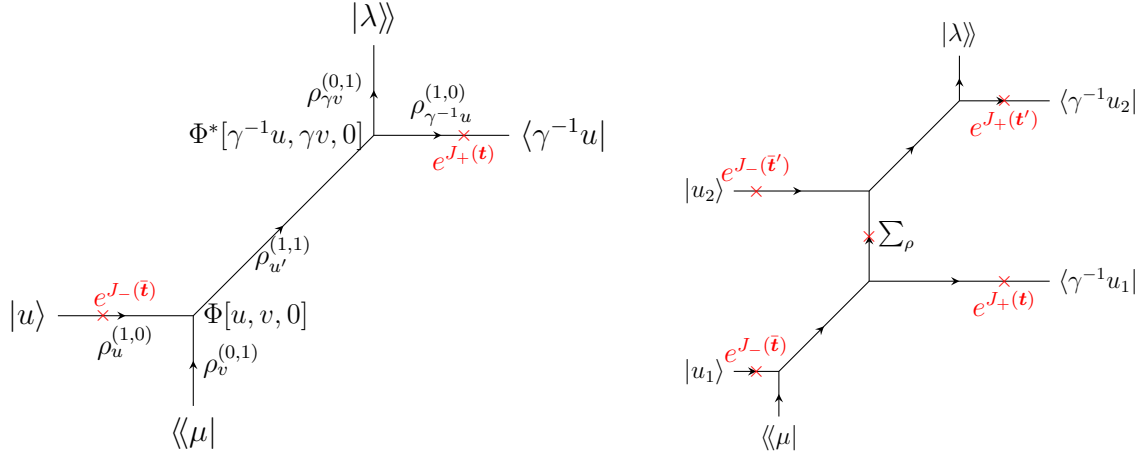
\begin{figure}
\begin{center}
\begin{tikzpicture}[scale=.6]
\draw[postaction={on each segment={mid arrow=black}}] (-3,0) -- (0,0) -- (4,4) -- (7,4);
\draw[postaction={on each segment={mid arrow=black}}] (0,-2) -- (0,0);
\draw[postaction={on each segment={mid arrow=black}}] (4,4) -- (4,6);
\node[scale=1,left] at (-3,0) {$\ket{u}$};
\node[scale=1,right] at (7,4) {$\bra{\g^{-1}u}$};
\node[scale=1,below] at (0,-2) {$\dbra{\mu}$};
\node[scale=1,above] at (4,6) {$\dket{\l}$};
\node[scale=.9,right] at (0,0) {$\Phi[u,v,0]$};
\node[scale=.9,left] at (4,4) {$\Phi^\ast[\g^{-1}u,\g v,0]$};
\node[scale=.9,below] at (-1.5,0) {$\rho^{(1,0)}_{u}$};
\node[scale=.9,above] at (5.5,4) {$\rho^{(1,0)}_{\g^{-1}u}$};
\node[scale=.9,below] at (2,2) {$\rho^{(1,1)}_{u'}$};
\node[scale=.9,left] at (4,5) {$\rho^{(0,1)}_{\g v}$};
\node[scale=.9,right] at (0,-1) {$\rho^{(0,1)}_{v}$};
\node[scale=.7 ,thick, red] at (-2,0) {$\times$};
\node[scale=.7 ,thick, red] at (6,4) {$\times$};
\node[scale=.9, red, below] at (6,4) {$e^{J_+(\bst)}$};
\node[scale=.9, red, above] at (-1.5,0) {$e^{J_-(\bbst)}$};
\end{tikzpicture}
\hspace{5mm}
\begin{tikzpicture}[scale=.3]
\draw[postaction={on each segment={mid arrow=black}}] (-2,0) -- (0,0) -- (4,4) -- (12,4);
\draw[postaction={on each segment={mid arrow=black}}] (-2,8) -- (4,8) -- (8,12) -- (12,12);
\draw[postaction={on each segment={mid arrow=black}}] (0,-2) -- (0,0);
\draw[postaction={on each segment={mid arrow=black}}] (4,4) -- (4,8);
\draw[postaction={on each segment={mid arrow=black}}] (8,12) -- (8,14);
\node[scale=.8,left] at (-2,0) {$\ket{u_1}$};
\node[scale=.8,right] at (12,4) {$\bra{\g^{-1}u_1}$};
\node[scale=.8,left] at (-2,8) {$\ket{u_2}$};
\node[scale=.8,right] at (12,12) {$\bra{\g^{-1}u_2}$};
\node[scale=.8,below] at (0,-2) {$\dbra{\mu}$};
\node[scale=.8,above] at (8,14) {$\dket{\l}$};
\node[scale=.8,right] at (4,6) {$\sum_\rho$};
\node[scale=.7 ,thick, red] at (-1,0) {$\times$};
\node[scale=.7 ,thick, red] at (10,4) {$\times$};
\node[scale=.9, red, below] at (10,4) {$e^{J_+(\bst)}$};
\node[scale=.9, red, above] at (-1,0) {$e^{J_-(\bbst)}$};
\node[scale=.7 ,thick, red] at (-1,8) {$\times$};
\node[scale=.7 ,thick, red] at (10,12) {$\times$};
\node[scale=.9, red, below] at (10,12) {$e^{J_+(\bst')}$};
\node[scale=.9, red, above] at (-1,8) {$e^{J_-(\bbst')}$};
\node[scale=.7 ,thick, red] at (4,6) {$\times$};
\node[scale=.7, red, left] at (4,6) {$ $};
\end{tikzpicture}
\end{center}
\caption{Network of representations corresponding to the tau function $\t_u^{\l,\mu}(\bst,\bbst)$, and the configuration entering in the bilinear identities (right).}
\label{fig_tau}
\end{figure}

The {\it vertical refined tau function} is defined as the evaluation of the universal refined tau function \eqref{eq:tau_bb} in the vertical Fock representation. In this section we will show directly that this refined tau function satisfies the bilinear equations \eqref{Hirota_univ}. 

The vertical-horizontal R-matrix is obtained from the evaluation of the universal R-matrix $\CR$ in the tensor product of representations $\rho_v^{(0,1)}\otimes\rho_u^{(1,0)}$. This R-matrix has the remarkable property to be expressible as a product of the intertwiners introduced by Awata--Feigin--Shiraishi (AFS) in \cite{AFS}\footnote{The normalization factor $\CG(q_3^{-1})$ plays no role here, it is expressed using the function $\CG(z)=\prod_{i,j=1}^\infty(1-q_1^{i-1}q_2^{j-1}z)$ (assuming $|q_1|,|q_2|<1$).}
\begin{equation}
\left(\rho_v^{(0,1)}\otimes\rho_u^{(1,0)}\right)\CR=\CG(q_3^{-1})\Phi^\ast[\g^{-1}u,\g v,0]\Phi[u,v,0].
\end{equation}
We refer to Appendix \ref{app:Intertwiners} for the definition of the vertical representation and a brief reminder on the AFS intertwining operators. We note that, due to the presence of the grading elements in \eqref{def_R}, explicitly
\begin{equation}
\left(\rho_v^{(0,1)}\otimes\rho_u^{(1,0)}\right)(\g^{d\otimes c+c\otimes d} \g^{\bc\otimes\bd+\bd\otimes\bc})=\g^{-v\p_v+u\p_u},
\end{equation}
this R-matrix should be considered as a map $\CF_v\otimes\CF_u\to\CF_{\g v}\otimes\CF_{\g^{-1}u}$. Expanding the intertwiners on their vertical components, we deduce a very explicit expression of the R-matrix
\begin{equation}\label{RVH}
\left(\rho_v^{(0,1)}\otimes\rho_u^{(1,0)}\right)\CR=\CG(q_3^{-1})\sum_{\l,\mu}n_\l n_\mu\dket{\l,\g v}\dbra{\mu,v}\otimes \Phi_\l^\ast[\g^{-1}u,\g v,0]\Phi_\mu[u,v,0].
\end{equation}
The summation is over all pairs of Young diagrams $\l,\mu$, and the product $\Phi_\l^\ast[\g^{-1}u,\g v,0]\Phi_\mu[u,v,0]$ is a vertex operator that can be deduced from the expressions \eqref{def_AFS}. 

\paragraph{Derivation of the vertical refined tau function} These tau functions are simply defined as the evaluation of the universal objects \eqref{def_hat_tau} in the vertical Fock representation $\rho_v^{(0,1)}$, we denote them
\begin{equation}
\hat\t_{u,v}^{(V)}(\bst,\bbst)=\rho_v^{(0,1)}(\hat\t_u(\bst,\bbst)):\CF_v\to\CF_{\g v}.
\end{equation} 
These objects have been represented graphically on figure \ref{fig_tau} (left). From the explicit form \eqref{RVH} of the R-matrix, we can compute the matrix elements $\t_{u,v}^{\l,\mu}(\bst,\bbst)$ of these operators in the vertical basis $\dket{\l,v}$
\begin{align}
\begin{split}\label{def_tVH_lm}
&\hat \t_{u,v}^{(V)}(\bst,\bbst)=\sum_{\l,\mu}n_\l n_\mu \t_{u,v}^{\l,\mu}(\bst,\bbst)\ \dket{\l,\g v}\dbra{\mu,v},\\
&\t_{u,v}^{\l,\mu}(\bst,\bbst)=\CG(q_3^{-1})\bra{\vac,\g^{-1}u}e^{J_+(\bst)}\Phi_\l^\ast[\g^{-1}u,\g v,0]\Phi_\mu[u,v,0]e^{J_-(\bbst)}\ket{\vac,u}.
\end{split}
\end{align}
The matrix elements $\t_{u,v}^{\l,\mu}(\bst,\bbst)$ can be seen as a class of (scalar) refined tau functions indexed by two partitions $\l,\mu$, and depending on two complex parameters $u,v$ in addition to the set of time parameters $\bst,\bbst$. They obey the following bilinear identities obtained by inserting the previous expansion in the universal formula \eqref{Hirota_univ} evaluated in the vertical representation
\begin{align}\label{Hirota_VH}
\begin{split}
&\sum_\rho n_\rho\oint_\infty \dfrac{dz}{2i\pi}z^{-n-1}e^{-\sum_{k>0}(\g^k t_k-t_k')\frac{1-t^k}{1-q^k}z^k}\t_{uq^{-n}t^{-1},\g v}^{\l,\rho}(\bst'-[\g^{-2}z^{-1}],\bbst')\t_{u,v}^{\rho,\mu}(\bst+[\g^{-1}z^{-1}],\bbst)\\
+&\sum_\rho n_\rho\oint_0\dfrac{dz}{2i\pi}z^{-n-1}e^{\sum_{k>0}(\g^kt_{-k}-t_{-k}')p^kz^{-k}}\t_{uq^{-n},\g v}^{\l,\rho}(\bst',\bbst'+[z]_{q,t})\t_{ut^{-1},v}^{\rho,\mu}(\bst,\bbst-[\g z]_{q,t})=0.
\end{split}
\end{align}
For simplicity, we only wrote down the first bilinear identity obtained from the screening charge $\Psi_+$, and the second one has a similar structure. To illustrate this formula, the integrands have been represented schematically on figure \ref{fig_tau} (right), with $J_+(\bst)=\sum_kt_kJ_k$ and $J_-(\bbst)=\sum_kt_{-k}J_{-k}$.

The correlation functions entering in the expression of the matrix elements $\t_{u,v}^{\l,\mu}(\bst,\bbst)$ in \eqref{def_tVH_lm} can be evaluated using the expressions \eqref{def_AFS} of the vertex operators $\Phi_\l$ and $\Phi_\l^\ast$. The resulting expression involves the Nekrasov factor $\CN_{\l,\mu}(\a)$ defined in \eqref{def_Nek}
\begin{align}\label{calc_tau_VH}
\begin{split}
\t_{u,v}^{\l,\mu}(\bst,\bbst|\bQ)=&(-\g)^{|\mu|}u^{|\mu|-|\l|}\prod_{\sAbox\in\mu}\chi_{\sAbox}^{-1}\ \CN_{\mu,\l}(1)\times e^{\sum_{k>0}kt_kt_{-k}}e^{-\sum_{k>0}\frac{1-q_3^{k}}{1-q_2^k}t_k v^k}\\
&\times  e^{\sum_{k>0}(1-q_1^k)t_k v^k\left(S_\mu^{(k)}-q_3^kS_\l^{(k)}\right)}e^{-\sum_{k>0}(1-q_2^k)t_{-k} v^{-k}\left(S_\mu^{(-k)}-S_\l^{(-k)}\right)},\\
\text{with}\quad &S_\l^{(k)}=\sum_{\sAbox\in\l}\chi_{\sAbox}^k=\dfrac1{1-q_2^k}\sum_{i=1}^\infty q_1^{(i-1)k}(1-q_2^{\l_ik})=\dfrac1{1-q_1^k}\sum_{j=1}^\infty q_2^{(j-1)k}(1-q_1^{\l'_jk}).
\end{split}
\end{align}
We observe that the dependence in the vertical weight $v$ can be absorbed in the simple rescaling $\t_k\to v^{-k}t_k$ of the time variables.

It is possible to verify directly that the expressions for the tau functions obtained in \eqref{calc_tau_VH} satisfy the bilinear identity. Inserting this expression in the equation \eqref{Hirota_VH}, it takes the form
\begin{align}
\begin{split}
&\sum_\rho n_\rho\t_{uq^{-n}t^{-1},\g v}^{\l,\rho}(\bst',\bbst')\t_{u,v}^{\rho,\mu}(\bst,\bbst)q^{n(|\l|-|\rho|)}\left(t^{|\l|-|\rho|}\CI^{(\infty)}_{\l,\r,\m}+t^{|\rho|-|\mu|}\CI^{(0)}_{\l,\r,\m}\right)=0,
\end{split}
\end{align}
with
\begin{align}
\begin{split}
&\CI^{(\infty)}_{\l,\r,\m}=\oint_\infty \dfrac{dz}{2i\pi}z^{-n-1}K(z)\prod_{\sAbox\in\l}\dfrac{z-q_1\g v\chi_{\sAbox}}{z-\g v\chi_{\sAbox}}
\prod_{\sAbox\in\mu}\dfrac{z-q_1\g^{-1} v\chi_{\sAbox}}{z-\g^{-1} v\chi_{\sAbox}}\prod_{\sAbox\in\rho}\dfrac{(z-\g^{-1}v\chi_{\sAbox})(z-\g v\chi_{\sAbox})}{(z-q_1\g^{-1}v\chi_{\sAbox})(z-q_1\g v\chi_{\sAbox})}\\
&\CI^{(0)}_{\l,\r,\m}=t^{|\l|+|\mu|-2|\rho|}\oint_0 \dfrac{dz}{2i\pi}z^{-n-1}K(z)\prod_{\sAbox\in\l}\dfrac{z-q_1\g v\chi_{\sAbox}}{z-\g v\chi_{\sAbox}}
\prod_{\sAbox\in\mu}\dfrac{z-q_1\g^{-1} v\chi_{\sAbox}}{z-\g^{-1} v\chi_{\sAbox}}\prod_{\sAbox\in\rho}\dfrac{(z-\g^{-1}v\chi_{\sAbox})(z-\g v\chi_{\sAbox})}{(z-q_1\g^{-1}v\chi_{\sAbox})(z-q_1\g v\chi_{\sAbox})},
\end{split}
\end{align}
where we introduced the kernel
\begin{equation}\label{def_K}
K(z)=K(z|\bst,\bbst|\bst',\bbst')=e^{-\sum_{k>0}(\g^k t_k-t_k')\frac{1-t^k}{1-q^k}z^k}e^{\sum_{k>0}(\g^kt_{-k}-t_{-k}')p^kz^{-k}}.
\end{equation}
Note that the rational function entering as the integrand of $\CI_\infty$ is obtained as the summation of a power series in $\chi_{\sAbox}/z$, while the same rational function enters in the integrand of $\CI_0$ as the summation of a series in $z/\chi_{\sAbox}$. Since these two rational functions coincide, the quantity $t^{|\l|-|\rho|}\CI_\infty+t^{|\rho|-|\mu|}\CI_0$ entering in the bilinear identity has indeed a vanishing expansion in $\e_k$.

\paragraph{Remark.} The tau functions will have the same structure if we take $\rho=\rho_v^{(0,N)}$ with $N>1$. The case of $\rho_v^{(0,2)}$ is actually particularly interesting as it is again connected to the deformed Virasoro algebra. This type of quantities were introduced in \cite{Jimbo2017} in the context of q-Painlev\'e equation and it would be interesting to see if their results can be recovered as a reduction of our hierarchy.

\subsection{Horizontal refined tau function}
The {\it horizontal refined tau function} is defined as the evaluation of the universal refined tau function \eqref{eq:tau_bb} in the horizontal Fock representation. In this section we will compute this object using the expression of the horizontal-horizontal R-matrix obtained in \cite{G_Ng}, and show that the resulting expression  obeys the bilinear equations derived in \eqref{Hirota_univ}.

The horizontal-horizontal R-matrix \eqref{eq:R_hh} can be written in the following form \cite{G_Ng}\footnote{In our current conventions the coproduct corresponds to the opposite coproduct of \cite{G_Ng} and the spectral parameters are inverted therefore the horizontal R-matrix \eqref{eq:R-hor} equals to the inverse of the R-matrix in \cite{G_Ng} with inverted spectral parameter.}
\begin{align}\label{eq:R-hor}
   R(x) = 
    \gamma^{L_0\otimes 1+1\otimes L_0}
    e^{-\left(
\sum_{k>0}\frac{\gamma^{k}-\gamma^{-k}}{k}  J_{k}\otimes J_{-k} \right)}
     \sum_{n = 0}^\infty 
\frac {(-x)^{n}}{n!} \oint \prod_{i=1}^n \frac{d w_i}{2i\pi w_i } 
 :\prod_{i=1}^n \eta^-(w_i) \otimes \eta^+(w_i):F_n(w),
\end{align}
where the integration is around $|w_1|=\dots=|w_n|=1$, the $q,t$ parameters are such that $|q|<1$ and $|t|>1$ and
\begin{align}
    \label{eq:Fn}
    F_n(w) = F_n(w_1,\dots,w_n) =\kappa^{-n}
    \prod_{1\leq i\neq j\leq n} \zeta\left(\frac{w_i}{w_j}\right),
\end{align}
with $\zeta(x)$ defined in \eqref{eq:zeta}. The horizontal R-matrix is an operator $R(x)\in\End(\CF_u\otimes\CF_{ux})$ which satisfies the Yang--Baxter equation as a consequence of \eqref{eq:YB}
\begin{align}\label{eq:YB-hor}
      R_{23}(x_3/x_2) R_{13}(x_3/x_1) R_{12}(x_2/x_1)=    R_{12}(x_2/x_1) R_{13}(x_3/x_1)  R_{23}(x_3/x_2) .
\end{align}
In addition to this equation, $R(x)$ satisfies two quadratic relations coming from the evaluation of the relations \eqref{eq:X-tilde} and \eqref{eq:Y-tilde}. More precisely, if we apply the horizontal representation in the first tensor space in \eqref{eq:X-tilde} and \eqref{eq:Y-tilde} we will find
\begin{align}\label{eq:RPsi+}
(\gamma^{n}\otimes \Phi^{+}_n)  R_{13}(q^{-n}t^{-1}u)R_{12}(u)&=R_{13}(q^{-n} u)R_{12}(t^{-1} u)(1\otimes \Phi^{+}_n),   \\
\label{eq:RPsi-}
(\gamma^{m}\otimes \Phi_m^{-})  R_{13}(t^m q u)R_{12}(u)&= R_{13}(t^m u)R_{12}(q u) (1\otimes \Phi_m^{-}) .
\end{align}
We note that the normalization of $R(x)$, i.e. its vacuum matrix element $R_{\vac,\vac}^{\vac,\vac}(x)$, plays a role in these equation, in contrast with the Yang--Baxter equation \eqref{eq:YB-hor} from which it drops out.

\paragraph{Derivation of the horizontal refined tau function} The horizontal refined tau function is defined by evaluating its universal counterpart $\hat\tau_{ux}(\bst,\bbst)$ in the horizontal Fock representation
\begin{equation}
\hat\t_x^{(H)}(\bst,\bbst)=\rho_{u}^{(1,0)}\left(\hat\tau_{ux}(\bst,\bbst)\right)\in\End(\CF_u).
\end{equation}
Using the expression \eqref{def_hat_tau} we get\footnote{We replaced $\ket{u x}$ and $\bra{u x}$  with $\ket{\vac}$ and $\bra{\vac}$ since the dependence on the spectral parameters has no effect.}
\begin{equation}\label{eq:tau_hor}
\hat\tau_x^{(H)}(\bst,\bbst)=
\left(1\otimes\bra{\vac}e^{\sum_{k>0} t_kJ_k}\right)
R(x)\left(1\otimes e^{\sum_{k>0}t_{-k}J_{-k}}\ket{\vac}\right).
\end{equation}
Let us now examine the structure of this tau function given the expression  \eqref{eq:R-hor} for the R-matrix. After inserting \eqref{eq:R-hor} into \eqref{eq:tau_hor}, evaluating the expectations and ordering the operators we get
\begin{align}
\label{eq:tau_factors}
    \hat\tau_x^{(H)}(\bst,\bbst) = 
    e^{\sum_{k>0} k \g^{k}t_k t_{-k}}
    \tau_0(\bst) 
    \varphi_x(\bst,\bbst)
\end{align}
where $\tau_0(\bst) =\hat \tau^{(H)}_0(\bst,0)$, or more explicitly
\begin{align}\label{eq:tau0}
    \tau_0(\bst) =  \g^{L_0}  e^{\sum_{k>0}(1-\gamma^{2k}) t_k  J_{k}},
\end{align}
and the operator $\varphi_x(\bst,\bbst)$ is defined by
\begin{align}
\label{eq:phi}
    &\varphi_x(\bst,\bbst) :=
    \sum_{n = 0}^\infty 
\frac {(-x)^{n}}{n!} \oint \prod_{i=1}^n \frac{d w_i}{2i\pi w_i }  :\prod_{i=1}^n \eta^-(w_i;\bst,\bbst):F_n(w),\\
\label{eq:eta_t}
    \text{with}\quad &\eta^-(w;\bst,\bbst):= \eta^-(w) 
    e^{\sum_{k>0}(1-t^{-k})\g^k  w^k t_k}
e^{-\sum_{k>0}(1-q^k)w^{-k} t_{-k}}.
\end{align}
We note that $\varphi_x=\varphi_x(0,0)$ is the generating function of a certain set of commuting Macdonald operators (see \cite{Feigin2009a,FJMM_BA,G_Ng}).  Let us turn to the properties of the operators $\tau_0(\bst)$ and $\varphi_x(\bst,\bbst)$, we have
\begin{align}
\label{eq:tau0_tau0}    &\tau_0(\bst)\tau_0(\bsg\bst')=\tau_0(\bst')\tau_0(\bsg\bst),\\
\label{eq:phi_phi}    
    &[\varphi_x(\bst,\bbst),\varphi_y(\bst,\bbst)] = 0,\\
\label{eq:phi_tau0}     
    &\tau_0(\bst) \varphi_x(\bst,\bbst)=
    \varphi_x(\gamma \bst,\gamma\bbst)\tau_0(\bst),\\
\label{eq:phi_shift}
    &\varphi_x(\bst,\bbst) |_{\eta^-(w;\bst,\bbst)\rightarrow c \eta^-(w;\bst,\bbst)}
=\varphi_{c x}(\bst,\bbst).
\end{align}
The first property is immediate from \eqref{eq:tau0}. The second property \eqref{eq:phi_phi} follows from the fact that
the operator $\eta^-(w;\bst,\bbst)$, entering the definition of $\varphi_x(\bst,\bbst)$ \eqref{eq:phi}, has the same normal ordering relations as $\eta^-(w)$ \eqref{eq:eta_ord}, so for fixed $\bst$ and $\bbst$ the operator $\varphi_x(\bst,\bbst)$ behaves like $\varphi_x$ and the equation follows from the commutativity of the expansion coefficients of $\varphi_x$. The third relation \eqref{eq:phi_tau0} follows from the exchange  relations between \eqref{eq:tau0} and vertex operators \eqref{eq:eta_t} and from rescaling integration variables in \eqref{eq:phi}. Finally, the shift property \eqref{eq:phi_shift} follows from the explicit form of $\varphi_x(\bst,\bbst)$ in \eqref{eq:phi}. 

Now we can recall the bilinear equations \eqref{Hirota_univ} and write them in the horizontal representation
\begin{align}\label{Hirota_hor}
\begin{split}
&\g^{n}\oint_\infty \dfrac{dz}{2i\pi}z^{-n-1}e^{-\sum_{k>0}(\g^k t_k-t_k')\frac{1-t^k}{1-q^k}z^k}\hat\t_{xq^{-n}t^{-1}}^{(H)}(\bst'-[\g^{-2}z^{-1}],\bbst')\hat\t_x^{(H)}(\bst+[\g^{-1}z^{-1}],\bbst)\\
+&\oint_0\dfrac{dz}{2i\pi}z^{-n-1}e^{\sum_{k>0}(\g^kt_{-k}-t_{-k}')\g^{-2k}z^{-k}}\hat\t_{xq^{-n}}^{(H)}(\bst',\bbst'+[z]_{q,t})\hat\t_{xt^{-1}}^{(H)}(\bst,\bbst-[\g z]_{q,t})=0,\\
&\g^{m}\oint_\infty\dfrac{dz}{2i\pi}z^{-m-1}e^{\sum_{k>0}(\g^k t_k-t'_k)z^k}\hat\t_{xqt^m}^{(H)}(\bst'+[z^{-1}]_{t,q},\bbst')\hat\t_{x}^{(H)}(\bst-[\g z^{-1}]_{t,q},\bbst)\\
+&\oint_0\dfrac{dz}{2i\pi}z^{-m-1} e^{-\sum_{k>0}(\g^k t_{-k}-t'_{-k})\frac{1-q^k}{1-t^k}z^{-k}}\hat\t^{(H)}_{xt^m}(\bst',\bbst'-[z])\hat\t_{xq}^{(H)}(\bst,\bbst+[\g z])=0.
\end{split}
\end{align}
The non-trivial part in \eqref{eq:tau_factors} is the operator $\varphi_x(\bst,\bbst)$, therefore we would like to rewrite \eqref{Hirota_hor} in terms of $\varphi_x( \bst,\bbst)$. We take \eqref{Hirota_hor}, substitute \eqref{eq:tau_factors} for the first $\hat \tau^{(H)}$ factor. For the second $\hat \tau^{(H)}$ factor we use \eqref{eq:tau_factors} but this time we commute the $\tau_0$ part to the right using \eqref{eq:phi_tau0}. Then we observe that all factors $\tau_0$ can be removed from the equation. As a consequence of this computation we get
\begin{align}\label{eq:Hirota_phi}
\begin{split}
&
\oint_\infty \dfrac{dz}{2i\pi}z^{-n-1}
K_+(z)
\varphi_{xq^{-n}t^{-1}}(\bst'-[z^{-1}],\bbst')
\varphi_x(\bsg \bst+[\g^{-2}z^{-1}],\bsg\bbst)
\\
+
&\oint_0\dfrac{dz}{2i\pi}z^{-n-1}
K_+(z)
\varphi_{xq^{-n}}(\bst',\bbst'+[\g^{-1}z]_{q,t})
\varphi_{xt^{-1}}(\bsg \bst,\bsg \bbst-[\g z]_{q,t})=0,\\
&\oint_\infty\dfrac{dz}{2i\pi}z^{-m-1}
K_-(z)
\varphi_{xqt^m}(\bst'+[\g^2 z^{-1}]_{t,q},\bbst')
\varphi_{x}(\bsg \bst-[z^{-1}]_{t,q},\bsg\bbst)\\
+
&\oint_0\dfrac{dz}{2i\pi}z^{-m-1} 
K_-(z)
\varphi_{xt^m}(\bst',\bbst'-[\g^{-1}z])
\varphi_{xq}(\bsg \bst,\bsg \bbst+[\g z])=0,
\end{split}
\end{align}
where we rescaled $z\rightarrow z \g^{-1}$ in the second integrals and introduced
\begin{align}\label{eq:K-factors}
\begin{split}
    &K_+(z)=K_+(z;\bst,\bbst,\bst',\bbst')= 
    e^{
    -\sum_{k>0}(\g^k t_k-t_k')\frac{1-t^k}{1-q^k}z^k}
        e^{
    \sum_{k>0}(\g^k t_{-k}-t'_{-k})\g^{-k}z^{-k}},
    \\
    &K_-(z)=K_-(z;\bst,\bbst,\bst',\bbst')=  e^{\sum_{k>0}(\g^k t_k-t_k')z^k}
    e^{
    -\sum_{k>0}(\g^k t_{-k}-t'_{-k})\g^{k} \frac{1-q^k}{1-t^k}z^{-k}}.
\end{split}
\end{align}
In order to simplify further \eqref{eq:Hirota_phi} we need to understand the shifts by the $z$ variable in the arguments of $\varphi$. We compute the shifted operator $\eta^-(w;\bst+\cdots,\bbst+\cdots)$ with the shifts occurring in the first equation in \eqref{eq:Hirota_phi}
\begin{align}\label{eq:eta_shifts}
\begin{split}
&\eta^-(w;\bst-[z^{-1}],\bbst)=t\frac{z-\g w }{t z-\g w} \eta^-(w;\bst,\bbst),\quad \eta^-(w;\bst+[\g^{-2} z^{-1}],\bbst)=t^{-1} \frac{w-t \g z }{w-\g z}\eta^-(w;\bst,\bbst), \\
&\eta^-(w;\bst,\bbst+[\g^{-1} z]_{q,t})=  \frac{z-\g w }{t z-\g w} \eta^-(w;\bst,\bbst),\quad \eta^-(w;\bst,\bbst-[\g z]_{q,t})=\frac{w-t \g z }{w-\g z}\eta^-(w;\bst,\bbst).
\end{split}
\end{align}
Using the shift property \eqref{eq:phi_shift} of $\varphi_x(\bst,\bbst)$ together with \eqref{eq:eta_shifts} gives us 
\begin{align}\label{eq:phiphi}
\begin{split}
& \varphi_{xq^{-n}t^{-1}}(\bst-[z^{-1}],\bbst) = \left[\CA_{xq^{-n}}(\g^{-1}z;\bst,\bbst)\right]_-,
\quad
\varphi_{xq^{-n}}(\bst,\bbst+ [\g^{-1}z]_{q,t})= \left[\CA_{xq^{-n}}(\g^{-1}z;\bst,\bbst)\right]_+,
\\
&\varphi_x(\bst+[\g^{-2}z^{-1}],\bbst)=\left[\CB_{xt^{-1}}(\g z;\bst,\bbst)\right]_-
,\quad 
\varphi_{xt^{-1}}(\bst,\bbst-[\g z]_{q,t})=\left[\CB_{xt^{-1}}(\g z;\bst,\bbst)\right]_+,\\
\end{split}
\end{align}
where the subscripts $\pm$ denote the expansion in powers of $z^{\pm 1}$ of the quantities
\begin{align}
\begin{split}
&\CA_x(z;\bst,\bbst)=\sum_{n = 0}^\infty\dfrac{(-x)^{n}}{n!}\oint\prod_{i=1}^n \frac{d w_i}{2i\pi w_i } \dfrac{z-w_i}{tz-w_i} \times :\prod_{i=1}^n \eta^-(w_i;\bst,\bbst):F_n(w),\\
&\CB_x(z;\bst,\bbst)=\sum_{n = 0}^\infty\dfrac{(-x)^{n}}{n!}\oint\prod_{i=1}^n \frac{d w_i}{2i\pi w_i }\dfrac{tz-w_i}{z-w_i} \times :\prod_{i=1}^n \eta^-(w_i;\bst,\bbst):F_n(w).
\end{split}
\end{align}
Using the observation \eqref{eq:phiphi}, the first reduced bilinear identity in \eqref{eq:Hirota_phi} can be recast in the familiar form
\begin{align}\label{eq:phi_eps_X}
\begin{split}
&
\left(\oint_0 + \oint_\infty\right)  \dfrac{dz}{2i\pi}z^{-n-1}K_+(z)\CA_{xq^{-n}}(\g^{-1}z;\bst,\bbst)\CB_{xt^{-1}}(\g z;\bst,\bbst)=0.
\end{split}
\end{align}
We arrive at the same type of expression as in the previous examples, and this equation holds for the same reason. The second bilinear identity in \eqref{eq:Hirota_phi} can be shown to hold using similar arguments.

\paragraph{Difference-differential equations} The difference-differential equations of the $(q,t)$-deformed hierarchy can be analysed perturbatively in $x$. Using the explicit expression of $\t_0(\bst)$ given in \eqref{eq:tau0}, it is easy to check that the equations derived in Section \ref{sec_equ_diff} indeed hold. In fact, we have checked explicitly that they hold at subleading order as well. We note that simpler difference-differential equations involving $\vphi_x(\bst,\bbst)$ instead of $\hat\t_x^{(H)}(\bst,\bbst)$ can be derived from the reduced bilinear equations \eqref{eq:Hirota_phi}. The equations obtained at order zero in $\e_k$ for $n=0,\pm1$ have the form
\begin{align}
\begin{split}
&\Eq_0^{(\vac,\vac)}:\quad [\vphi_x(\bst,\bbst),\vphi_{xt^{-1}}(\bst,\bbst)]=0,\\
&\Eq_{\mp1}^{(\vac,\vac)}:\quad \p_{\pm1}\vphi_x(\bst,\bbst)\vphi_{p^{\mp1}x}(\bst,\bbst)=p^{\pm1}\vphi_x(\bst,\bbst)\p_{\pm 1}\vphi_{p^{\mp1} x}(\bst,\bbst).
\end{split}
\end{align}
The $(q,t)$-deformed Toda equations also take a simplified form (suppressing the time dependence)
\begin{align}
\begin{split}
&\Eq_{-1}^{(\vac,\sAbox)}:\quad \p_1\p_{-1}\vphi_{xp}\vphi_x-p\p_{-1}\vphi_{xp}\p_1\vphi_x=\g^{-1}\left(\vphi_{xq}\vphi_{xt^{-1}}-\vphi_{xp}\vphi_x\right),\\
&\Eq_{1}^{(\sAbox,\vac)}:\quad \p_1\p_{-1}\vphi_{x}\vphi_{xp}-p^{-1}\p_{1}\vphi_{x}\p_{-1}\vphi_{xp}=\g\left(\vphi_{xt^{-1}}\vphi_{xq}-\vphi_{x}\vphi_{xp}\right).
\end{split}
\end{align}

\section{Discussion}
In this paper, we have introduced a $(q,t)$-deformation of the 2d Toda integrable hierarchy by extending the underlying symmetry algebra q-$W_{1+\infty}$ to the quantum toroidal $\gl(1)$ algebra. Refined tau functions are defined by replacing the elements of the group $\widehat{GL(\infty)}$ in the bosonic correlation function with an element of the quantum toroidal algebra evaluated in the horizontal Fock representation. A $(q,t)$-deformed basic bilinear condition has been written in \eqref{def_Psi_pm} in which the role of the Casimir operator is played by the screening charges of the q-deformed Virasoro algebra, exploiting here the known isomorphism with the representation of level $(2,0)$. From this relation, we derived the two bilinear identities \eqref{Ref_Hirota_2dToda}, which were then expanded to obtain the first difference-differential equations of the hierarchy. Two equations refining the 2d Toda equation were found in this way \eqref{Ref_Toda}, as well as several new equations that trivialize in the self-dual limit $t\to q$. Due to the non-trivial nature of the coproduct of the quantum toroidal $\gl(1)$ algebra, these equations relate a set of tau functions built from different elements of the algebra, instead of a single tau function as it is the case in the original hierarchy. At the moment, it is not clear how the integrability manifests itself in these equation. We expect this question to be answered by the development of a refined Lax formalism which might be derived from the study of the $(q,t)$-Baker-Akhiezer wave functions defined in Appendix \ref{app:Fermions}. This work is currently in progress and we hope to be able to report on it soon. 

Two main classes of tau functions were analyzed as examples of application of the general formalism. The first class is built from Cartan generators which are (almost) co-commutative, it leads to simplified differential equations but the corresponding solution remains deceptively simple. The other class of refined tau functions is built using the R-matrix, exploiting the relation $(1\otimes\D)\CR=\CR_{13}\CR_{12}$ to simplify the bilinear identities. More specifically, two examples were considered, corresponding to the vertical-horizontal and the horizontal-horizontal Fock R-matrices, for which the explicit expressions of the refined tau functions were obtained from known expressions of the R-matrices.

Another important remaining task is to understand the $(q,t)$-deformation of the tau functions known to provide solutions of the original 2d Toda hierarchy. One major difficulty comes from the fact that the algebraic deformation involves the quantum $W_{1+\infty}$ algebra generated by the operators $W_{m,n}=\sum_r q^{-n(s+1/2)}\vdots\bpsi_{m-r}\psi_r\vdots$, instead of the $\widehat{\gl(\infty)}$ generators $\vdots\bpsi_r\psi_s\vdots$. The former can be mapped to elements of the elliptic Hall algebra \cite{Burban2012,FJMM2}, in agreement with the two gradings $(d,\bd)$, but the deformation of the latter is more challenging. In particular, it is not clear at this stage how the soliton solutions built upon the elements\footnote{In this exponential, modes $\psi_r$ are ordered on the right of modes $\bpsi_r$.}
\begin{equation}
G=\exp'\left(\sum_{k=1}^N\a_k\bpsi(z_k)\psi(w_k)\right)
\end{equation} 
can be deformed. However, for some tau functions the situation is better and a deformation can be proposed, like the $(q,t)$-deformation of hypergeometric tau functions introduced in Section \ref{sec_ref_tau}. In this case, the main difficulty lies in the complexity of the coproduct of the operators $b_k$ which takes the form of infinite series, even in the simplest case of $b_1\propto x_0^+$. A possible approach could be to use a stable map to twist the coproduct which effectively performs the rotation $\CS$ \cite{MO2012,Hernandez2022}. We note that a different approach to the $(q,t)$-deformation of hypergeometric tau functions has been initiated recently in \cite{Liu2023} (see also \cite{Alexandrov2011,Mironov2023}), and it would be interesting to understand how this approach fits in our framework.

Our original motivation came from the correspondence between integrable hierarchies and topological strings amplitudes. In this context, the vertical-horizontal R-matrix can be interpreted as the matrix elements of an operator associated to the conifold. More general amplitudes could be studied along the same lines, and in particular those involved in the algebraic construction of non-stationary Ruijsenaars functions (and its various limits) \cite{Shiraishi2019,Fukuda2020,Awata2022}. Besides, similar considerations should also apply to other quantum algebras, like quantum toroidal $\gl(n)$ algebras, or the new Hopf algebras introduced in \cite{Bourgine:2019phm}.

Finally, it would be interesting to study certain limit of the parameters $(q,t)$, and in particular the Hall–Littlewood limit $q\to 0$ with $t$ fixed which should involve the twisted Fock representation of the shifted quantum affine $\sl(2)$ algebra defined recently in \cite{Bourgine2022}. However, it is not clear which operator plays the role of the Casimir in this context. Specializing further the parameters to $(q,t)=(0,-1)$, Macdonald functions reduce to Q-Schur functions which are involved in the BKP hierarchy. Algebraically, the horizontal Fock representation of the quantum toroidal $\gl(1)$ algebra used in this paper appears to reduce to the enveloping algebra of a Majorana fermion through the specialization of the twisted Fock representation of the shifted quantum affine $\sl(2)$ algebra \cite{Bourgine2022}. In this way, we may be able to show that our $(q,t)$-deformed hierarchies interpolate between KP hierarchies for $q=t$ and BKP hierarchies for $(q,t)=(0,-1)$. We hope to be able to come back to this problem in a future publication.\footnote{We would like to thank Sasha Alexandrov for suggesting it to us.} Another important limit is the q-Whittaker limit $t\to0$ with $q$ fixed, in which the deformed hierarchy is expected to be related to the q-Toda system.

\section*{Acknowledgments}
We would like to acknowledge Sasha Alexandrov, Jan de Gier, Yutaka Matsuo, Andrei Mironov, Junichi Shiraishi, and Artem Stoyan for discussions at various stages of the preparation of this paper. We thank the mathematical research institute MATRIX (Australia), program ``Integrable Probability, Combinatorics and Representation Theory'', where part of this research was performed. 

\appendix

\section{Baker-Akhiezer wave functions and $(q,t)$-fermions}\label{app:Fermions}
In this Appendix, we introduce two $(q,t)$-deformations of the vertex operators defining the free fermionic fields $\bpsi(z)$, $\psi(z)$ through the boson-fermion correspondence. To be consistent with the terminology used in this paper, we call these operators the \textit{refined fermions} for short. The idea is to deform the fermionic expression of the Casimir operator
\begin{equation}
\Psi=\oint S(z)dz,\quad S(z)=\bpsi(z)\otimes\psi(z).
\end{equation} 
As explained in Section \ref{sec_qVir}, after the $(q,t)$-deformation the current $S(z)$ is replaced by one of the screening currents $S_\pm(z)$ of the q-Virasoro algebra \eqref{def_Spm}. In fact, these screening currents can also be constructed as the tensor product of the AFS intertwining operators defined in Appendix \ref{app:Intertwiners} \cite{Kimura2015}. Indeed, we can rewrite the vertical components in \eqref{def_AFS} as the following normal-ordered products
\begin{align}
\begin{split}
&\Phi_\l[u,v,n]=t_\l[u,v,n]:\prod_{i=1}^\infty V(vq_1^{i}q_2^{\l_i}):,\quad \Phi_\l^\ast[u,v,n]=t_\l^\ast[u,v,n]:\prod_{i=1}^\infty V^\ast(vq_1^{i}q_2^{\l_i}):,\\
\text{with}\quad
&V(z)=e^{\sum_{k>0}\frac{z^k}k\frac{1-t^k}{1-q^k}J_{-k}}e^{-\sum_{k>0}\frac{z^{-k}}{k}p^kJ_k},\quad  V^\ast(z)=e^{-\sum_{k>0}\frac{z^k}k\frac{1-t^k}{1-q^k}\g^kJ_{-k}}e^{\sum_{k>0}\frac{z^{-k}}{k}p^k\g^k J_k}.
\end{split}
\end{align}
Then, the screening current $S_+(z)$ can be written in the form
\begin{equation}
S_+(z)=\psi_{q,t}(z)\otimes\bpsi_{q,t}(z),
\end{equation} 
with the vertex operators
\begin{equation}
\bpsi_{q,t}(z)=t^{-\hd} z^{\frac{\b-1}{2}+\frac{\log \hat u}{\log q}}V(z),\quad \psi_{q,t}(z)=t^{\hd}z^{\frac{\b-1}2-\frac{\log\hat u}{\log q}}V^\ast(z).
\end{equation}
The identification of zero modes follows from the remark made at the end of Section \ref{sec_qVir}. In the self-dual limit \eqref{sd_lim}, we recover indeed the free fermion $\bpsi_{q,t}(z)\to\bpsi(z)$, $\psi_{q,t}(z)\to\psi(z)$. It is relatively easy to compute the normal-ordering relations for the deformed fermions
\begin{align}
\begin{split}
&\bpsi_{q,t}(z)\bpsi_{q,t}(w)::z^{\b}\Pi_{q,t}(pw/z)^{-1},\quad\psi_{q,t}(z)\psi_{q,t}(w)::z^{\b}\Pi_{q,t}(w/z)^{-1},\quad \bpsi_{q,t}(z)\psi_{q,t}(w)::z^{-\b}\Pi_{q,t}(\g w/z),\\
&\psi_{q,t}(z)\bpsi_{q,t}(w)::z^{-\b}\Pi_{q,t}(\g w/z),\quad \text{with}\quad \Pi_{q,t}(z)=e^{\sum_{k>0}\frac{1-t^k}{1-q^k}\frac{z^k}{k}}=\dfrac{(tz;q)_\infty}{(z;q)_\infty}.
\end{split}
\end{align}
As a result, the anti-commutation relations obeyed by the Dirac fermion are replaced with complicated exchange relations involving theta-functions.


A second $(q,t)$-deformation of the Dirac fermions can be defined starting from the other screening current, namely $S_-(z)$. Its relation with AFS intertwiners follows from a different factorization where the Young diagram $\l$ is replaced by its transposed
\begin{align}
\begin{split}
&\Phi_\l[u,v,n]=t_\l[u,v,n]:\prod_{j=1}^\infty V'(vq_1^{\l'_j}q_2^{j-1}):,\quad \Phi_\l^\ast[u,v,n]=t_\l^\ast[u,v,n]:\prod_{j=1}^\infty V^{\prime\ast}(vq_1^{\l'_j}q_2^{j-1}):,\\
\text{with}\quad
&V'(z)=e^{-\sum_{k>0}\frac{z^k}kJ_{-k}}e^{\sum_{k>0}\frac{z^{-k}}{k}\frac{1-q^k}{1-t^k}J_k},\quad V^{\prime\ast}(z)=e^{\sum_{k>0}\frac{z^k}k\g^kJ_{-k}}e^{-\sum_{k>0}\frac{z^{-k}}{k}\frac{1-q^k}{1-t^k}\g^k J_k}.
\end{split}
\end{align}
Then, the screening current takes the form
\begin{equation}
S_-(z)=\bpsi'_{q,t}(z)\otimes\psi'_{q,t}(z),
\end{equation} 
with the vertex operators
\begin{equation}
\bpsi'_{q,t}(z)=q^{-\bd}z^{\frac{1-\b}{2\b}+\frac{\log\hat u}{\b\log q}}V^{\ast\prime}(z),\quad \psi_{q,t}(z)=q^{\bd}z^{\frac{1-\b}{2\b}-\frac{\log\hat u}{\b\log q}}V'(z).
\end{equation}
Note that the identification of the zero modes is the same as in the case of $S_+(z)$. In the self-dual limit, these refined fermions also reduce to the usual ones, namely $\bpsi'_{q,t}(z)\to\bpsi(z)$ and $\psi'_{q,t}(z)\to\psi(z)$.

\paragraph{Baker-Akhiezer wave function} Using the refined fermions obtained from $S_+(z)$, we define the correlation functions
\begin{align}
\begin{split}
&W^+_u(z|\bst,\bbst|G)=\bra{q u,\vac}e^{\sum_{k>0}t_kJ_k}\psi_{q,t}(z)\rho_u^{(1,0)}(G)e^{\sum_{k>0}t_{-k}J_{-k}}\ket{\vac,u},\\
&\bar W^+_u(z|\bst,\bbst|G)=\bra{q^{-1}u,\vac}e^{\sum_{k>0}t_kJ_k}\bar\psi_{q,t}(z)\rho_u^{(1,0)}(G)e^{\sum_{k>0}t_{-k}J_{-k}}\ket{\vac,u},\\
&W^-_u(z|\bst,\bbst|G)=\bra{qu,\vac}e^{\sum_{k>0}t_kJ_k}\rho_u^{(1,0)}(G)\psi_{q,t}(z)e^{\sum_{k>0}t_{-k}J_{-k}}\ket{\vac,u},\\
&\bar W^-_u(z|\bst,\bbst|G)=\bra{q^{-1}u,\vac}e^{\sum_{k>0}t_kJ_k}\rho_u^{(1,0)}(G)\bar\psi_{q,t}(z)e^{\sum_{k>0}t_{-k}J_{-k}}\ket{\vac,u}.
\end{split}
\end{align}
A short calculation gives
\begin{align}
\begin{split}
&W^+_u(z|\bst,\bbst|G)=z^{\frac{\b-1}{2}-\frac{\log u}{\log q}}e^{-\sum_{k>0}t_k\frac{1-t^k}{1-q^k}\g^{k}z^k}\t_u(\bst+[\g^{-1}z^{-1}],\bbst|G),\\
&\bar W^+_u(z|\bst,\bbst|G)=z^{\frac{\b-1}{2}+\frac{\log u}{\log q}}e^{\sum_{k>0}t_k\frac{1-t^k}{1-q^k}z^k}\t_u(\bst-[\g^{-2} z^{-1}],\bbst|G),\\
&W^-_u(z|\bst,\bbst|G)=z^{\frac{\b-1}{2}-\frac{\log u}{\log q}}e^{\sum_{k>0}t_{-k}\g^{-k}z^{-k}}\t_{ut^{-1}}(\bst,\bbst-[\g z]_{q,t}|G),\\
&\bar W^-_u(z|\bst,\bbst|G)=z^{\frac{\b-1}{2}+\frac{\log u}{\log q}}e^{-\sum_{k>0}p^kt_{-k}z^{-k}}\t_{ut}(\bst,\bbst+[z]_{q,t}|G).
\end{split}
\end{align}
These quantities can be used to rewrite the bilinear identity \eqref{Hir_I} in  the form
\begin{align}
\begin{split}
&\sum_\a\oint_\infty \dfrac{dz}{2i\pi}W_u^+(z|\bst,\bbst|G_\a)\bar W_{uq^{-n}t^{-1}}^+(z|\bst',\bbst'|G_\a')+\sum_\a\oint_0\dfrac{dz}{2i\pi}W_{u}^-(z|\bst,\bbst|G_\a)\bar W_{uq^{-n}t^{-1}}^-(z|\bst',\bbst'|G_\a')=0.
\end{split}
\end{align}
The Baker-Akhiezer wave functions are usually defined as a ratio of the form
\begin{equation}
\dfrac{W_u(z|\bst,\bbst|G)}{\t_u(\bst,\bbst|G)},
\end{equation}
but it is not clear how to introduce these quantities in the general setting with $G_\a\neq G$. However, in the case of the horizontal refined tau function, the summation in the previous equation is replaced by a product of operators,
\begin{align}
\begin{split}
\oint_\infty \dfrac{dz}{2i\pi}\bar W_{uq^{-n}t^{-1}}^+(z|\bst',\bbst')W_u^+(z|\bst,\bbst)+\oint_0\dfrac{dz}{2i\pi}\bar W_{uq^{-n}t^{-1}}^-(z|\bst',\bbst')W_{u}^-(z|\bst,\bbst)=0.
\end{split}
\end{align}
In this case, we can multiply on the right by the operator $\t_u(\bst,\bbst)^{-1}$ and on the left by $\t_{uq^{-n}t^{-1}}(\bst',\bbst')^{-1}$ to write
\begin{equation}
\oint_\infty \dfrac{dz}{2i\pi}\bar\Psi_{uq^{-n}t^{-1}}^+(z|\bst',\bbst')\Psi_u^+(z|\bst,\bbst)+\oint_0\dfrac{dz}{2i\pi}\bar\Psi_{uq^{-n}t^{-1}}^-(z|\bst',\bbst')\Psi_{u}^-(z|\bst,\bbst)=0,
\end{equation} 
with the refined wave functions
\begin{equation}
\Psi_u^\pm(z|\bst,\bbst)=W_u^\pm(z|\bst,\bbst)\t_u(\bst,\bbst)^{-1},\quad \bar\Psi_u^\pm(z|\bst,\bbst)=\t_u(\bst,\bbst)^{-1}\bar W_u^\pm(z|\bst,\bbst).
\end{equation} 

\section{Derivation of the refined bilinear identities}\label{app:Bilin}
\subsection{2d Toda hierarchy}
We reproduce here the derivation of the bilinear identities \eqref{Hirota_2dToda} starting from the basic bilinear condition \eqref{bbc}, and mostly follow the review \cite{Alexandrov2012} (subsection 3.2.2). The basic bilinear condition implies
\begin{align}
\begin{split}\label{equ_H_step1}
&\bra{n+1,n'-1}\left(e^{J_+(\bst)}\otimes e^{J_+(\bst')}\right)\Psi(G\otimes G)\left(e^{J_-(\bbst)}\otimes e^{J_-(\bbst')}\right)\ket{n,n'}\\
=&\left(\bra{n+1,n'-1}\right)\left(e^{J_+(\bst)}\otimes e^{J_+(\bst')}\right)(G\otimes G)\Psi\left(e^{J_-(\bbst)}\otimes e^{J_-(\bbst')}\right)\ket{n,n'},
\end{split}
\end{align}
where we wrote for short $J_+(\bst)=\sum_{k>0} t_kJ_k$, $J_-(\bbst)=\sum_{k>0}t_{-k}J_{-k}$ and $\ket{n,n'}=\ket{n}\otimes\ket{n'}$. The first step is to replace the Casimir $\Psi$ with its expression as a contour integral. In agreement with radial ordering, we choose a contour circling the origin in the r.h.s. and infinity in the l.h.s.
\begin{align}
\begin{split}
&\oint_\infty\dfrac{dz}{2i\pi}\bra{n+1,n'-1}\left(e^{J_+(\bst)}\otimes e^{J_+(\bst')}\right)(\bpsi(z)\otimes\psi(z))(G\otimes G)\left(e^{J_-(\bbst)}\otimes e^{J_-(\bbst')}\right)\ket{n,n'}\\
=-&\oint_0\dfrac{dz}{2i\pi}\bra{n+1,n'-1}\left(e^{J_+(\bst)}\otimes e^{J_+(\bst')}\right)(G\otimes G)(\bpsi(z)\otimes\psi(z))\left(e^{J_-(\bbst)}\otimes e^{J_-(\bbst')}\right)\ket{n,n'}.
\end{split}
\end{align}
Then, the exponentials of times are commuted through the fermionic fields to produce
\begin{align}
\begin{split}\tiny
&\oint_\infty\dfrac{dz}{2i\pi}e^{\sum_{k>0} (t_k-t'_k)z^{k}}\bra{n+1,n'-1}(\bpsi(z)\otimes\psi(z))\left(e^{J_+(\bst)}\otimes e^{J_+(\bst')}\right)(G\otimes G)\left(e^{J_-(\bbst)}\otimes e^{J_-(\bbst')}\right)\ket{n,n'}\\
=&-\oint_0\dfrac{dz}{2i\pi}e^{-\sum_{k>0} (t_{-k}-t'_{-k})z^{-k}}\bra{n+1,n'-1}\left(e^{J_+(\bst)}\otimes e^{J_+(\bst')}\right)(G\otimes G)\left(e^{J_-(\bbst)}\otimes e^{J_-(\bbst')}\right)(\bpsi(z)\otimes\psi(z))\ket{n,n'}.
\end{split}
\end{align}
The action of the fermionic fields on the charged vacua is computed using the bosonization formulas \eqref{bosonization} which gives
\begin{align}
\begin{split}
&\psi(z)\ket{n}=z^{-n}e^{-\sum_{k>0}\frac{z^k}kJ_{-k}}\ket{n-1},\quad \bpsi(z)\ket{n}=z^{n}e^{\sum_{k>0}\frac{z^k}kJ_{-k}}\ket{n+1},\\
&\bra{n}\psi(z)=z^{-n-1}\bra{n+1}e^{\sum_{k>0}\frac{z^{-k}}kJ_{k}},\quad \bra{n}\bpsi(z)=z^{n-1}\bra{n-1}e^{-\sum_{k>0}\frac{z^{-k}}kJ_{k}}.
\end{split}
\end{align}
As a result, the previous integrals reduce to 
\begin{align}
\begin{split}\tiny
&\oint_\infty\dfrac{dz}{2i\pi}z^{n-n'}e^{\sum_{k>0} (t_k-t'_k)z^{k}}\\
&\quad\bra{n,n'}\left(e^{J_+(\bst-[z^{-1}])}\otimes e^{J_+(\bst'+[z^{-1}])}\right)(G\otimes G)\left(e^{J_-(\bbst)}\otimes e^{J_-(\bbst')}\right)\ket{n,n'}\\
=&-\oint_0\dfrac{dz}{2i\pi}z^{n-n'}e^{-\sum_{k>0} (t_{-k}-t'_{-k})z^{-k}}\\
&\quad\bra{n+1,n'-1}\left(e^{J_+(\bst)}\otimes e^{J_+(\bst')}\right)(G\otimes G)\left(e^{J_-(\bbst+[z])}\otimes e^{J_-(\bbst'-[z])}\right)\ket{n+1,n'-1}.
\end{split}
\end{align}
From this expression, it is just a matter of identifying the tau-functions using their definition \eqref{tau_2dToda} to recover the bilinear equations \eqref{Hirota_2dToda}.

\subsection{Refined Toda hierarchy}
To derive the refined bilinear identities \eqref{Ref_Hirota_2dToda}, we sandwich the commutation relation \eqref{com_Psi_rho} between specific states
\begin{align}
\begin{split}
&\bra{u_1', u_2'}\left(e^{J_+(\bst)}\otimes e^{J_+(\bst')}\right)\Psi_\pm\rho_{u_1,u_2}^{(2,0)}(G)\left(e^{J_-(\bbst)}\otimes e^{J_-(\bbst')}\right)\ket{u_1, u_2}\\
=&\bra{u_1', u_2'}\left(e^{J_+(\bst)}\otimes e^{J_+(\bst')}\right)\rho_{u_1',u_2'}^{(2,0)}(G)\Psi_\pm\left(e^{J_-(\bbst)}\otimes e^{J_-(\bbst')}\right)\ket{u_1, u_2},
\end{split}
\end{align}
where we denoted the tensored vacua $\ket{u_1,u_2}=\ket{u_1}\otimes\ket{u_2}$, and $\bra{u_1, u_2}$ the corresponding dual state. The weights $u_1,u_2,u'_1,u'_2$ take the values specified in equation \eqref{weights}. Inserting the expression of the screening charge as an integral of the screening current \eqref{def_Psi_pm} with the contour chosen according to radial ordering, and then normal-ordering the resulting expression, we arrive at
\begin{align}\small
\begin{split}
&-\oint_\infty z^{-n-1}dz e^{-\sum_{k>0}(\g^kt_k-t'_k)r_kz^k}\bra{u_1, u_2}\left(e^{J_+(\bst+[\g^{-1}z^{-1}])}\otimes e^{J_+(\bst'-[\g^{-2}z^{-1}])}\right)\rho_{u_1,u_2}^{(2,0)}(G)\left(e^{J_-(\bbst)}\otimes e^{J_-(\bbst')}\right)\ket{u_1, u_2}\\
=&\oint_0 z^{-n-1}dz e^{\sum_{k>0}(\g^kt_{-k}-t'_{-k})p^kz^{-k}}\bra{u_1', u_2'}\left(e^{J_+(\bst)}\otimes e^{J_+(\bst')}\right)\rho_{u_1',u_2'}^{(2,0)}(G)\left(e^{J_-(\bbst-[\g z]_{q,t})}\otimes e^{J_-(\bbst'+[z]_{q,t})}\right)\ket{u_1', u_2'},
\end{split}
\end{align}
for $\Psi_+$ and
\begin{align}\small
\begin{split}
&-\oint_\infty z^{-m-1}dz e^{\sum_{k>0}(\g^kt_k-t'_k)z^k}\bra{u_1, u_2}\left(e^{J_+(\bst-[\g z^{-1}]_{t,q})}\otimes e^{J_+(\bst'+[z^{-1}]_{t,q})}\right)\rho_{u_1,u_2}^{(2,0)}(G)\left(e^{J_-(\bbst)}\otimes e^{J_-(\bbst')}\right)\ket{u_1, u_2}\\
=&\oint_0 z^{-m-1}dz e^{-\sum_{k>0}(\g^kt_{-k}-t'_{-k})r_k^{-1}z^{-k}}\bra{u_1', u_2'}\left(e^{J_+(\bst)}\otimes e^{J_+(\bst')}\right)\rho_{u_1',u_2'}^{(2,0)}(G)\left(e^{J_-(\bbst+[\g z])}\otimes e^{J_-(\bbst'-[z])}\right)\ket{u_1', u_2'},
\end{split}
\end{align}
for $\Psi_-$. In these expressions, we used the shortcut notation $r_k$ introduced in \eqref{def_rk}. To find the integral equations \eqref{Ref_Hirota_2dToda}, it only remains to insert the expression \eqref{coproduct_G} of the coproduct, and the values \eqref{weights} for the weights.

\section{Vertical Fock representation and AFS intertwiners}\label{app:Intertwiners}
In this Appendix, we provide the information needed for the study of the refined tau functions built from the R-matrix in the Vertical $\otimes$ Horizontal Fock representations.

\subsection{Vertical representation}
In the vertical Fock representation, the algebra $\CE$ acts on a module $\CF_v$ spanned by states $\dket{\l}$ labeled with partitions $\l$. The representation has levels $(0,1)$ and depends on a weight $v\in\mC^{\times}$, it will be denoted $\rho_v^{(0,1)}$. It is defined by the following action of the Drinfeld currents on the basis $\dket{\l}$
\begin{align}\label{def_vert_rep}
\begin{split}
&\rho_v^{(0,1)}(x^+(z))\dket{\l}=\sum_{\sAbox\in A(\l)}\delta(v\chi_\sAbox/z)\res_{w=v\chi_\sAbox}\dfrac1{w\CY_{\lambda}(w)}\dket{\l+\Abox},\\
&\rho_v^{(0,1)}(x^-(z))\dket{\l}=\g^{-1}\sum_{\sAbox\in R(\l)}\delta(v\chi_\sAbox/z)\res_{w=v\chi_\sAbox}w^{-1}\CY_{\l}(q_3^{-1}w)\dket{\l-\Abox},\\
&\rho_v^{(0,1)}(\psi^\pm(z))\dket{\l}=\g^{-1}\left[\dfrac{\CY_{\l}(q_3^{-1}z)}{\CY_{\l}(z)}\right]_\pm\dket{\l},
\end{split}
\end{align}
where the subscript $\pm$ indicates an expansion in powers of $z^{\mp1}$, and $A(\l)$ (resp. $R(\l)$) denotes the set of boxes that can be added to (removed from) the Young diagram of partition $\l$. To each box $\Abox\in\l$ of coordinates $(i,j)$ is associated the box content $\chi_{\sAbox}=q_1^{i-1}q_2^{j-1}\in\mC^{\times}$. The rational function $\CY_\l(z)$ is defined as 
\begin{equation}
\CY_{\l}(z)=\frac{\prod_{\sAbox\in A(\l)}1-\chi_\sAbox/z}{\prod_{\sAbox\in R(\l)}1-\chi_\sAbox/(q_3z)}.
\end{equation} 

Using the automorphism $\CS$ introduced by Miki in \cite{Miki2007}, it can be shown that the horizontal and vertical Fock representations are isomorphic, $\rho_u^{(1,0)}\circ\CS\simeq\rho_v^{(0,1)}$, with the relation $u=-\g v$ between the weights. The isomorphism between the horizontal and vertical modules $\CF_u$ and $\CF_v$ is defined by the identification of the states $\dket{\l}$ with the Macdonald states $\ket{P_\l,u}\in\CF_u$ defined in \ref{sec_ref_tau} up to a normalization factor (see \cite{Bourgine2018a}).

The vertical representation of the grading operator is the opposite of the horizontal one since Miki's automorphism acts as $\CS(d,\bd)\to (-\bd,d)$
\begin{equation}
\rho_v^{(0,1)}(\bd)\dket{\l}=|\l|\dket{\l},\quad \rho_v^{(0,1)}(d)=-v\p_v.
\end{equation} 

The contragredient action of the algebra on the states $\dbra{\l}$ is obtained from the scalar product
\begin{equation}
\dbra{\l}\!\!\!\dket{\mu}=(n_\l)^{-1}\d_{\l,\mu},\quad n_\l=(-\g)^{-|\l|}\CN_{\l,\l}(1)^{-1}\prod_{\sAbox\in\l}\chi_\sAbox,
\end{equation}
where the normalization factor $n_\l$ is expressed in using the Nekrasov factor 
\begin{equation}\label{def_Nek}
\CN_{\l,\mu}(\a)=\prod_{\superp{(i,j)\in\l}{(i',j')\in\mu}}\dfrac{(1-\a q_1^{i-i'+1}q_2^{j-j'})(1-\a q_1^{i-i'}q_2^{j-j'+1})}{(1-\a q_1^{i-i'}q_2^{j-j'})(1-\a q_1^{i-i'+1}q_2^{j-j'+1})}\times\prod_{(i,j)\in\l}(1-\a q_1^iq_2^j)\times\prod_{(i',j')\in\mu}(1-\a q_1^{1-i'}q_2^{1-j'}).
\end{equation}

\subsection{Intertwiners}
The AFS intertwiners introduced in \cite{AFS} (see also \cite{Bourgine2017b,Bourgine2018a}) are determined uniquely, up to an overall normalization factor, by the intertwining properties for $e\in\CE$
\begin{align}\label{prop_intw}
\begin{split}
&\Phi\left(\rho_{v}^{(0,1)}\otimes\rho_u^{(1,n)}\ \D(e)\right)=\rho_{u'}^{(1,n+1)}(e)\Phi,\\
&\Phi^\ast\rho_{u'}^{(1,n+1)}(e)=\left(\rho_{v}^{(0,1)}\otimes\rho_u^{(1,n)}\ \D^{op}(e)\right)\Phi^\ast,
\end{split}
\end{align}
with the constraint $u'=-\g uv$ and $\D^{op}$ denoting the opposite coproduct. The solution of these equations can be expanded in the vertical basis as follows
\begin{equation}\label{def_Phi}
\Phi[u,v,n]=\sum_{\l} n_\l \dbra{\l}\otimes \Phi_{\l}[u,v,n],\quad \Phi^\ast[u,v,n]=\sum_{\l}n_\l\dket{\l}\otimes\Phi_{\l}^\ast[u,v,n].
\end{equation}
The vertical components are vertex operators built as normal-ordered products of the vertex operators $\eta^\pm(z)$ defined in \eqref{def_eta}
\begin{align}\label{def_AFS}
\begin{split}
&\Phi_\l[u,v,n]=t_{\l}[u,v,n]\ :\Phi_\vac(v)\prod_{\sAbox\in\l}\eta^+(v\chi_\sAbox ):,\quad \Phi_{\l}^\ast[u,v,n]=t_{\l}^\ast[u,v,n]\ :\Phi_\vac^\ast(v)\prod_{\sAbox\in\l}\eta^-(v\chi_\sAbox ):,\\
\text{with:}\quad &t_\l[u,v,n]=(-\g u)^{|\l|}v^{-n|\l|}\prod_{\sAbox\in\l}\chi_\sAbox^{-n-1},\quad t_{\l}^\ast[u,v,n]=u^{-|\l|}\g^{-|\l|}v^{n|\l|}\prod_{\sAbox\in\l}\chi_\sAbox^{n},\\
&\Phi_\vac(v)=e^{-\sum_{k>0}\frac{v^k}{k(1-q_2^k)}J_{-k}}e^{\sum_{k>0}\frac{v^{-k}q_3^{-k}}{k(1-q_1^{k})}J_k},\quad
\Phi_\vac^\ast(v)=e^{\sum_{k>0}\frac{\g^k v^k}{k(1-q_2^k)}J_{-k}}e^{-\sum_{k>0}\frac{v^{-k}\g^{-k}}{k(1-q_1^{k})}J_k}.
\end{split}
\end{align}

\paragraph{Gradings and zero modes} The intertwining relations also apply to the grading elements $d$ and $\bd$. Once projected on the vertical components, we deduce the following relations that can also be derived directly using the expressions \eqref{def_AFS}
\begin{align}
\begin{split}
&\a^{L_0}\Phi_\l[u,v,n]\a^{-L_0}=\a^{n|\l|}\Phi_\l[u,pv,n],\quad \a^{L_0}\Phi_\l^\ast[u,v,n]\a^{-L_0}=\a^{-n|\l|}\Phi_\l^\ast[u,pv,n],\\
&\a^{\hd}\Phi_\l[u,v,n]\a^{-\hd}=\a^{|\l|}\Phi_\l[u,v,n],\quad \a^{\hd}\Phi_\l^\ast[u,v,n]\a^{-\bd}=\a^{-|\l|}\Phi_\l^\ast[u,v,n].
\end{split}
\end{align}
Promoting horizontal weights to operators, we can replace the normalization factors $t_\l[u,v,n]$ and $t^\ast_\l[u,v,n]$ by the following operators
\begin{equation}
\hat t_\l[v,n]=(-\g v)^{\hd}(-\g \hat u)^{|\l|}v^{-n|\l|}\prod_{\sAbox\in\l}\chi_\sAbox^{-n-1},\quad \hat t_{\l}^\ast[v,n]=\hat u^{-|\l|}(-\g v)^{-\hd}\g^{-|\l|}v^{n|\l|}\prod_{\sAbox\in\l}\chi_\sAbox^{n},\\
\end{equation}
where the factors $(-\g v)^{\pm \hd}$ are inserted to fulfill the condition $u'=-\g uv$.

\section*{Declarations}
\paragraph{Fundings}  The authors gratefully acknowledge support from the Australian Research Council.

\paragraph{Competing interests} The authors have no competing interests to declare that are relevant to the content of this article.

\paragraph{Data} Data sharing not applicable to this article as no datasets were generated or analysed during the current study.

\bibliographystyle{utphys}
\bibliography{Refined_IH}
\end{document}